\documentclass[12pt]{article}
\usepackage[utf8]{inputenc}
\usepackage[english]{babel}
\usepackage[top = 2 cm, bottom = 2 cm, left = 2.5 cm, right = 1.5 cm]{geometry}
\usepackage[]{amsmath}
\usepackage[pdftex]{graphicx}
\graphicspath{ {images/} }
\usepackage{float}
\restylefloat{figure} 
\usepackage{subfig}
\usepackage{indentfirst}
\usepackage{amsthm}
\usepackage{amsfonts}
\usepackage{mathrsfs}
\usepackage{mathtools}
\usepackage{cmap}
\usepackage{exercise}
\usepackage{xcolor}
\usepackage{breqn}
\usepackage{cancel}
\usepackage{amsfonts}
\usepackage{comment}
\usepackage{tcolorbox}
\usepackage[pdftex,unicode,
colorlinks=true,
linkcolor = blue]{hyperref}
\usepackage[offset=1.2em]{simpler-wick}
\usepackage[compat=1.1.0]{tikz-feynman}
\usepackage{tikz-3dplot}
\usetikzlibrary{arrows}
\usetikzlibrary{arrows.spaced}
\usetikzlibrary{decorations.pathmorphing}
\usetikzlibrary{snakes}
\usetikzlibrary{shadows}
\usetikzlibrary{decorations.pathreplacing,decorations.markings}
\usetikzlibrary {fadings,patterns}
\usepackage[bulletsep]{collref}

\def\period{\,.}
\def\comma{\,,}
\def\pmatrix#1#2{\left( 
\begin{array}{#1}
#2\end{array} 
\right)}

\usepackage[colorinlistoftodos]{todonotes}%allows to do colorful comments
\newcommand{\mathbbm}[1]{\text{\usefont{U}{bbm}{m}{n}#1}}

\usepackage{marginnote}

\definecolor{nicklethistwink}{rgb}{0.69, 0.19, 0.38}
\definecolor{ticklemepink}{rgb}{0.99, 0.54, 0.67}
\definecolor{mistyrose}{rgb}{1.0, 0.89, 0.88}
\definecolor{lolololol}{rgb}{0.25, 0.5, 0.75}
\definecolor{upsdellred}{rgb}{0.68, 0.09, 0.13}
\definecolor{champagne}{rgb}{0.97, 0.91, 0.81}
\definecolor{ghostwhite}{rgb}{0.97, 0.97, 1.0}
\definecolor{ivory}{rgb}{1.0, 1.0, 0.94}
\definecolor{bubblegum}{rgb}{0.99, 0.76, 0.8}
\definecolor{mintcream}{rgb}{0.96, 1.0, 0.98}
\definecolor{honeydew}{rgb}{0.94, 1.0, 0.94}
\definecolor{magnolia}{rgb}{0.97, 0.96, 1.0}
\definecolor{isabelline}{rgb}{0.96, 0.94, 0.93}
\definecolor{bubbles}{rgb}{0.91, 1.0, 1.0}
\definecolor{floralwhite}{rgb}{1.0, 0.98, 0.94}
\definecolor{turquoise}{rgb}{0.19, 0.84, 0.78}
\definecolor{bittersweet}{rgb}{1.0, 0.44, 0.37}
\definecolor{ultramarineblue}{rgb}{0.25, 0.4, 0.96}

\numberwithin{equation}{section}

\renewcommand{\O}{\mathcal{O}}
\def\tr{{\rm tr}\,}

\def\beq{\begin{equation}}
\def\eeq{\end{equation}}

\newcommand{\dk}[1]{\frac{d^d#1}{(2\pi )^d}\,}

\begin{document}
	\title{\vspace{0.1cm}{\Large {\bf 
				Vacuum Condensates on the Coulomb Branch}}}\vspace{10pt}
		\author{{\small Vyacheslav Ivanovskiy,\,\, Shota Komatsu,\,\, Victor Mishnyakov,}\\ {\small  Nikolay Terziev,\,\, Nikita Zaigraev,\,\, Konstantin Zarembo}\vspace{10pt}\\		
			{}\date{ }
		}
	\maketitle
	
	\vspace{-5.9cm}
	
	\begin{center}
		\hfill \\
	\end{center}
	
	\vspace{3.5cm}
	
	\begin{center}
		
		 %{\small {\it $^1$ Moscow Institute of Physics and Technology, 141701, Dolgoprudny, Moscow region, Russia}}\\
 %{\small {\it $^2$ Institute for Theoretical and Mathematical Physics, Lomonosov Moscow State University, 119991 Moscow, Russia}}\\
 %{\small {\it $^3$ Department of Theoretical Physics, CERN, 1211 Meyrin, Switzerland}}\\ 
 %{\small {\it $^4$Nordita, KTH Royal Institute of Technology and Stockholm University,}}\\
 %{\small {\it Hannes Alfv\'ens v\"ag 12, SE-106 91 Stockholm, Sweden}}\\
 %{\small {\it $^5$ ITEP, Moscow 117218, Russia}}\\
  %{\small {\it $^6$ Bogoliubov Laboratory of Theoretical Physics, JINR, 141980 Dubna, Moscow region, Russia}}\\
   %{\small {\it $^7$ Niels Bohr Institute, Copenhagen University, Blegdamsvej 17, 2100 Copenhagen, Denmark}}\\
% {\small {\it $^6$ Lebedev Physics Institute, Moscow 119991, Russia}}\\

	\end{center}
	
	\vspace{1cm}
	
	\begin{abstract}
		We study correlation functions on the Coulomb branch
		of planar $\mathcal{N}=4$ super-Yang-Mills theory (SYM), and their relationship with integrability, the operator product expansion (OPE), the sum rule, the large charge expansion, and holography. First, we compute one-point functions of arbitrary scalar operators at weak coupling and derive a compact
		spin-chain representation. We next study the two-point functions of chiral primaries at one loop and find that the radius of convergence of OPE in position space is infinite. We  estimate the asymptotic growth of the OPE data based on this finding. Finally, we propose a concrete nonperturbative formula that connects the correlation functions on the Coulomb branch with the correlation functions with large charge insertions at the conformal point and provide a holographic interpretation based on a large D3-brane in AdS. The formula extends the known connection between the large charge sector and the Coulomb branch for rank-1 theories to the large $N$ limit. %The results shed light on the interplay between various aspects of planar $\mathcal{N}=4$ SYM and provides new insights into the correlation functions on the Coulomb branch.
	\end{abstract}
	
	\vspace{.5cm}  
	
	\newpage
	
	\tableofcontents
	\setlength{\parskip}{0.5em}
	\section{Introduction}
	
	Vacuum condensates --- or equivalently one-point functions of local operators --- are an important attribute of any field theory. The QCD sum rules, for instance, rely on parametrizing
	non-perturbative effects by vacuum condensates through the operator product expansion \cite{Shifman:1978bx,Shifman:1978by,Shifman:1978bw}\footnote{See \cite{Caron-Huot:2023tpw} for a recent application of the sum rule to three-dimensional Yang-Mills theory, and \cite{Marino:2024uco} for the analysis of the interplay between the sum rule and the resurgence in two dimensions.}. They also determine the large order behavior of a class of Feynman diagrams called infrared renormalons (see \cite{Beneke:1998ui,Dunne:2015eoa} and references therein), for which a new interest has emerged in recent years because of their possible relation to the theory of resurgence. For reviews of the subject, see \cite{Marino:2012zq,Dorigoni:2014hea,Aniceto:2018bis}.
	
	Despite their importance, the calculation of vacuum condensates in interacting quantum field theories (QFTs) is notoriously difficult. Even in perturbation theory, it requires careful renormalization of operators. At the non-perturbative level, one needs to understand the structure of the vacuum first, which is by itself a challenging problem.
	
	One way of making progress is to study QFTs with spontaneously broken conformal symmetry. 
	The vacuum structure of such theories is more controllable thanks to the (broken) conformal invariance. For instance, the breaking of the conformal symmetry predicts the existence of a massless particle, namely the dilaton, whose coupling to local operators is determined by the Ward identity. Moreover the renormalization of operators in these theories is simpler: As we will see in this paper, once the operators are renormalized at the ultraviolet (UV) fixed point, no extra renormalization along the renormalization group (RG) flow is necessary.
	
	Theories with spontaneously broken conformal symmetry are ideal testing grounds for other fundamental questions on QFTs as well. One example is the convergence of operator product expansion (OPE): In conformal field theory, the OPE is a powerful tool because it converges at finite radius. In massive QFTs, however, an analogous statement has not been established. In fact it is nontrivial even to formulate this question since, in generic massive QFTs, operators themselves flow under the change of the scale. In contrast, the question can be sharply posed in theories with spontaneously broken conformal symmetry, owing to the absence of extra renormalizations as mentioned earlier.
	
	With these motivations in mind, in this paper we study a special, but arguably the simplest quantum field theory with spontaneous conformal symmetry breaking---namely planar $\mathcal{N}=4$ supersymmetric Yang-Mills (SYM) theory on the Coulomb branch. In the following we provide four additional reasons for this particular choice.
	
	Firstly it provides the simplest example of non-conformal holography beyond AdS/CFT. Once the gauge symmetry of $\mathcal{N}=4$ SYM
is broken down to its subgroup, the Higgs condensate sets the mass scale and gives expectation values to generic operators that are not protected by unbroken global symmetries. Previously, such one-point functions were studied using holography for the maximally broken gauge symmetry: $U(N)\rightarrow U(1)^N$ \cite{Skenderis:2006uy,Skenderis:2006di}. In this paper, we focus on a different symmetry-breaking pattern, $U(N)\rightarrow U(N-1)\times U(1)$, which
	entails a different large-$N$ counting and leads to a different holographic setup (a single brane as opposed to a smooth distribution of branes), requiring a separate analysis on the gravity side.
	
	Secondly observables on the Coulomb branch of $\mathcal{N}=4$ SYM are potentially amenable to the integrability machinery. Establishing the integrability on the Coulomb branch is an important first step towards the application of the integrability techniques to non-conformal setups. There have already been a plenty of indirect evidence for this coming from string theory \cite{Dekel:2011ja}, from the analysis of scattering amplitudes \cite{Alday:2009zm,Loebbert:2020hxk,Loebbert:2020tje} and correlation functions \cite{Caron-Huot:2021usw}, and from a remarkable insight into the nature of the bound-state spectrum of W-bosons \cite{Caron-Huot:2014gia}. More recently, it was shown that mathematical structures found in perturbative amplitudes of massless gluons in $\mathcal{N}=4$ SYM admit a deformation which can describe amplitudes on the Coulomb branch \cite{Arkani-Hamed:2023epq}. In addition, integrability tools proved highly efficient for one-point functions in related setups of defect CFT \cite{deLeeuw:2015hxa,Buhl-Mortensen:2015gfd,Buhl-Mortensen:2017ind,Komatsu:2020sup,Gombor:2020kgu} and heavy-heavy-light three-point functions where heavy operators effectively act as a background for the light insertion \cite{Jiang:2019xdz,Jiang:2019zig,Yang:2021hrl}.  Implications for vacuum condensates are currently under investigation \cite{progress:2021}, and we hope that our results serve as a stepping stone for future developments.
	
	Thirdly there have been a series of interesting recent works on the relationship between the Coulomb-branch effective action and operators with large charge in $\mathcal{N}=2$ superconformal field theory (SCFT) \cite{Hellerman:2017veg,Hellerman:2017sur,Hellerman:2018xpi}. The underlying physical mechanism is that the operators with large charge effectively act as a source for scalar fields and put the theory on the Coulomb branch. This intuitive picture was substantiated in \cite{Hellerman:2017sur}, in which they systematically analyzed the Coulomb-branch effective field action and demonstrated a precise agreement between the results from the effective field theory and the results from supersymmetric localization. However, the analysis in the literature is limited so far to rank-$1$ theories and mostly to BPS operators except for \cite{Caetano:2023zwe}, which analyzed non-BPS operators in rank-1 $\mathcal{N}=4$ SYM. In this paper, we discuss generalization to non-BPS operators in large $N$ $\mathcal{N}=4$ SYM. In particular, we propose a concrete formula relating the correlation functions with two large charge insertions at the conformal point and the correlation functions on the Coulomb branch.
	
	Finally, studying correlation functions and OPE on the Coulomb branch potentially sheds light on the structure of the vacuum manifold in (super)conformal field theories. To appreciate this point, let us consider the two-point function on the vacuum manifold $\langle \mathcal{O}_i(x)\mathcal{O}_j(0)\rangle_{\rm vac}$. In the long distance, the two-point function can be approximated by a product of one-point functions and the dilaton exchange,
	\begin{align}
	    \langle \mathcal{O}_i(x)\mathcal{O}_j(0)\rangle_{\rm vac}=\langle \mathcal{O}_i\rangle_{\rm vac}\langle \mathcal{O}_j\rangle_{\rm vac}+\frac{\#}{x^2}+\cdots\comma
	\end{align}
	while in the short distance it can be expanded into the OPE
	\begin{align}
	    \langle \mathcal{O}_i(x)\mathcal{O}_j(0)\rangle_{\rm vac}=\sum_{k}c_{ijk}(vx)^{\Delta_k-\Delta_i-\Delta_j}\langle \mathcal{O}_k\rangle_{\rm vac}\comma
	\end{align}
	where $c_{ijk}$ is a structure constant.
	As was pointed out in \cite{Karananas:2017zrg}, the equality between the two expressions imposes an infinite set of constraints on possible values of $\langle \mathcal{O}_i\rangle_{\rm vac}$. A priori, it is not clear if a solution to these constraints exists at all. In fact, it is likely that a solution does not exist for generic CFTs since all the known CFTs with a vacuum manifold are rather special; namely they are either free or supersymmetric. At present we do not have a clear understanding of what specific features of the CFT data ($c_{ijk}$ and $\Delta_j$) allow for a solution to exist. As a first step in answering this question, we derive a necessary condition on the asymptotic growth of the OPE data in section \ref{sec:OPE}.
	
	Let us also emphasize that the interplay between the OPE and the vacuum manifold is not fully understood even in supersymmetric theories. Take $\mathcal{N}=1$ SCFTs in four dimensions as an example. It is known that the coordinates of the vacuum manifold in $\mathcal{N}=1$ SCFTs are given by the chiral rings. This fact, although well-known, is a highly nontrivial and profound statement since it implies that the solutions to the infinitely many OPE constraints are completely parametrized by a finite set of data provided by the chiral ring operators. A clear understanding of this mechanism is still lacking and we hope that the detailed analysis of the correlation functions on the Coulomb branch in $\mathcal{N}=4$ SYM could help to make progress in the future.
	
	The rest of the paper is organized as follows: In {\bf section} \ref{sec:general}, we explain the setup and discuss the absence of extra renormalizations of operators along the RG flow. In {\bf section} \ref{sec:weak}, we compute the one-point functions of single-trace operators at weak coupling. We derive a general expression up to one-loop for arbitrary scalar operators and compute them explicitly for a class of operators (including the Konishi operator), for which the mixing with fermions at two loops is well-understood. In {\bf section} \ref{sec:strong}, we compute the one-point function of chiral primary operators (CPOs) at strong coupling using a probe brane and confirm the non-renormalization of one-point functions of CPOs. In {\bf section} \ref{sec:twopnt}, we compute the two-point functions on the Coulomb branch at the leading and the next-leading orders. We mostly focus on the two-point functions of CPOs but we also present some results for Konishi operators. We then check the consistency of the result with the OPE and the dilaton Ward identity, and find that the dilaton decay constant is not renormalized at one loop. In {\bf section} \ref{sec:largecharge}, we discuss the connection to the large charge expansion and propose a concrete formula relating the correlation functions on the Coulomb branch and the correlation functions with large charge insertions. We also discuss general properties of OPE on the Coulomb branch based on the results for the two-point function. In {\bf section} \ref{sec:OPE},  we show that the radius of convergence of OPE is infinite in position space, up to the order we performed the computation. Based on this, we provide an estimate for the asymptotic growth of the OPE data. Finally in {\bf section} \ref{sec:conclusion}, we conclude and discuss future directions. Several appendices are included to explain technical details.
	
	\section{Generalities}\label{sec:general}
	%\kt{move into appendix conventions:
	
	%where $(\Gamma ^\mu ,\Gamma ^i)$ form %ten-dimensional Clifford algebra; and their %explicit form?
	%}
	\subsection{Action, gauge fixing and operators}
	Let us explain the general setup and set the conventions used in this paper.
	\paragraph{Action and propagators.}
	The action of $\mathcal{N}=4$ SYM (in the Minkowski signature) is given by
	\begin{equation}
		\begin{split}
			\mathcal{L}=&\frac{1}{g_{\mathrm{YM}}^{2}} \operatorname{tr}\left(-\frac{1}{2} F_{\mu \nu} F^{\mu \nu}+\mathrm{D}_{\mu} \phi_{i} \mathrm{D}^{\mu} \phi_{i}+\frac{1}{2}\left[\phi_{i}, \phi_{j}\right]\left[\phi_{i}, \phi_{j}\right]+i \bar{\psi} \Gamma^{\mu} \mathrm{D}_{\mu} \psi +\bar{\psi}\Gamma^{i}\left[\phi_{i}, \psi\right]\right)\period
		\end{split}
	\end{equation}
	Here $\mu, \nu = 0,\ldots,3$, $i,j = 1,\ldots,6$, $(\Gamma ^\mu ,\Gamma ^i)$ form ten-dimensional Clifford algebra, $\psi$ is a ten-dimensional Majorana-Weyl fermion. 
	
	Since the scalar potential is given by the commutator squared, one can give constant vacuum expectation values (VEVs) to $\phi_i$'s as long as they commute with each other. This provides a canonical example of the spontaneous breaking of conformal symmetry. In the rest of this paper, we consider the following VEVs,
	\begin{equation}\label{higgsvev}
		\phi^{\rm cl}_i =
		\left(\begin{matrix}
			v_i & 0 & \ldots & 0 \\
			0 & 0 &  &0 \\
			\vdots &  & \ddots &0 \\
			0 & 0 & 0 &0
		\end{matrix}\right) \in U(N), \qquad  \sum\limits_{i=1}^6 v_i^2=v^2
	\end{equation}
	which put the theory on the Coulomb branch with $U(N)$  Higgsed to $U(N-1)\times U(1)$. In addition, we take the large $N$ limit, in which the rank of the gauge group is sent to infinity while the 't Hooft coupling $\lambda$ is kept finite:
	\begin{align}
	    \lambda \equiv g_{\rm YM}^2 N :\text{ fixed}\comma\qquad N\to \infty\period
	\end{align}
	
	In order to analyze the correlation functions on this background, we expand the action around the classical Higgs VEV,
	\begin{equation}
		\phi_i = \phi_i^{\rm cl}+\tilde{\phi}_i\comma
	\end{equation}
	and gauge-fix the Lagrangian using the $R_{\xi=1}$ gauge with the background gauge-fixing condition,
	\begin{align}\label{eq:gaugefix}
	    G=\partial_{\mu}A^{\mu}-i[\phi^{\rm cl}_i,\tilde{\phi}_i]\period
	\end{align}
	Let us make two comments. First, as we see shortly, the gauge-fixing condition \eqref{eq:gaugefix} kills various unwanted terms in the Lagrangian and drastically simplifies the perturbative analysis. This was already realized in \cite{Alday:2009zm}, which analyzed scattering amplitudes on the Coulomb branch. Second one can rewrite the condition in a more suggestive form 
	\begin{align}\label{eq:IKKT}
	G=-i\left([A_{\mu}^{\rm cl},A^{\mu}]+[\phi_i^{\rm cl},\tilde{\phi}_i]\right) \period    
	\end{align}
	with $A_{\mu}^{{\rm cl}}\equiv i\partial_{\mu}$.
	In this notation, the combination $A_{\mu}^{\rm cl}+A_{\mu}$ becomes proportional to the covariant derivative ${\rm D}_{\mu}$ and transforms as the adjoint representation of the gauge group, much like $\phi_i=\phi_i^{\rm cl}+\tilde{\phi}_i$. This structure is reminiscent of the Eguchi-Kawai mechanism \cite{Eguchi:1982nm}, which relates a higher-dimensional gauge theory to a matrix model\footnote{See \cite{Young:2014jma,Shaghoulian:2016xbx} for recent discussions on the Eguchi-Kawai mechanism in the context of holography.}.  In fact, the condition \eqref{eq:IKKT} can be viewed as a natural background gauge condition for the IKKT matrix model \cite{Ishibashi:1996xs}.

	The resulting Lagrangian contains the cubic and quartic interaction vertices,
	\begin{equation}\label{eq:cubicvert}
		\begin{split}
			S&{}_{\text {cubic }}=\frac{2}{g_{\mathrm{YM}}^{2}} \int  \mathrm{d}^{4} x  \operatorname{tr}\left[ i\left[A^{\mu}, A^{\nu}\right] \partial_{\mu} A_{\nu}+\left[\phi_{i}^{\rm cl}, \tilde{\phi}_{j}\right]\left[\tilde{\phi}_{i}, \tilde{\phi}_{j}\right]+i\left[A^{\mu}, \tilde{\phi}_{i}\right] \partial_{\mu} \tilde{\phi}_{i}+\right.
			\\
			&\left. +\left[A_{\mu}, \phi_{i}^{\mathrm{cl}}\right]\left[A^{\mu}, \tilde{\phi}_{i}\right]+\frac{1}{2} \bar{\psi} \gamma^{\mu}\left[A_{\mu}, \psi\right]+\dfrac{1}{2}\bar{\psi} \Gamma^i\left[\tilde{\phi}_i , \psi \right]+i\left(\partial_{\mu} \bar{c}\right)\left[A^{\mu}, c\right]-\bar{c}\left[\phi_{i}^{\mathrm{cl}},\left[\tilde{\phi}_{i}, c\right]\right]   \right]\comma
		\end{split}
	\end{equation}
	\begin{equation}
		S_{\text {quartic }}=\frac{2}{g_{\mathrm{YM}}^{2}} \int \mathrm{d}^{4} x \operatorname{tr}\left[\frac{1}{4}\left[A_{\mu}, A_{\nu}\right]\left[A^{\mu}, A^{\nu}\right]+\frac{1}{2}\left[A_{\mu}, \tilde{\phi}_{i}\right]\left[A^{\mu}, \tilde{\phi}_{i}\right]+\frac{1}{4}\left[\tilde{\phi}_{i}, \tilde{\phi}_{j}\right]\left[\tilde{\phi}_{i}, \tilde{\phi}_{j}\right]\right]\comma
	\end{equation}
	while the gauge-fixing term gives
	\begin{equation}\label{eq:gaugefixterms}
		\mathcal{L}_{\mathrm{gf}}=\frac{2}{g_{\mathrm{YM}}^{2}}  \operatorname{tr}
		\left(\bar{c}\,\partial_{\mu} \mathrm{D}^{\mu} c-\bar{c}\left[\phi_{i}^{\mathrm{cl}},\left[{\tilde{\phi}}_{i}+\phi_{i}^{\mathrm{cl}}, c\right]\right]-\frac{\left(\partial_{\mu} A^{\mu}\right)^{2}}{2} +i\left[A^{\mu}, \partial_{\mu} {\tilde{\phi}}_{i}\right] \phi_{i}^{\mathrm{cl}}+\frac{1}{2}\left[\phi_{i}^{\mathrm{cl}}, {\tilde{\phi}}_{i}\right]^{2}\right)\period
	\end{equation}
	
	Owing to the scalar VEV and the gauge-fixing term, the ``off-diagonal" fields, namely fields with color indices $1k$ or $k1$  with $k\neq 1$ become massive. For instance, among the $6\times N^2$ scalar fields, 
	$12(N-1)$ scalar fields become massive:
	\begin{equation}\label{eq:structuremass}
		\mathcal{L}_{\mathrm{scalar \ quadratic}} = \frac{1}{2}\sum_{k=1}^N \tilde{\phi}_{kk}^{*j} (\partial_\nu\partial^\nu) \tilde{\phi}_{kk}^j+ \sum_{k=2}^N \tilde{\phi}_{1k}^{*j} (\partial_\nu\partial^\nu - v^2) \tilde{\phi}_{1k}^j + \sum_{1<l<k}^N \tilde{\phi}_{lk}^{*j} (\partial_\nu\partial^\nu) \tilde{\phi}_{lk}^j\period
	\end{equation}
	This is the case also for vectors, fermions and ghosts. The fields with indices $11$ remain massless\footnote{The distinction between different indices will be important also for the large $N$ counting. For example, a loop of massless fields will come with an additional factor of $N$ compared to a loop of massive fields.} and constitute the dilaton multiplet of the spontaneously broken scale symmetry. In particular, among the six scalar fields with indices $11$, the one along the direction of the VEV is the dilaton while the remaining five are the Goldstone bosons of the spontaneously broken R-symmetry.
	
	Let us also emphasize the importance of the gauge-fixing term \eqref{eq:gaugefixterms}. Without it, the ghosts and some of the off-diagonal scalars would remain massless and the scalar propagators would break the $SO(6)$ symmetry. What is worse, there would be a kinetic mixing between the scalars and the gauge fields. All these unwanted features disappear and the scalar propagators recover the $SO(6)$ symmetry once we choose the gauge-fixing condition \eqref{eq:gaugefix} and include the gauge-fixing term \eqref{eq:gaugefixterms} in the Lagrangian. These are the simplifications we alluded to above.

	The propagator of the massless scalars is
	\begin{eqnarray}
		D(x)=\dfrac{\Gamma \left(\frac{d}{2}-1\right)}{4\pi ^{\frac{d}{2}} x^{d-2}}\comma
	\end{eqnarray}
	while that of the massive scalars is
	\begin{equation}\label{massiveGF}
		D_{v}(x) = \frac{1}{2\pi }\left(\dfrac{v}{2\pi x} \right)^{\frac{d}{2}-1} K_{\frac{d}{2}-1}\left( v x \right)\comma
	\end{equation}
	with $K_{\nu}\left( z \right)$ being the modified Bessel function. Here we wrote expressions in arbitrary  dimensions in anticipation of using the dimensional regularization. In the actual computation, we will use manifestly supersymmetric scheme of dimensional reduction (DR), in which gluons and scalar have $d=4-2\epsilon $ and $10-d=6+2\epsilon $ components each, while fermions form a 10D Majorana-Weyl spinor in any dimensions.

	\paragraph{Chiral primaries and Konishi.}Our goal is to compute correlation functions of various single-trace operators. In particular, operators of primary interest are chiral primary operators (CPO)
		\begin{equation}\label{eq:cpo}
		\mathcal{C}_L(x,y)=\frac{\left(8\pi ^2\right)^{\frac{L}{2}}}{\sqrt{L}\,\lambda ^{\frac{L }{2}}}\,\mathop{\mathrm{tr}}\left((y\cdot \phi )^{L}\right)(x),
	\end{equation}
% 	\begin{equation}\label{eq:cpo}
% 		\mathcal{O}_\Delta=\frac{\left(8\pi ^2\right)^{\frac{\Delta }{2}}}{\sqrt{\Delta }\,\lambda ^{\frac{\Delta }{2}}}\,C_A^{i_1\ldots i_\Delta }\mathop{\mathrm{tr}}\phi _{i_1}\ldots \phi _{i_\Delta },
% 	\end{equation}
where $y$ is a six-dimensional null vector $y\cdot y=0$ which specifies the $SO(6)$ polarization. Another operator of interest is
 the Konishi operator, which is the simplest non-BPS operator in $\mathcal{N}=4$ SYM:
	\begin{equation}\label{Konishi}
		K(x)=Z_K\, \sum_i\tr{\phi_i \phi_i}(x)\period
	\end{equation}
Here $Z_K$ is the normalization constant at tree level equal to 
$$Z_K^{\rm tree}=\frac{4\pi^2}{\sqrt{3}\,\lambda}\,.$$ 
Unlike CPOs, the Konishi normalization constant receives loop corrections which are moreover UV divergent and in general scheme-dependent. The UV divergences remove infinities in the correlation functions and
generate the anomalous dimension. Fortunately, we need not worry proper normalization and can start with bare (=un-renormalized) operators. To distinguish them from the renormalized ones we put a superscript $b$:
	\begin{align}
	    K^{b}(x)=\sum_i{\rm tr}\,\phi_i\phi_i(x)\period
	\end{align}
	The correlation functions of $K^{b}$ are divergent at the loop level. The UV finite one-point function of the properly normalized operator can be constructed as a ratio of two correlators as we shall see later. 
	
	With these definitions, the two-point function of CPOs is canonically normalized,
	\begin{align}
	    \langle\mathcal{C}_{L_1}(x_1,y_1)\mathcal{C}_{L_2}(x_1,y_1)\rangle=\delta_{L_1,L_2}\frac{(y_1\cdot y_2)^{L_1}}{|x_1-x_2|^{2L_1}}\period
	\end{align}
% 	Here $C_A$ are appropriately normalized 
% 	\begin{equation}
% 		C^{i_{1} \ldots i_{\Delta}}_{A} C^{i_{1} \ldots i_{\Delta}}_{B}=\delta_{A B}
% 	\end{equation}
% 	traceless symmetric tensors that also define spherical harmonics on $S^{5}:$
% 	\begin{equation}
% 		Y^{A}(\mathbf{n})=C_{i_{1} \ldots i_{\Delta}}^{A} n^{i_{1}} \ldots n^{i_{\Delta}}.
% 	\end{equation}
% 	With these conventions the two-point functions of CPOs are unit-normalized at short distances. The constant $Z_K$ in the definition of Konishi serves the same purpose, but in addition removes all UV divergences and so is scheme-dependent. This multiplicative renormalization renders all correlation functions finite, but for technical reasons  we will also consider the bare operator $K_0=\mathop{\mathrm{tr}}\phi ^2$ whose correlation functions are  divergent.
	The tree-level (classical)  one-point functions can be computed by substituting the Higgs VEV into the explicit expressions for the operators:
	\begin{equation}\label{classical}
		\text{Tree level:}\qquad\left\langle\mathcal{C}_{L}\right\rangle_v=\frac{(y\cdot v)^{L}}{\sqrt{L}}\left(\frac{8\pi^2}{\lambda}\right)^{\frac{L}{2}} \comma\qquad \qquad
		\left\langle K\right\rangle_v=\frac{4\pi ^2v^2}{\sqrt{3}\,\lambda }\,.
	\end{equation}
	Here and below $\langle \mathcal{O}\rangle_v$ means the expectation value of the operator $\mathcal{O}$ on the Coulomb branch \eqref{higgsvev}.
	\paragraph{General structure of one-point functions.} The conformal invariance of the underlying theory dictates the structure of one-point functions on the Coulomb branch to be of the following form:
	\begin{align}\label{eq:generalconformal}
	    \langle \mathcal{O}\rangle_{v}=c_{\mathcal{O}}\,v^{\Delta_{\mathcal{O}}}\,.
	\end{align}
	Here $\Delta_{\mathcal{O}}$ is the conformal dimension of the operator. The factor $v^{\Delta_{\mathcal{O}}}$ is purely kinematical and is required in order to match the scalings on both sides. By contrast, $c_{\mathcal{O}}$ contains genuine dynamical information about the operator. This is a schematic formula, since for non-singlet operators, the R symmetry structure has to be also taken into account.  Later in the paper, we will confirm that the perturbative results reproduce this expected structure \eqref{eq:generalconformal}.
	 \subsection{Finiteness of one-point functions}
	 As mentioned already several times, one important property of theories with spontaneously broken conformal symmetry is that the correlation functions are finite once we renormalize the operators at the UV fixed point. 
	 
	 Before explaining why this is so, let us point out that this does not hold in generic massive QFTs. Take a composite operator $\varphi^2$ in free massive scalar theory as an example. At the UV fixed point, we can define the renormalized operator $:\varphi^2:$ simply by subtracting the massless propagator,
	 \begin{align}
	     :\varphi^2(x):=\lim_{x\to y}\left[\varphi(x)\varphi(y)-G_{m^2=0}(x-y)\right]\period
	 \end{align}
	 Here $G_{m^2}(x-y)$ is a propagator with mass $m$,
	 \begin{align}
	     G_{m^2}(x-y)=\int\frac{d^{d}k}{(2\pi)^d}\frac{e^{ik (x-y)}}{k^2+m^2}\period
	 \end{align}
	 If we use this definition and compute the one-point function in free massive scalar theory, we obtain
	 \begin{align}
	     \left\langle:\varphi^2(x):\right\rangle=\lim_{x\to y}  \left[G_{m^2}(x-y)-G_{m^2=0}(x-y)\right]=\lim_{x\to y}\int\frac{d^{d}k\, e^{ik(x-y)}}{(2\pi)^d}\left(\frac{1}{k^2+m^2}-\frac{1}{k^2}\right)\period
	 \end{align}
	 In $d=4$, this integral is logarithmically divergent indicating that the extra renormalization is necessary to define finite correlation functions. This reflects the  mixing between $:\varphi^2(x):$ and the identity operator, which is allowed once a dimensionful parameter (mass $m^2$) is introduced.
	 
	 On the other hand we do not need such extra renormalization in theories with spontaneously broken conformal symmetry. This can be seen explicitly in the one-loop computation performed in the next section, but an intuitive reason for this is as follows. When the theory spontaneously breaks the conformal symmetry, the theory itself is not modifed. Instead what we are doing is simply evaluating the operators on a different state, namely a state on the vacuum manifold. It is the state not the theory that supplies the mass scale. Since the renormalization of operators should not depend on the states we consider, this implicitly guarantees the finiteness of correlation functions on the spontaneously broken vacuum.
	 
	 This argument however is rather indirect and somewhat abstract. With a Lagrangian description at hand, we can also give a more direct and concrete justification. Correlation functions in any renormalizable QFT remain finite in an arbitrary background field, once all necessary counterterms are added to the Lagrangian (in the $\mathcal{N}=4$ SYM those are absent) and to local operators.  Demanding finiteness in the general classical background is a convenient practical tool to compute the counterterms, the essence of the background field method. A vacuum expectation value that breaks the conformal symmetry is just an example of constant background field. Switching it on, consequently, does not require any new counterterms beyond those already present at the conformal point (at zero vev).
	 
	\section{One-point functions at one loop}\label{sec:weak}
	In this section, we study the loop corrections to the one-point functions of single-trace operators. 
	To compute them at one loop, we need to take into account the multiplicative renormalization. As discussed above, for theories with spontaneous conformal symmetry breaking, the required renormalization is the same as the one at the conformal point. It is therefore convenient to consider the following ratio of correlation functions of bare operators (taking Konishi as an example):
	\begin{equation}\label{ratio-to-compute}
		\left\langle K\right\rangle_v=
		\frac{\left\langle K^{b}(0)\right\rangle_v}{|x|^\Delta \left\langle K^{b}(x)K^{b}(0)\right\rangle^{\frac{1}{2}}}\,,
	\end{equation}
	The denominator is computed at the conformal point or, equivalently, at $x\rightarrow 0$.
	
Notice that any rescaling $K^b\rightarrow ZK^b$ cancels in the ratio. If the operator, like Konishi, is multiplicatively renormalizable, we can drop the renormalization factor altogether and deal with bare opeartors - hence the superscript $b$ in $K^b$. Both the numerator and the denominator on the right hand side then diverge but the divergences will cancel upon taking the ratio.  Below we will use this formula to compute the renormalized one-point functions.

	\subsection{Konishi and chiral primaries}
	\paragraph{One-loop diagrams.}For any scalar operator, there are two Feynman diagrams which contribute to the one-point function at one loop:
	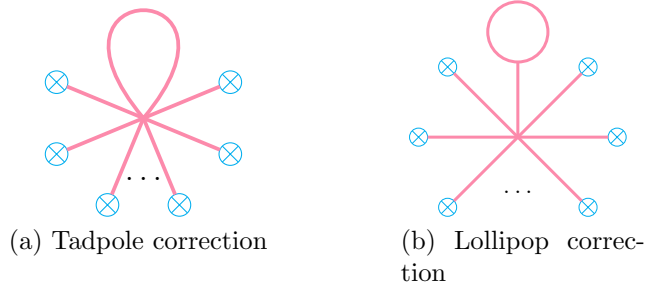
\begin{figure}[H]
		\centering
		\subfloat[Tadpole correction]{  \label{fig:tadpole}
			\begin{tikzpicture} 
				\begin{feynman} 
					\vertex (a); 
					\vertex [crossed dot, cyan] (b) at (22.5:1.25cm) {}; 
					\vertex [inner sep=0pt] (c) at (67.5:1.1cm) {}; 
					\vertex [inner sep=0pt] (d) at (112.5:1.1cm) {};
					\vertex [crossed dot, cyan] (f) at (157.5:1.25cm) {};
					\vertex [crossed dot, cyan] (d1) at (202.5:1.25cm) {};
					\vertex [crossed dot, cyan] (f1) at (247.5:1.25cm) {};
					\vertex [crossed dot, cyan] (d2) at (292.5:1.25cm) {};
					\vertex [crossed dot, cyan] (f2) at (337.5:1.25cm) {};
					\vertex [] (qq) at (0, -0.8) {$\ldots$};
					% \vertex[inner sep=0pt] (cd) at (90:1.5cm) {};
					
					\diagram* { (b) --[line width=0.5mm, ticklemepink]  (a), 
						%(a) --[line width=0.5mm, ticklemepink]  (c),
						%(a) --[line width=0.5mm, ticklemepink]  (d),
						(f) --[line width=0.5mm, ticklemepink]  (a),
						(d1) --[line width=0.5mm, ticklemepink]  (a),
						(a) --[line width=0.5mm, ticklemepink]  (f1),
						(a) --[line width=0.5mm, ticklemepink]  (f2),
						(a) --[line width=0.5mm, ticklemepink]  (d2), 
						%(c) --[line width = 0.5mm,quarter right] (cd),
						%(cd) --[line width = 0.5mm,quarter right] (d)
					};
					\draw [line width=0.5mm, ticklemepink] (a) to[out=130, in=50, loop, min distance=2.5cm] (a);
					%\draw[line width=0.5mm, ticklemepink] (d) ++(0,-1pt) let \p1 = ($(d)-(c)$) in 
					%arc [start angle =180, end angle = 0, x radius = {veclen(\x1/2,\y1/2)}, y radius = {veclen(\x1/2,\y1/2)}];
				\end{feynman} 
			\end{tikzpicture}
			
		}\qquad \qquad
		\subfloat[Lollipop correction]{\scalebox{0.8}{\label{fig:lollipop}
				\centering
				\begin{tikzpicture}
					\begin{feynman} 
						\vertex (a); 
						\vertex [left=of a, crossed dot, cyan] (b) {}; 
						\vertex [right=of a, crossed dot, cyan] (c) {}; 
						\vertex [below left=of a, crossed dot, cyan] (d) {};
						\vertex [below right=of a, crossed dot, cyan] (f) {};
						\vertex [above left=of a, crossed dot, cyan] (d1) {};
						\vertex [above right=of a, crossed dot, cyan] (f1) {};
						\vertex [above =of a] (g) at (0, -0.25);
						\vertex [above =of g] (k) at (0, 0.25);
						\vertex [below =of a] (qq) at (0, 0.75) {$\ldots$};
						
						\diagram* { (b) --[line width=0.5mm, ticklemepink]  (a), (c) --[line width=0.5mm, ticklemepink]  (a), (d) --[line width=0.5mm, ticklemepink]  (a), (f) --[line width=0.5mm, ticklemepink]  (a), (g) --[line width=0.5mm, ticklemepink]  (a), (d1) --[line width=0.5mm, ticklemepink]  (a), (a) --[line width=0.5mm, ticklemepink]  (f1)}; 
						\draw (k) let
						\p1 = ($ (k) - (g) $)
						in
						circle ({veclen(\x1,\y1)})[ticklemepink, line width=0.5mm];
					\end{feynman} 
				\end{tikzpicture}
		} }
		\caption{Loop correction to operator vev}	
	\end{figure} 
	\noindent Among the two, the lollipop diagram actually cancels. The effective vertex (external field renormalization) is proportional to the scalar propagator evaluated at zero distance with the combinatorial coefficient
	\begin{equation}
		N_{\rm fermion}+2N_{\rm ghost}-2(N_{\rm scalar}-1)-2N_{\rm gauge}=
		16+2-2(9-d)-2d=0.
	\end{equation}
	This cancellation is exact in any dimension in the DR scheme. Thus the one loop correction to the one point function comes entirely from the tadpole. 
	
	\paragraph{Konishi operator.}To compute the one-point functions at $\ell$ loops, in general we need to diagonalize the $(\ell+1)$-loop dilatation operator and resolve the operator mixing, as is the case with the degenerate perturbation theory\footnote{See \cite{Escobedo:2010xs} for more discussions on this point in the context of three-point functions.} in quantum mechanics. This leads to additional contributions to the one-point functions which we will discuss in the next subsection.
	
	However, for the CPOs and the Konishi operator, the operator mixing is known to be absent at weak coupling. For the former, this is because of the BPS condition while for the latter this is due to the non-existence of other operators with the same quantum numbers. Hence for these operators, the one-point functions at one loop can be computed simply by evaluating the diagrams discussed above.
	For instance, for the bare Konishi operator \eqref{Konishi} the one-loop one-point function reads
	\begin{equation}\label{numerator}
		\langle K^{b}\rangle_v
		=v^2+\left.(\text{tadpole})\right|_{d=4-2\epsilon}
		=
		v^2\left[1+\frac{3\lambda}{8\pi ^2} \left(-\frac{1}{\epsilon} 
		-\frac{4}{3}+\gamma - \ln\frac{4\pi}{v^2}\right)\right].
		%\label{2.13}
	\end{equation}
	where the tadpole integral in $d$ dimensions is given by
	\begin{align}
	    (\text{tadpole})=\lambda N_\phi \int\frac{d^{d}k}{(2\pi)^{d}}
		\frac{1}{k^2+v^2}=\lambda(10-d)v^2\left(\frac{v^2}{4\pi}\right)^{\frac{d}{2}}\Gamma \left(1-\tfrac{d}{2}\right)\period
	\end{align}
	
	To extract a finite contribution, we also need to compute the one-loop two-point function at the conformal point. This computation has been done in the literature, in different regularization schemes. Below, we outline the calculation in the DR scheme, carefully keeping track of the finite part. There are three types of diagrams that contribute:
	\begin{figure}[H]
		\centering
		\subfloat[$\phi^4$ vertex]{  \label{fig:phi4vertex}
			\begin{tikzpicture}
				\begin{feynman}
					\node[crossed dot,cyan] (a);
					\node[empty dot, right=of a] (b);
					\node[crossed dot, right=of b,cyan] (c);
					\diagram* {(a)--[out=50, in=40+90, thick, momentum=\(k\),ticklemepink,text=black] (b), (b)--[out=50, in=40+90, thick, momentum=\(q\),ticklemepink,text=black] (c)--[%out=-40-90, in=-50, thick, rmomentum=\(p-q\),ticklemepink,text=black%
						bend left, thick, rmomentum=\(p-q\),ticklemepink,text=black
						] (b)--[out=-40-90, in=-50, thick, rmomentum=\(p-k\),ticklemepink,text=black] (a),};
				\end{feynman}
		\end{tikzpicture}}
		\qquad
		\subfloat[Gluon exchange]{  \label{fig:gluonexchange}
			\begin{tikzpicture}
				\begin{feynman}
					\tikzstyle{derArrowEnd} = [thick,
					decoration={markings, 
						mark=at position 0.7 with {\node[dot,red];}},
					preaction = {decorate},
					postaction = {draw,shorten >= 4.5pt}]
					\tikzstyle{derArrowBeg} = [thick,
					decoration={markings, 
						mark=at position 0.3 with {\node[dot,red];}},
					preaction = {decorate},
					postaction = {draw,shorten >= 4.5pt}]
					\vertex [crossed dot,cyan] (a1) {};
					\vertex [above right = 0.5cm and 1 cm of a1] (a2);
					\vertex [below =1 cm of a2] (a3);
					\vertex [crossed dot,cyan, right =2cm of a1] (a4) {};
					\diagram*{ (a1) -- [derArrowEnd, bend left, thick, momentum=\(k\),ticklemepink,text=black] (a2), (a2) -- [bend left, thick, momentum=\(k-q\),ticklemepink,text=black] (a4), (a2) --[gluon, thick, momentum=\(q\),ticklemepink,text=black] (a3) , (a4)-- [derArrowEnd, bend left, thick, rmomentum=\(p-k+q\),ticklemepink,text=black] (a3) --[ticklemepink, bend left, thick, rmomentum=\(p-k\),text=black] (a1) };
				\end{feynman}
		\end{tikzpicture}}
		\qquad
		\subfloat[Self-energy]{  \label{fig:self-energy}
			\begin{tikzpicture}
				\begin{feynman}
					\vertex [crossed dot,cyan] (a1) {};
					\vertex [above right = 0.5cm and 1 cm of a1, blob] (a2){};
					\vertex [below =1 cm of a2] (a3);
					\vertex [crossed dot,cyan, right =2cm of a1] (a4) {};
					\diagram*{ (a1) -- [ticklemepink, bend left, thick, momentum=\(k\),text=black] (a2)-- [ticklemepink, bend left, thick, momentum=\(k\),text=black] (a4),(a4)-- [ticklemepink, out=-40-90, in= -50, thick, momentum=\(p-k\),text=black] (a1)};
				\end{feynman}
	\end{tikzpicture}}
		\caption{\label{fig:loop2pt}
		%\begin{minipage}{7cm}
				One-loop corrections to the two-point function of Konishi at the conformal point.
		%\end{minipage}
		}
	\end{figure}
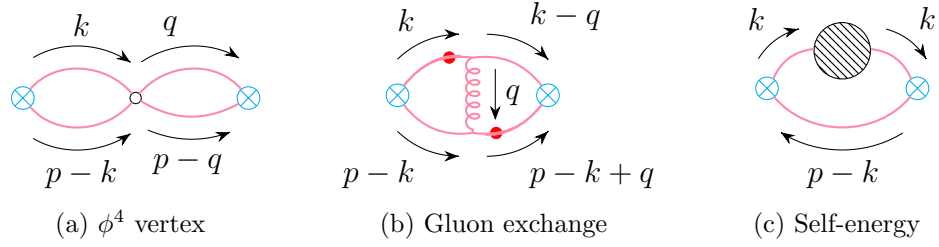
Combined with the tree-level contribution, they give:
	\begin{equation}\label{K0K0}
		\langle K^{b}(x) K^{b}(0) \rangle = \dfrac{ \lambda^2 N_\phi \Gamma\left(\frac{d}{2}-1\right)^2}{32 \pi^d x^{2d-4}}\left(
		1+\frac{\lambda \, A^K}{16\pi ^{\frac{d}{2}}x^{d-4}}
		\right)\,,
	\end{equation}
	\begin{equation}
		A^K=A^K_{\rm sc}+A^K_{\rm g}+A^K_{\rm s-e},
	\end{equation}
	where each of the three diagrams contributes
	\begin{itemize}
		\item $\phi^4$ vertex:\\
		$$A^K_{\rm sc}=\frac{\left(1-N_\phi \right) 
			\Gamma
			\left(2-\frac{d}{2}\right)^2 \Gamma
			\left(\frac{d}{2}-1\right)^2 \Gamma \left(\frac{3
				d}{2}-4\right)}{\Gamma (4-d) \Gamma (d-2)^2}$$
		\item Self-energy:   \\
		$$
		A^K_{\rm s-e}=\frac{4\Gamma \left(\frac{d}{2}-2\right)}{d-3}
		$$
		\item gluon exchange:\\
		$$
		A^K_{\rm g} =- \frac{4\Gamma \left(\frac{d}{2}-2\right)}{d-3} -\frac{\Gamma \left(2-\frac{d}{2}\right)^2\Gamma \left(\frac{d}{2}-1\right)^2\Gamma \left(\frac{3d}{2}-4\right)}{\Gamma (4-d)\Gamma (d-2)^2}\,
		$$
	\end{itemize}
	and they satisfy
	\begin{equation}\label{diagram-relation}
		A^K_{\rm sc}=(N_\phi -1)\left(A^K_{\rm s-e}+A^K_{\rm g}\right).
	\end{equation}
	
	Hence, near four dimensions,
	\begin{equation}
		\langle K^{b}(x) K^{b}(0) \rangle = \dfrac{ 3\lambda^2 }{16 \pi^4 x^4} \left\{1 -\dfrac{3 \lambda}{4 \pi^2 } \left[ \dfrac{1}{\epsilon} + \dfrac{4}{3}+\gamma + \ln(\pi x^2) \right]  \right\}.
	\end{equation}
	The logarithm reproduces the anomalous dimension of Konishi,
	\begin{equation}
		\Delta_K =2+\frac{3\lambda }{4\pi ^2}\,,
	\end{equation}
	and in the ratio (\ref{ratio-to-compute}) the $x$-dependence cancels as it should.
	The $1/\epsilon $ pole also cancels because it has precisely the same coefficient as does the numerator (\ref{numerator}).
	
	As a result, the renormalized 1-loop one-point function is given by 
	\begin{equation}\label{<K>}
		\left\langle K\right\rangle_v=\frac{4\pi ^2v^2}{\sqrt{3}\,\lambda }
		\left[1+\frac{3\lambda }{4\pi ^2}\left(\gamma +\ln\frac{v}{2}\right)\right].
	\end{equation}
	The remaining logarithm is a manifestation of the anomalous scaling and is consistent with the general form dictated by dimensional analysis:
	\begin{equation}\label{<K>gen}
		\left\langle K\right\rangle_v=c_Kv^{\Delta_K}.
	\end{equation}
	To the one-loop accuracy we find for the coefficient $c_K$,
	\begin{equation}\label{c_K}
		c_K=\dfrac{4 \pi^2}{\sqrt{3}\,\lambda}+\sqrt{3} \left( \gamma -\ln 2\right).
	\end{equation}
	Interestingly, the one-loop correction contains the Euler $\gamma $. Typically $\gamma $ goes along with $1/\epsilon $ and gets removed by renormalization, but here it survives and is part of the universal, renormalized one-point function. This also happens in D3-D5 dCFT \cite{Buhl-Mortensen:2016pxs} and is crucial for reproducing the integrability prediction \cite{Buhl-Mortensen:2017ind}. In that case, the magnon reflection phase
 off the D5-brane is a complex transcendental function  
which, when expanded at weak coupling, gives rise to the Euler $\gamma$ \cite{Komatsu:2020sup}. We expect a similar mechanism to work on the Coulomb branch as well \cite{progress:2021}.
	
	\paragraph{Chiral primaries.}
	The one-point functions of chiral primaries are protected and should not receive loop corrections \cite{Skenderis:2006uy}, as evidenced by the supergravity analysis at strong coupling \cite{Skenderis:2006uy,Skenderis:2006di}. Using the formulas above we can confirm that the one-loop corrections vanish. Cancellation of the tadpole diagram
is an obvious consequence of the trace condition and the residual $SO(6)$ invariance of the scalar propagator. Absence of quantum corrections to the two-point functions of chiral primaries at the conformal point is a well-known fact, on which we comment  later, in discussing more general scalar operators.
	
	\subsection{Arbitrary scalar operators}
	
	\paragraph{Tree level.}Consider next the most general single-trace scalar operator
	\begin{equation}\label{SO(6)-spin-chain-state}
		\mathcal{O}=\Psi ^{i_1\ldots i_L}\mathop{\mathrm{tr}}\phi _{i_1}\ldots \phi _{i_L}.
	\end{equation}
	Mixing of these operators is described by an integrable $SO(6)$ spin chain \cite{Minahan:2002ve}, and the prefactor $\Psi_{^{i_1,\ldots, i_L}}$ is given by a wave function of an energy eigenstate of the spin chain. To evaluate a one-point function at tree level, we simply  
substitute the scalar fields by the Higgs expectation value. In the spin-chain language this results in an overlap between an energy eigenstate associated with the operator and 
a boundary state defined by the Higgs vev,
	\begin{equation}
		\left\langle\mathcal{O}\right\rangle_v=
		\left(\frac{8\pi^2 v^2}{\lambda}\right)^{\frac{L}{2}}
		\frac{\langle B|\Psi\rangle}{\sqrt{L\langle\Psi|\Psi\rangle}}\,,
	\end{equation}
	where
	\begin{equation}
		B_{i_1\ldots i_L}=n_{i_1}\ldots n_{i_L},\qquad\qquad n_i\equiv \frac{v_i}{v}\period
	\end{equation}
	Here the prefactor and $1/\sqrt{L\langle \Psi|\Psi\rangle}$ come from taking the ratio \eqref{ratio-to-compute}. 
	
	The boundary state $\langle B|$ is known to be annihilated by infinitely many conserved charges and a closed-form expression for the overlap $\langle B|\Psi\rangle$ was derived in \cite{deLeeuw:2019ebw}. This gives the first evidence for integrability of the vacuum condensates on the Coulomb branch. We will address this question in more detail in the upcoming paper \cite{progress:2021}, in which we develop a non-perturbative integrability approach to the vacuum condensates.
	
	\paragraph{One loop.}At the next order, several new effects kick in. The first effect is the correction to the wavefunction due to higher-loop mixing. In particular, the $SO(6)$ sector no longer forms a closed subsector beyond one loop. A novel effect appearing at two loops is a length-changing mixings with fermions \cite{Beisert:2003ys}:
	\begin{equation}
		\mathcal{O}=\Psi_B ^{i_1\ldots i_L}\mathop{\mathrm{tr}}\phi _{i_1}\ldots \phi _{i_L}+\lambda \sum_{k=1}^{L-3}\Psi _F^{\alpha i_1\ldots \beta i_k\ldots i_{L-3}}\mathop{\mathrm{tr}}\psi _\alpha \phi _{i_1}\ldots \psi _\beta \phi _{i_k}\ldots \phi _{i_{L-3}}.
	\end{equation}
It is tacitly assumed that $\Psi _B$ contains both the one-loop $\mathcal{O}(1)$ term and a two-loop $\mathcal{O}(\lambda )$ correction:
\begin{equation}
 \Psi _B=\Psi _B^{(0)}+\lambda \Psi _B^{(1)}.
\end{equation}
The fermion contribution to the expectation value is suppressed by an additional loop factor and can be neglected to this order, but fermions do contribute indirectly, through an $\mathcal{O}(\lambda )$ correction to the norm. The fact that the fermionic term contains one less field operator is crucial for this effect. Indeed, the norm of the bosonic wavefunction is $\mathcal{O}(\lambda ^L)$, from $L$ propagators in the two-point correlator. The fermionic norm has one less propagator but an additional suppression from the explicit power of $\lambda $ makes it $\mathcal{O}(\lambda ^{L+1})$. This should be retained at the NLO accuracy.
	
	The second NLO effect is the correction to the boundary state, from the explicit Wick contraction between the fields in the operator, given by the tadpole diagram  in \eqref{fig:tadpole}. Its effect is encapsulated in a correction to the boundary state: 
	\begin{equation}
		\left\langle \mathcal{O}_0\right\rangle_v=v^L
		\left[\left\langle B\right.\!\left|\Psi_B  \right\rangle
		+\frac{\lambda }{32\pi ^2}
		\left(
		-\frac{1}{\epsilon }-1+\gamma +\ln\frac{v^2}{4\pi }
		\right)
		\left\langle B'\right.\!\left| \Psi_B \right\rangle
		\right],
	\end{equation}
	where
	\begin{equation}\label{eq:defBprime}
		B'_{{i_1\ldots i_L}}=\sum_{l=1}^{L}n_{i_1}\ldots \delta _{i_li_{l+1}}\ldots n_{i_L}.
	\end{equation}
	
	Further simplifications occur once the explicit structure of the one-loop mixing matrix \cite{Minahan:2002ve} is taken into account:
	\begin{equation}\label{mixingSO(6)}
		\Gamma =\frac{\lambda }{16\pi ^2}\sum_{l=1}^{L}\left(2-2P_{l,l+1}+K_{l,l+1}\right).
	\end{equation}
	Here $P$ and $K$ are permutation and trace operators acting on states
	of the spin chain:
	\begin{equation}\label{PK}
		P_{ij}^{kl}=\delta ^l_i\delta^k_j,\qquad K_{ij}^{kl}=\delta _{ij}\delta ^{kl}. 
	\end{equation}
	It is easy to show that
	\begin{equation}\label{eq:BprimeB}
		\frac{\lambda }{16\pi ^2}\,\left\langle B'\right|=\left\langle B\right|\Gamma.
	\end{equation}
	Since operators are eigenvectors of the mixing matrix with the eigenvalues\footnote{Since $\left\langle B'\right|$ is already $\mathcal{O}(\lambda )$, we can use the one-loop, leading-order mixing.} $\lambda \Delta^{(1)}$, the one-point function can be simplified to
	\begin{equation}\label{<O0>}
		\left\langle\mathcal{O}^{b}\right\rangle_v=v^L
		\left\langle B\right.\!\left|\Psi_B  \right\rangle
		\left[
		1+
		\frac{\lambda \Delta^{(1)}}{2}\left(
		-\frac{1}{\epsilon }-1+\gamma +\ln\frac{v^2}{4\pi }
		\right) 
		\right].
	\end{equation}
	
Finally, we need to compute the norm of the operator, i.e.~the two-point function at the conformal point. The two point function of arbitrary operators is given by inserting one-loop corrections at each pair of adjacent scalar propagators and is thus described by the same diagrams in fig.~\ref{fig:loop2pt}, now decorated with the external $SO(6)$ indices. Notice, that the relation \eqref{diagram-relation} holds independently of the external indices. 
One can see this by applying Passarino-Veltman reduction in momentum space:
	\begin{equation}
		\begin{split}
			A_{\rm sc}{}^{kl}_{ij} =&
			\left(2\delta ^l_i\delta ^k_j-\delta ^k_i\delta ^l_j
			-\delta _{ij}\delta ^{kl}\right)
			 \int 
			 \dk{k}\dk{q}\,
			 \frac{1}{k^2(p-k)^2q^2(p-q)^2}\, ,  \\
			A_{\rm g}{}^{kl}_{ij}=&  
			-\delta _i^k\delta _j^l
			\int 
			\dk{k}\dk{q}\,
			\frac{k(p-k+q)+(k-q)(p-k)+k(p-k)+(k-q)(p-k+q)}{q^2k^2(k-q)^2(p-k)^2(p-k+q)^2}\, ,\\
			A_{\rm s-e}{}^{kl}_{ij} =& 
			-2\delta _i^k\delta _j^l
			\int \dk{k}\dk{q}\,
			\frac{2 kq +q^2+k^2}{(k^2)^2q^2(p-k)^2(k-q)^2} \,.
		\end{split}
	\end{equation}
Rewriting the numerators as	
	\begin{equation}
		\begin{split}
		{\rm num}_{\rm s-e}=&
		2k^2+2q^2-(k-q)^2\\
	  {\rm num}_{\rm g}=&
			q^2 - k^2-(k-q)^2-(p-k)^2-(p-k+q)^2+2p^2,
		\end{split}
	\end{equation}
	we find that each term cancels a propagator leading to the following relations:\\
\begin{figure}[h]
 \centerline{\includegraphics[width=15cm]{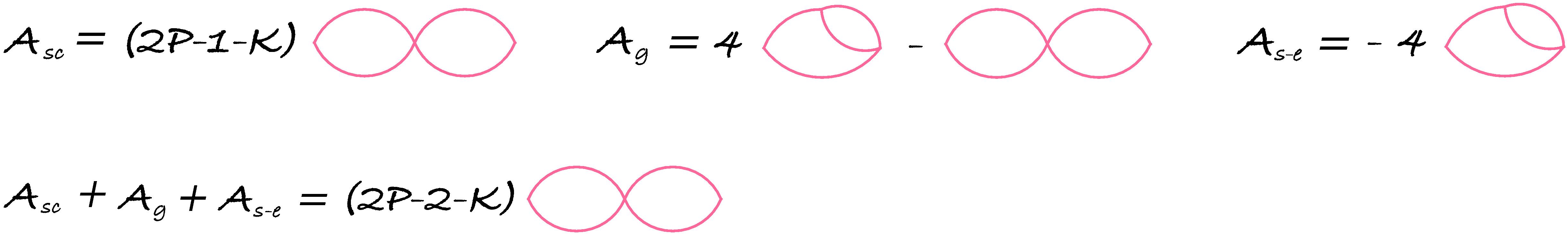}}
\end{figure}
\\
where (\ref{PK}) is used to represent the index structures. We have dropped the $q^2$ and $(k-q)^2$ terms in $A_{\rm s-e}$, as the ensuing integrals vanish in dimensional regularization, and also the $p^2$ term in $A_{\rm g}$. The latter results a convergent integral of the form $f(d)p^{2d-8}$ becoming a finite, momentum-independent constant in four dimensions and hence produces a contact term, which we can ignore.
	
As a result, the two-point function can be  expressed through the mixing matrix (\ref{mixingSO(6)}), once all pairs of lines are included\footnote{The formula (\ref{diagram-relation}) for Konishi is a particular case of this.
}:
	\begin{equation}\label{O0O0}
		\left\langle\mathcal{O}^{b}(x)\mathcal{O}^{b}(0)\right\rangle=
		L\left(\frac{\lambda }{8\pi^2x^2}\right)^L
		\left\langle \Psi \right|1-\left( \dfrac{1}{\epsilon} + 1+\gamma + \ln\pi x^2 \right)\Gamma \left|\Psi  \right\rangle.
	\end{equation}
Assuming the wavefunction is an eigenvector of $\Gamma $, we get:
	\begin{equation}\label{simplified-O0O0}
		\left\langle\mathcal{O}^{b}(x)\mathcal{O}^{b}(0)\right\rangle=
		L\left(\frac{\lambda }{8\pi^2x^2}\right)^L
		\left\langle \Psi \right.\!\left| \Psi\right\rangle
		\left[1-\lambda \Delta^{(1)} \left(\dfrac{1}{\epsilon} + 1+\gamma + \ln\pi x^2 \right)\right].
	\end{equation}
	
	To compute the normalized one-point function, we divide the answer above by the loop-corrected norm, which includes contributions from fermions\footnote{The norm defined by the two-point function differs from the natural spin-chain norm by the cyclicity factor $L$ and $(\lambda /8\pi ^2)^L$ from the $L$ propagators in the tree-level diagram. There is one less propagator in the fermion two-point function, and the fermion operator is one field shorter. The prefactor in the second term takes into account these two effects. }:
	\begin{equation}\label{BBFF}
		\left\langle \Psi \right.\!\left|\Psi  \right\rangle=
		\left\langle \Psi_B \right.\!\left|\Psi_B  \right\rangle+\frac{\lambda(L-1)}{8\pi ^2L} \left\langle \Psi_F \right.\!\left|\Psi_F  \right\rangle.
	\end{equation}
	Taking the ratio (\ref{ratio-to-compute}), and using (\ref{<O0>}), (\ref{simplified-O0O0}) and (\ref{BBFF}) we obtain the renormalized one-point function:
	\begin{eqnarray}\label{Orenormalized}
		\left\langle\mathcal{O}\right\rangle_v&=&
		\left(\frac{8\pi^2 v^2}{\lambda}\right)^{\frac{L}{2}}
		\frac{\langle B|\Psi\rangle}{\sqrt{L\langle\Psi|\Psi\rangle}}
		\left[1+  \lambda \Delta^{(1)} \left(\ln \frac{v}{2 }+ \gamma\right)\right]
		\nonumber \\
		& =&\left(\frac{8\pi^2v^2 }{\lambda}\right)^{\frac{L}{2}}
		\frac{\langle B|\Psi_B\rangle}{\sqrt{L\langle\Psi_B|\Psi_B\rangle}}
		\left\{1+\lambda \left[\left(\ln\frac{v}{2}+\gamma \right)\Delta _1-\frac{L-1}{16\pi ^2L}\,\,\frac{\langle\Psi_F|\Psi_F\rangle}{\langle\Psi_B|\Psi_B\rangle}\right]\right\}.
	\end{eqnarray}
	Here it is understood that $\left|\Psi _B\right\rangle$ includes $\mathcal{O}(\lambda)$ terms that come from the two-loop mixing. Those terms do not affect the norm $\langle\Psi_B|\Psi_B\rangle$ because perturbation theory preserves orthogonality of the wavefunctions, but they are relevant for the overlap $\langle B|\Psi_B\rangle$.
		There are thus three sources of $\mathcal{O}(\lambda )$ corrections:  (i) the universal piece proportional to the anomalous dimension, (ii) the two-loop operator mixing that effects encapsulated in $\langle B|\Psi_B\rangle$, and (iii) mixing with fermions.
	Below we illustrate the general formula on a few examples.
	\subsection{Examples}\label{subsec:example}
	\paragraph{Konishi.} The Konishi wavefunction is not renormalized, and only the universal correction contributes, recovering (\ref{<K>}) upon substitution of $\Delta ^{(1)}=3/4\pi ^2$ for the anomalous dimension.
	
	\paragraph{Chiral primaries.} Their wavefunctions and scaling dimensions are not renormalized, which guarantees the absence of one-loop corrections to the expectation values.
	
	\paragraph{Dimension three.} The operators of length three split into four  $SO(6)$ multiplets:
\begin{equation}
 \mathbf{6}\otimes\mathbf{6}\otimes\mathbf{6}
 =\mathbf{6}\oplus\mathbf{10}\oplus\bar{\mathbf{10}}\oplus\mathbf{50}
 {\color{gray}
 \oplus\mathbf{6}\oplus\mathbf{6}\oplus\mathbf{64}\oplus\mathbf{64}
 }.
\end{equation}
The shaded representations violate trace cyclicity and do not correspond to any operators. 

Among the four legible representations, the $\mathbf{50}$ is a chiral primary whose expectation value is  not renormalized. The $\mathbf{10}$ is a complex self-dual tensor of rank three:
\begin{equation}
 \mathcal{O}^{\mathbf{10}}_{ikj}=
 \mathop{\mathrm{tr}}\phi _i\left[\phi _j,\phi _k\right]
 +\frac{i}{3}\,\varepsilon _{ijklmn}\mathop{\mathrm{tr}}\phi _l\phi _m\phi _n.
\end{equation}
 The expectation value of this operator vanishes, because $\mathbf{10}$ has no singlet component under the $SO(6)\rightarrow SO(5)$ reduction.
 %, and the one-point function of $\mathcal{O}^{\mathbf{10}}$ is protected by the residual $SO(5)$ symmetry. 

The most interesting case is the vector representation,
corresponding to the operator
	\begin{equation}\label{L=3}
		\mathcal{V}_i=\mathop{\mathrm{tr}}\phi ^2\phi _i.
	\end{equation}
Operators of length three do not mix among themselves by group theory. The $\mathbf{6}$ in addition 
	does not mix with fermions \cite{Georgiou:2012zj}, for a fermion bilinear with the same quantum numbers does not exist. A possible candidate is the ``Kaluza-Klein current" $\bar{\psi }\gamma _i\psi $ but that lies in the adjoint of the color group. The Lorentz and color singlets of dimension three are $\epsilon ^{\alpha \beta }\mathop{\mathrm{tr}}\psi  _{\alpha A}\psi _{\beta B}$ and its conjugate (in the spin-chain notations), both  symmetric in the $SU(4)$ indices and hence belonging to $\mathbf{10}$ or  $\bar{\mathbf{10}}$.
	
	Therefore, the one-loop expectation value of $\mathbf{6}$ is again universal and is dictated by it anomalous dimension $\Delta _1=1/2\pi ^2$. Specifying to a particular component,
	\begin{equation}\label{length-3}
		\mathcal{V}_z=\mathop{\mathrm{tr}}\phi ^2Z,\qquad Z=\phi _1+i\phi _2,
	\end{equation}
	we find:
	\begin{equation}
		\left\langle \mathcal{V}_z\right\rangle_v=\frac{8\pi ^3v^3}{\sqrt{2\lambda ^3}}
		\left[1+\frac{\lambda }{2\pi ^2}\left(\gamma +\ln\frac{v}{2}\right)\right].
	\end{equation}
	
	\paragraph{BMN operators.} Applying the general formula to more complicated operators is hindered by operator mixing which has to be resolved to two loops, technically a rather difficult problem. Remarkable tour de force calculations of \cite{Georgiou:2012zj} provides a few explicit examples, namely for  BMN operators of low dimension. 
	
	The BMN operators are defined by replacing two $Z$s in a chiral primary $\mathop{\mathrm{tr}}Z^L$ by other scalars. This gives an infinite family of operators in one-to-one correspondence with quantized string modes
	\cite{Berenstein:2002jq}. When the two insertions form a singlet, the BMN operators have the following form
	\cite{Beisert:2002tn}:
	\begin{eqnarray}\label{Konishi-nL}
		\mathcal{O}_{nL}&=&2\cos\frac{\pi n}{L+1}\,\mathop{\mathrm{tr}}\bar{Z}Z^{L-1}
		-\sum_{l=0}^{L-2}\cos\frac{(2l+3)\pi n}{L+1}\sum_{a=3}^{6}\mathop{\mathrm{tr}}\phi _aZ^l\phi _aZ^{L-l-2} 
		\nonumber \\
		&=&-\sum_{l=0}^{L-2}\cos\frac{(2l+3)\pi n}{L+1}\,\mathop{\mathrm{tr}}\phi _iZ^l\phi _iZ^{L-l-2}.
	\end{eqnarray}
	These are exact eigenstates of the one-loop dilatation operator with the anomalous dimension 
	\begin{equation}
		\Delta _{nL}-L=\frac{\lambda }{\pi ^2}\,\sin^2\frac{\pi n}{L+1}\,.
	\end{equation}
	Konishi and the length-3 operator (\ref{length-3}) are particular cases.
	
	At two loops these operators start mixing with fermions: 
	\begin{equation}
		\mathcal{O}^{{\rm 2-loop}}_{nL}=\mathcal{O}^B_{nL}+\lambda \,\mathcal{O}^F_{nL}.
	\end{equation}
	The fermionic components are fixed by supersymmetry  \cite{Georgiou:2008vk,Georgiou:2009tp,Georgiou:2012zj}\footnote{Here $\Gamma _z$ is a combination of 10D Dirac matrices with the quantum numbers of $Z$: $\Gamma _z=\Gamma _4+i\Gamma _5$. The norms of the operators $\mathcal{O}_{nL}$ and $\mathcal{O}_{nL}^F$ are defined by their two-point functions. The fermionic operators are shorter by one field which enhances their norm by one power of $\lambda $.}:
	\begin{equation}
		\mathcal{O}^F_{nL}=-\frac{1}{8\pi ^2}\,\sin\frac{\pi n}{L+1}
		\sum_{l=0}^{L-3}\sin\frac{(2l+4)\pi n}{L+1}\,\mathop{\mathrm{tr}}\bar{\psi }Z^l\Gamma _z\psi Z^{L-l-3}.
	\end{equation}
	The mixing among purely bosonic operators, on the other hand, is a complex dynamical problem  \cite{Georgiou:2011xj} that has been resolved on a case-by-case basis up to dimension five \cite{Georgiou:2012zj}\footnote{Our notations differ slightly from those of \cite{Georgiou:2012zj} where the operators are parameterized by the R-charge rather than length: our $\mathcal{O}_{nL}$ corresponds to $\mathcal{O}_n^{L-2}$ in \cite{Georgiou:2012zj}.}:
	\begin{eqnarray}
		\mathcal{O}^B_{14}&=&\mathcal{O}_{14}+\frac{\lambda }{4\pi ^2}\,\,\frac{5+\sqrt{5}}{20}\,\mathcal{O}_{24},
		\nonumber \\
		\mathcal{O}^B_{24}&=&\mathcal{O}_{24}+\frac{\lambda }{4\pi ^2}\,\,\frac{5-\sqrt{5}}{20}\,\mathcal{O}_{14},
		\nonumber \\
		\mathcal{O}^B_{15}&=&\mathcal{O}_{15}+\frac{\lambda }{4\pi ^2}\,\,\frac{3\sqrt{3}}{16}\,\mathcal{O}_{25},
		\nonumber \\
		\mathcal{O}^B_{25}&=&\mathcal{O}_{25}+\frac{\lambda }{4\pi ^2}\,\,\frac{\sqrt{3}}{16}\,\mathcal{O}_{15}.
	\end{eqnarray}
	
	This mixing pattern can be compacted into a single equation
	\begin{eqnarray}
		\frac{\mathcal{O}^B_{nL}}{\cos\frac{\pi n}{L+1}}&=&\frac{\mathcal{O}_{nL}}{\cos\frac{\pi n}{L+1}}
		+\frac{\lambda }{4\pi ^2}\,\,\frac{1}{L+1}\,\sin^2\frac{\pi n}{L+1}
		\nonumber \\
		&&\times 
		\left[
		2L-1-(2L-4)\sin^2\frac{\pi n}{L+1}-4\sin^2\frac{2\pi n}{L+1}
		\right]
		\frac{\mathcal{O}_{3-n,L}}{\cos\frac{\pi (3-n)}{L+1}}\,.
	\end{eqnarray}
	We stress that this is not a general formula (which is not known) but rather a convenient representation valid for $L=4,5$ and $n=1,2$. 
	The rationale to normalize by cosine is to simplify the overlap:	
	\begin{equation}
		\left\langle B\right.\!\left| \Psi _{nL}\right\rangle=2\cos\frac{\pi n}{L+1}\,.
	\end{equation}
	Taking this into account, we immediately find the overlap of the true two-loop wavefunction:
	\begin{eqnarray}
		\left\langle B\right.\!\left| \Psi^B _{nL}\right\rangle&=&
		2\cos\frac{\pi n}{L+1}\left\{
		1+ \frac{\lambda }{4\pi ^2}\,\,\frac{1}{L+1}\,\sin^2\frac{\pi n}{L+1}
		\right.
		\nonumber \\
		&&\left.\times 
		\left[
		2L-1-(2L-4)\sin^2\frac{\pi n}{L+1}-4\sin^2\frac{2\pi n}{L+1}
		\right]
		\right\}.
	\end{eqnarray}
While for the norm we get:
\begin{equation}
 \left\langle \Psi^{B} _{nL}\right.\!\left|\Psi^{B} _{nL} \right\rangle
 =2^L\,\frac{L+1}{L}\,.
\end{equation}
		
	The last missing piece is the norm of the operator, which contains an  $\mathcal{O}(\lambda )$  contribution from fermions. This was also computed in
	\cite{Georgiou:2012zj}\footnote{We reverse the relative sign of the fermionic contribution compared to \cite{Georgiou:2012zj}. As a norm of a spin-chain state it cannot be negative.}:
	\begin{equation}
		\frac{\left\langle \Psi ^F_{nL}\right.\!\left| \Psi^F _{nL}\right\rangle}
		{\left\langle \Psi ^B_{nL}\right.\!\left| \Psi^B _{nL}
		\right\rangle}
		=\frac{4L}{L^2-1}\,\sin^2\frac{\pi n}{L+1}\left(
		L-3+4\cos^2\frac{2\pi n}{L+1}
		\right).
	\end{equation}
	
	Collecting the pieces together and adding the universal transcendental term, we find for the one-point functions:
	\begin{equation}\label{av-diag}
		\left\langle \mathcal{O}_{nL}\right\rangle_v=\frac{2\cos\frac{\pi n}{L+1}}{\sqrt{L+1}}\left(\frac{2\pi v}{\sqrt{\lambda }}\right)^L
		\left(
		1+\frac{\lambda }{\pi ^2}\,C_{nL}\sin^2\frac{\pi n}{L+1}
		\right).
	\end{equation}
	with
	\begin{eqnarray}
		C_{nL}&=&\gamma +\ln\frac{v}{2}+\frac{1}{4}\,\,\frac{1}{L+1}
		\left[
		2L-1-(2L-4)\sin^2\frac{\pi n}{L+1}-4\sin^2\frac{2\pi n}{L+1}
		\right.
		\nonumber \\
		&&\left.
		-L+3-4\cos^2\frac{2\pi n}{L+1}
		\right]
		\nonumber \\
		&=&\gamma +\ln\frac{v}{2}+\frac{1}{4}\,\,\frac{L-2}{L+1}\,\cos\frac{2\pi n}{L+1}\,.
	\end{eqnarray}
	Once again, this is not a general result. The formula is valid for $L$ up to 5 and $n=1,2$.

\paragraph{Dimention-four singlets.} 	There are two $SO(6)$ singlets at length four, discussed recently from the integrability point of view \cite{Eden:2023ygu}, which at one loop mix according to
\begin{equation}
  \mathcal{O}_\pm=\mathop{\mathrm{tr}}\phi _{i}\phi _i\phi _j\phi _j+
  \frac{5\mp\sqrt{41}}{4}\,\mathop{\mathrm{tr}}\phi _i\phi _j\phi _i\phi _j,
\end{equation}
and have the anomalous dimensions 
\begin{equation}
 \Delta _1^\pm=\frac{13\pm\sqrt{41}}{16\pi ^2}\,.
\end{equation}
It is straightforward to calculate
\begin{equation}
 \frac{\left\langle B\right.\!\left|\Psi _\pm \right\rangle}{\left\langle \Psi _\pm\right.\!\left|\Psi _\pm \right\rangle}
 =\sqrt{\frac{1}{32}\mp\frac{7}{96\sqrt{41}}}\,.
\end{equation}
At two loops these operators start mixing with fermions and among themselves. All the mixing coefficients are in principle known
\cite{Georgiou:2012zj} enabling the use of the formalism developed above, but the resulting expressions are rather lengthy and we will not show them here. 

	\section{One-point functions at strong coupling}\label{sec:strong}
	
	In this section we calculate the one-point correlator of a chiral primary operator from the gravity side. For the $U(N)$ broken down to $U(1)^N$ the holographic one-point functions were computed in \cite{Skenderis:2006uy,Skenderis:2006di} and were found to agree with the tree-level weak-coupling expressions. The VEV in \cite{Skenderis:2006uy,Skenderis:2006di}   was modeled by a continuous distribution of D3 branes, which backreact on the geometry. The one-point functions arise as a response to a macroscopic curvature the D-branes induce. 
	We consider a different regime, where only one D3 brane is separated from the stack and placed in the bulk of $AdS_5\times S^5$. The one-point function is then obtained by connecting the  operator insertion on the boundary to a point on the brane by a bulk-to-boundary propagator and integrating over the propagator's endpoint  \cite{Berenstein:1998ij}. 
	
	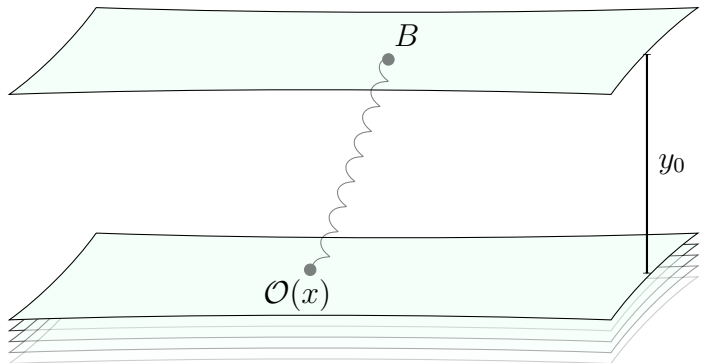
\begin{figure}[H]
		\begin{center}
			\begin{tikzpicture}
				\draw[copy shadow={shadow yshift=-0.8ex,shadow xshift=-0ex ,opacity=0.8}
				,copy shadow={shadow yshift=-1.6ex,shadow xshift=-0ex ,opacity=0.6}
				,copy shadow={shadow yshift=-2.4ex,shadow xshift=-0ex ,opacity=0.4}
				,copy shadow={shadow yshift=-3.2ex,shadow xshift=-0ex ,opacity=0.2}
				,fill=mintcream,decorate,decoration={bent,aspect=.3,amplitude=0.2em}] (-4,-1.5,-1.5) -- (-4,-1.5,1.5) -- (4,-1.5,1.5) -- (4,-1.5,-1.5) -- cycle;
				\draw[fill=mintcream,overlay,decorate,decoration={bent,aspect=0.3,amplitude=0.2em}] (-4,1.5,-1.5) -- (-4,1.5,1.5) -- (4,1.5,1.5) -- (4,1.5,-1.5) -- cycle;
				%\fill[black](0,0,0) circle (2pt);
				%\fill[black](0,3,0) circle (2pt);
				\begin{scope}[transparency group, opacity=0.5]
					\draw[snake=coil,arrows={spaced *-spaced *}]  (-0.5,-1.5,0) --  (0.5,1.5,0);
				\end{scope}
				\node at ([xshift=-0.6em, yshift=-0.6em]-0.5,-1.5,0) {$\mathcal{O}(x)$};
				\node at ([xshift=0.5em, yshift=0.5em]0.5,1.5,0) {$B$};
				\draw[arrows={spaced serif cm-spaced serif cm },thick]  (3.9,-1.5,0) -- node[midway, right] {$y_0$} (3.9,1.5,0) ;
				%\node at ([xshift=0.5em, yshift=0.5em]4.7,0.6,0) {$1/v$};
			\end{tikzpicture}
		\end{center}
		\caption{Gravity setup}\label{gravity-picture}
	\end{figure}
	
	The ``boundary" part of the calculation is determined by supergravity modes dual to the chiral primary as given by the standard AdS/CFT dictionary. Meanwhile the ``bulk" is given by the D-brane action, which determines the coupling of the supergravity modes to the brane. This determines the measure of integration for the propagator.  This formalism has been applied to giant gravitons in $AdS_5\times S^5$, which are also D3 branes albeit of a different shape \cite{Bissi:2011dc,Yang:2021kot,Caputa:2012yj,Kristjansen:2015gpa}, and these results can be utilized in the computation of the Coulomb-branch one-point functions.

	\subsection{Action for a probe D3 brane}
	The probe D3 brane is embedded in the $AdS_5 \times S_5$ background with metric $G_{MN}$ with the components:
	\begin{equation}
		G_{mn}=G^{\mathrm{AdS}}_{mn}  \qquad ds^2_{\mathrm{AdS}} =\frac{dx^2+dz^2}{z^2}\, ,
	\end{equation}
	and $G_{\alpha\beta}= G^{S^5}_{\alpha\beta}$, where $M,N$  run from 0 to 9, and $m,n$ and $\alpha,\beta$ run through $AdS_5$ and the sphere respectively. 
	The symmetry breaking pattern (\ref{higgsvev}) corresponds to a D3 brane placed parallel to the boundary of $AdS_5$ at the radial position\footnote{This identification can be justified by comparing the energy of a string stretched to the horizon $\sqrt{\lambda }/2\pi z_0$ with the mass of the W-boson $m_W=v$.}
	\begin{equation}\label{eq:coulombandz0}
		z_0=\frac{\sqrt{\lambda }}{2\pi v}
	\end{equation}
	and at the point on a sphere given by $n_i=v_i/v$. This corresponds to detaching a single brane from the stack of $N$ $D3$ branes, which produce the usual unbroken AdS/CFT scenario, and ignoring any backreaction.
	Supergravity fields couple to the probe brane via the DBI action with a Wess-Zumino term:
	\begin{equation}\label{eq:gravaction}
		S_{D3} =  -\frac{N}{2\pi ^2}\left(\int_{D3} d^4 x \sqrt{g} + \int_{D3} C^{(4)}\right),
	\end{equation}
	where $g_{\mu \nu }=\eta _{\mu \nu }/z_0^2$ and $C^{(4)}=-1/z_0^4  \, dx^1 \wedge dx^2 \wedge dx^3 \wedge dx^4$ are the induced metric and the RR-form on the brane and $N/2\pi ^2$ is the D3-brane tension in the units of the AdS radius.\\\\
	The supergravity scalar field dual to the chiral primary operator $\mathcal{C}_A$ is a mixture of the metric perturbation and that of the RR 4-form  \cite{Lee:1998bxa}:
	\begin{equation}\label{fluc} 
		\begin{split}
			h_{m n}^{\mathrm{AdS}} 
			&=\frac{4}{\Delta+1} \,Y^A\nabla_{m} \nabla_{n} s_A 
			-\frac{2 \Delta(\Delta-1)}{\Delta+1}\,Y^A  G_{m n} s_A,
			\\
			h_{\alpha \beta}^{S^5} 
			&=2 Y^A\Delta \,  G^{S^5}_{\alpha \beta}  s_A,
			\\ 
			a_{mnlk}^{\mathrm{AdS}} &= -4Y^A\sqrt{G^{\rm AdS}}\,\epsilon_{mnlkp} \nabla^{p}  s_A,
			\\
			a_{\alpha \beta \gamma \delta }^{S^5} &=4Y^A \sqrt{G^{ S^5}}\,\epsilon_{\alpha \beta \gamma \delta\eta } \nabla^{\eta}  s_A.
		\end{split}
	\end{equation}
	Here $h^{AdS}_{mn}$, $h^S_{\alpha\beta}$, $a_{mnlk}^{\mathrm{AdS}}$ and $a_{\alpha \beta \gamma \delta }^{\mathrm{S}}$ are $AdS_5$ and $S^5$ projections of the metric and 4-form perturbations:
	\begin{equation}
		\begin{split}
			G_{MN} &\rightarrow G_{MN}+h_{NM}, \\
			C^{(4)}&\rightarrow C^{(4)}+a^{(4)}.
		\end{split}
	\end{equation}
	The dimension of the dual operator is $\Delta $ and the spherical harmonics  $Y^A(n)$ describes its $SO(6)$ polarization. Finally, $s_A$ is a Klein-Gordon field on $AdS_5$ with the bulk-to-boundary propagator
	\begin{equation}\label{b-to-b}
		K(x,y;x')=\frac{2^{\frac{\Delta }{2}-2}(\Delta +1)}{N\sqrt{\Delta }}\,\,
		\frac{z^\Delta }{\left[\left(x-x'\right)^2+z^2\right]^\Delta }\,.
	\end{equation}
	
	The action of the D3-brane \eqref{eq:gravaction} expanded to the linear order in perturbations (\ref{fluc}) is
	\begin{equation}
		\delta S_{D3} = \frac{N}{2\pi ^2}\Bigl(\frac{1}{2}\int d^{4}x\,\sqrt{ g}\,g^{\mu \nu } \partial_{\mu } X^{M} \partial_{\nu } X^{N} h_{M N} - \int a^{(4)}\Bigr),
	\end{equation}
	For the metric fluctuation and the 4-form projected on the brane we get:
	\begin{eqnarray}
		h_{\mu \nu }&=&-\frac{2Y^A\eta _{\mu \nu }}{(\Delta +1)z^2}\left[
		2z\,\frac{\partial }{\partial z}+\Delta (\Delta -1)
		\right]s_A+{\rm total~derivative},
		\nonumber \\
		a_{0123}&=&-\frac{4Y^A}{z^3}\,\partial _zs_A.
	\end{eqnarray}
	and the action variation becomes
	\begin{equation}
		\delta S_{D3}=\frac{2(\Delta -1)NY^A}{(\Delta +1)\pi ^2z^4}\,\int_{}^{}
		d^4x\,\left(z\,\frac{\partial }{\partial z}-\Delta \right)s_A.
	\end{equation}
	
	\subsection{One-point function}
	
	To compute the one-point function of the chiral primary $\left\langle \mathcal{C}^A(0)\right\rangle$, which corresponds to the spherical harmonics $Y^{A}$, we need to evaluate the supergravity path integral in the presence of the source for $s_A$ at the origin on the boundary of $AdS_5$. To the leading order, we just expand the D-brane action  $\exp(-S-\delta S)$ to the linear order in $s_A$ and replace $s_A$ by the bulk-to-boundary propagator  (\ref{b-to-b}). The propagator connects the point $(x,z_0)$ on the brane to the point $(0,0)$ on the boundary, as in fig.~\ref{gravity-picture}. This  leads to
	\begin{equation}
		\left\langle \mathcal{C}^A\right\rangle=-\frac{2^{\frac{\Delta }{2}-1}(\Delta -1)z_0^{\Delta -3}}{\pi ^2\sqrt{\Delta }}\,\,\left.\frac{\partial }{\partial z}\int_{}^{}\frac{d^4x}{\left(x^2+z^2\right)^\Delta }\right|_{z=z_0}\,.
	\end{equation}
	Integration gives
	\begin{equation}
		\left\langle \mathcal{C}^A\right\rangle=\frac{\left(8\pi ^2v^2\right)^{\frac{\Delta }{2}}}{\sqrt{\Delta }\,\lambda ^{\frac{\Delta }{2}}}\,Y^A(n).
	\end{equation}
	For $Y^{A}(n)=(y\cdot n)^{\Delta}$, this
	coincides precisely with the tree-level expression (\ref{classical}) (after setting $\Delta=L$), confirming that one-point functions of chiral primaries do not receive quantum corrections.
	
	\section{Two-point functions}\label{sec:twopnt}
	We now compute the two-point functions on the Coulomb branch, which allow us to address various physically interesting questions, such as the radius of convergence of OPE and the interplay between the short-distance and the long-distance expansions. In addition, the OPE expansion of the two-point functions produces infinitely many data on the one-point functions, some of which we check against the direct computation in the previous section.  
	\subsection{Warm up: one loop without loops}
	Before proceeding with the general setup, we consider a simple example, the two-point function of the length-two chiral primary:
	\begin{equation}
	    \mathcal{O}=\mathop{\mathrm{tr}}Z^2,
	\end{equation}
    where $Z=\phi_1+i\phi_2$.
	We do not worry about normalization of the operator, postponing this to a later stage in the calculation. In this subsection we assume that $v^i=\delta^{i1}v$ for simplicity. 
	
	Up to order $\mathcal{O}(\lambda^2)$ the two-point function is given by three tree-level diagrams:\\
\begin{figure}[h]
 \centerline{\includegraphics[width=7cm]{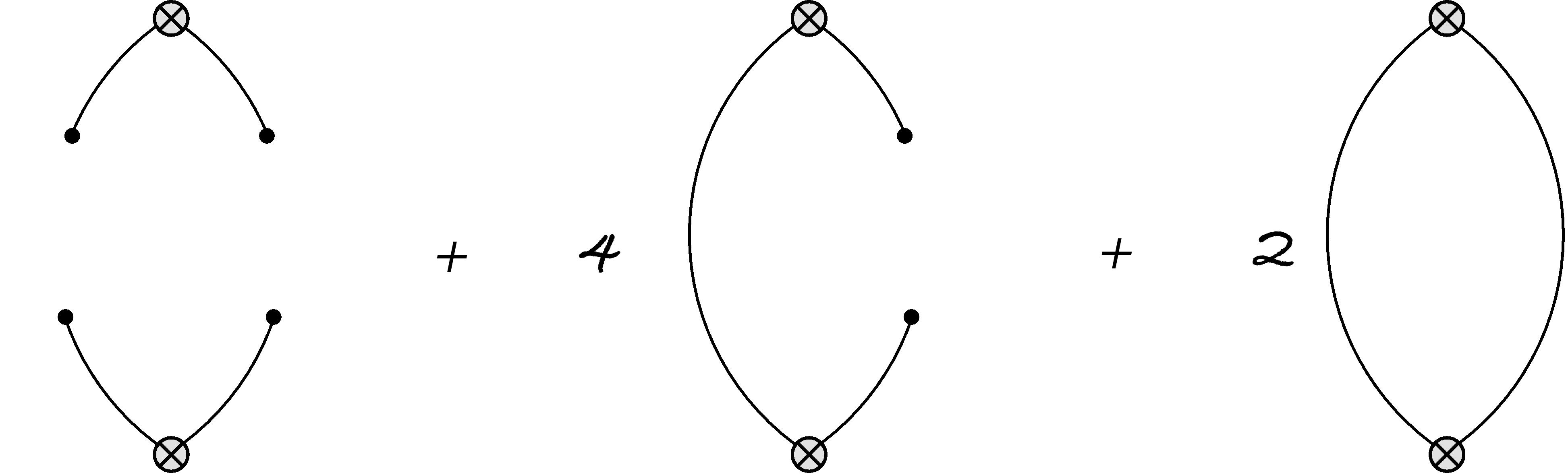}}
\end{figure}
\\
which give:
\begin{equation}
    \left\langle\mathcal{O}^\dagger(x)\mathcal{O}(0)\right\rangle
    =v^4+\frac{\lambda v^2}{\pi^2 x^2N}+\frac{\lambda^2}{8\pi^4x^4}
    +\frac{4\lambda^2}{N}\,D_v(x)^2,
\end{equation}
where $D_v(x)$ is the massive propagator (\ref{massiveGF}) in the coordinate representation. 

From its large-distance asymptotics,
\begin{equation}
    D_v(x)\stackrel{x\rightarrow\infty}{\simeq}
    \sqrt{\frac{v}{32\pi^3x^3}}\,\,{\rm e}\,^{-vx},
\end{equation}
we infer the behavior of the two-point function at infinity:
\begin{equation}
     \left\langle\mathcal{O}^\dagger(x)\mathcal{O}(0)\right\rangle
     \stackrel{x\rightarrow\infty}{\simeq}
     v^4+\frac{\lambda^2}{8\pi^4x^4}
     +\frac{1}{N}\left(\frac{\lambda v^2}{\pi^2 x^2}
     +\frac{\lambda^2v}{8\pi^3x^3}\,\,{\rm e}\,^{-2vx}
     \right).
\end{equation}
The first term violating the cluster decomposition can be taken as a formal definition of the Higgs condensate. This term  is not renormalized by loop corrections due to supersymmetry. 

The second term is the contribution of the unbroken conformal sector and is also tree-level exact. Interestingly, the $1/N$ correction is not entirely exponential either and actually defines the leading behavior of the connected correlator at large distances. The algebraic decay of the correlator beyond the conformal sector is due to the dilaton exchange. As we detail later the coefficient is actually fixed by the Ward identity of the broken conformal symmetry. Finally, the exponential term comes from the exchange of two massive particles. A more refined analysis would have revealed an infinite series of bound states with masses $M_n=2v-E_n$, where $E_n=\mathcal{O}(\lambda^2v)$ is the binding energy, but seeing those requires resummation of infinitely many diagrams.

In the opposite, short-distance limit:
\begin{equation}
\label{eq:massiveshort}
D_v(x)\stackrel{x\rightarrow 0}{\simeq}
    \frac{1}{4\pi^2x^2}+\left(\ln\frac{vx}{2}+\gamma-\frac{1}{2}\right)\frac{v^2}{8\pi^2}
\end{equation}
and for the first two terms of the short-distance expansion we get:
\begin{equation}\label{OOexp}
     \frac{8\pi^4}{\lambda^2}\left\langle\mathcal{O}^\dagger(x)\mathcal{O}(0)\right\rangle
     \stackrel{x\rightarrow 0}{\simeq}
     \frac{1}{x^4}
     +\frac{2v^2}{Nx^2}
     \left(
     \frac{4\pi^2}{\lambda}+\ln\frac{vx}{2}+\gamma-\frac{1}{2}
     \right).
\end{equation}

On the other hand, the short-distance behavior of the correlator is dictated
by the OPE:
\begin{equation}\label{OOth}
  \mathcal{O}^\dagger(x)\mathcal{O}(0)=
  {\rm const}\left[
  \frac{\mathbbm{1}}{x^4}
  +\frac{4}{\sqrt{6}\,Nx^2}\,\widetilde{\mathcal{O}}(0)
  +\frac{2}{\sqrt{3}\,Nx^{4-\Delta_K}}\left(1-\frac{3\lambda}{8\pi^2}\right)K(0)+\ldots\right].
\end{equation}
Here $\widetilde{\mathcal{O}}$ is another chiral primary of dimension two:
\begin{equation}
    \widetilde{\mathcal{O}}=\frac{4\pi^2}{\sqrt{6}\,\lambda}
    \mathop{\mathrm{tr}}\left(3\bar{Z}Z-\phi_i\phi_i
    \right),
\end{equation}
and $K$ is the Konishi operator, all properly normalized. The normalization of $\mathcal{O}$ does not really matter, in other words it is fixed by requiring that the OPE coefficient of the unit operator is one. For Konishi and the new chiral primary we used the known OPE coefficients \cite{Arutyunov:2000ku,Arutyunov:2000im}:
\begin{equation}
    C_{\mathcal{O}^\dagger\mathcal{O}\widetilde{\mathcal{O}}}
    =\frac{4}{\sqrt{6}\,N}\,,\qquad
    C_{\mathcal{O}^\dagger\mathcal{O}K}
    =\frac{2}{\sqrt{3}\,N}\left(1-\frac{3\lambda}{8\pi^2}+\ldots\right).
\end{equation}
The OPE of three chiral primaries is not renormalized and can be easily extracted from the Wick contractions in the three-point function. The OPE coefficient of Konishi receives loop corrections and is actually known to a very high loop order \cite{Georgoudis:2017meq,Goncalves:2016vir}, but for our purposes the one-loop result of \cite{Arutyunov:2000im} will suffice.  

The vacuum expectation value of the chiral primary is again tree-level exact:
\begin{equation}
    \left\langle\widetilde{\mathcal{O}}\right\rangle
    =\frac{8\pi^2v^2}{\sqrt{6}\,\lambda}\,.
\end{equation}
We are now ready to compare the explicit form of the correlator (\ref{OOexp}) to the structure expected from the OPE (\ref{OOth}). Keeping the anomalous dimension of Konishi and its expectation value as unknowns, we find:
\begin{equation}
    \frac{2}{\sqrt{3}}\left[1+\left(\Delta_K-2\right)\ln x-\frac{3\lambda}{8\pi^2}\right]
    \left\langle K\right\rangle
    =2v^2\left(
     \frac{4\pi^2}{3\lambda}+\ln\frac{vx}{2}+\gamma-\frac{1}{2}
     \right),
\end{equation}
where the contribution of $\widetilde{\mathcal{O}}$ has been
subtracted from the right-hand side.
Requiring cancellation of $\ln x$ fixes the anomalous dimension:
\begin{equation}
    \Delta_K-2=\frac{3\lambda}{4\pi^2}\,.
\end{equation}
The expectation value then follows, with the one-loop accuracy:
\begin{equation}
     \left\langle K\right\rangle=\frac{4\pi^2v^2}{\sqrt{3}\,\lambda}
     \left[1+\frac{3\lambda}{4\pi^2}\left(\ln\frac{v}{2}+\gamma\right)\right].
\end{equation}
This agrees well with the explicit calculation (\ref{<K>})-(\ref{c_K}). But we should stress the anomalous dimension and the one-loop expectation value of Konishi are now extracted from the tree-level diagrams without ever computing a single loop integral.

	\subsection{General structures}\label{subsec:generalstructure}
	Let us turn to the general structures of two-point functions on the Coulomb branch. We are mostly interested in the two-point functions of CPOs
	\begin{align}
	    \left\langle\mathcal{C}_{L_1}(x_1,y_1)\mathcal{C}_{L_2}(x_2,y_2)\right\rangle\comma
	\end{align}
	but we will discuss the two-point functions of Konishi operators as well.
	
	The two-point function of CPOs is given by a polynomial of the $R$-symmetry invariants, $(y_1\cdot y_2)$, $(y_1\cdot v)$ and $(y_2\cdot v)$, and a coefficient of each term is a nontrivial function of the distance $|x_{12}|$. As we see shortly, keeping track of the R-symmetry invariants is important in order to disentangle contributions from various diagrams that come with different powers of $\lambda$.
	\paragraph{Large N expansion.}
	Topological classification of the Coulomb branch diagrammatics at large $N$ has three elements: handles, punctures and holes. Punctures correspond to insertions of single-trace operators and holes represent strings ending on the D-brane. In the double-line notation, a hole is a facet with the color index 1 assigned to it, making adjacent propagators massive. A handle costs $1/N^2$, while a puncture or a hole costs $1/N$.
	
	There are two topologies that can contribute to the two-point function at the leading order at large $N$:  a sphere with two punctures and the disconnected diagram with a hole and a puncture in each component. Both diagrams are $\mathcal{O}(1)$. The sphere topology gives the two-point correlator at the conformal point and the disconnected diagram gives the product of the one-point functions:
	\begin{equation}\label{eq:leadinglargeN}
		\left\langle \mathcal{O}_1 (x)\mathcal{O}_2(0)\right\rangle_v
		\overset{N\to\infty}{=}
		\underbrace{\left\langle\mathcal{O}_1\mathcal{O}_2 \right\rangle_{v=0}}_{\text{sphere}}+\underbrace{\left\langle \mathcal{O}_1\right\rangle_v
		\left\langle\vphantom{\mathcal{O}_1} \mathcal{O}_2\right\rangle_v}_{\text{disconnected}}
		\period
	\end{equation}
	When the operators are chiral primaries, these two terms have the definite R-symmetry and spacetime dependence:
	\begin{align}
	    \begin{aligned}
	        \left\langle\mathcal{O}_1\mathcal{O}_2 \right\rangle_{v=0}\propto \delta_{L_1,L_2}\frac{(y_1\cdot y_2)^{L_1}}{|x_{12}|^{2L_1}}\comma\qquad \qquad
	        \left\langle \mathcal{O}_1\right\rangle_v\left\langle \mathcal{O}_2\right\rangle_v\propto (y_1\cdot v)^{L_1}(y_2\cdot v)^{L_2}\period
	    \end{aligned}
	\end{align}

	The leading large $N$ answers \eqref{eq:leadinglargeN} are not particularly interesting from the point of view of the OPE, since they entirely come from the identity operator and the double traces respectively.
	The main nontrivial information in the two-point function sits in the $\mathcal{O}(1/N)$ term, which we call the ``disk correlator" (since they correspond to the worldsheet with a disk topology) and denote by $\left\langle \mathcal{O}_1 (x)\mathcal{O}_2(0)\right\rangle_{v,{\rm disk}}$. This  comes from the sphere with two punctures and one hole, and contains the contributions from single-trace operators in the OPE.
	%This signifies a special role of the unit operator and double traces at large-$N$ which both have $\mathcal{O}(1)$ OPE coefficients. 
	%We want to subtract those, to which end we write:
	\paragraph{R-symmetry structure.} Let us take a close look at the structure of the disk correlator for CPOs. Unlike the leading large $N$ answers \eqref{eq:leadinglargeN}, the disk correlator is given by a sum of different R-symmetry structures:
	\begin{align}\label{eq:rsymstructure}
	    \langle \mathcal{C}_{L_1}\mathcal{C}_{L_2}\rangle_{v,\text{disk}}=(y_1\cdot v)^{L_1}(y_2\cdot v)^{L_2}\sum_{\ell=1}^{{\rm min}(L_1,L_2)}\left(\frac{(y_1\cdot y_2)(v\cdot v)}{(y_1\cdot v)(y_2\cdot v)}\right)^{\ell}g_{L_1,L_2|\ell}\period
	\end{align}
	Here $g_{L_1,L_2|\ell}$ is a nontrivial function of a product $|v||x_{12}|$.
	
	By inspecting the structure of perturbation theory, we find that $g_{L_1,L_2|\ell}$ admits the following expansion
\begin{align}
    g_{L_1,L_2|\ell}=\lambda^{\ell-\frac{L_1+L_2}{2}}\left(\# + \lambda \# +\cdots\right)\period 
\end{align}
As can be seen from this expression, different $g_{L_1,L_2|\ell}$'s start to contribute at different orders of perturbation theory. However, by keeping track of the R-symmetry dependence, we can in principle disentangle them and isolate the leading and the sub-leading contributions to each $g_{L_1,L_2|\ell}$.

\paragraph{Dilaton Ward identity.}The $\ell=1$ contribution deserves a special mention since it contains an exchange of the dilaton multiplet, which gives the leading connected contribution in the large distance limit ($|x|\to \infty$). The structure of the dilaton exchange can be determined by the Ward identity as shown in \cite{Karananas:2017zrg}. In the case at hand, we need a generalization of the formula to the full dilaton multiplet including the exchanges of the R-symmetry Goldstone boson. Following the argument in \cite{Karananas:2017zrg}, it is straightforward to show that the conformal and the R-symmetry Ward identities fix the structure of the two-point functions of scalar operators up to a constant $c$:
\begin{align}\label{eq:dilatonward}
\langle \mathcal{O}_{1}\mathcal{O}_{2}\rangle_{v,{\rm disk}}\overset{|x|\to\infty}{\to}\frac{\left(\Delta_1\Delta_2+c\sum_{I=1}^{5}\hat{R}_I^{(1)}\hat{R}_{I}^{(2)}\right)\langle \mathcal{O}_1\rangle_v\langle \mathcal{O}_2\rangle_v}{f_{\pi}^2v^2|x|^2}\period
\end{align}
Here $f_{\pi}$ is the dilaton decay constant defined by the dilaton form factor of the stress tensor
\begin{align}
    {}_{v}\langle 0|T_{\mu\nu}(p)|\pi\rangle_{v}=\frac{f_{\pi}}{3}vp_{\mu}p_{\nu}\comma
\end{align}
and $\hat{R}_I^{(j)}$ is a broken $R$-symmetry generator acting on the operator $\mathcal{O}_j$. For SO(6) singlet operators such as the Konishi operator, the action of $\hat{R}_I^{(i)}$ vanishes. For chiral primaries they are given by differential operators acting on the polarizations $y$:
\begin{align}
    \hat{R}_I=\frac{(v\cdot y) \partial_{y^{I}}-y^{I} (v\cdot \partial_{y})}{\sqrt{(v\cdot v)}}\qquad \text{for chiral priamries}.
\end{align}
The constant $c$ can be fixed by imposing the superconformal Ward identities. Here instead we take a short cut and make use of the fact that it is determined purely by the symmetry and does not depend on the coupling constants of the theory. Because of these, we can simply check the relation at tree level and read off $c$. This fixes $c$ to be
\begin{align}
    c=1\period
\end{align}

The dilaton Ward identity \eqref{eq:dilatonward} leads to a non-perturbative prediction on the two-point function of chiral primaries\footnote{Here we used the non-renormalization of one-point functions of chiral primaries.}:
\begin{align}
    \langle \mathcal{O}_{L_1}\mathcal{O}_{L_2}\rangle_{v,{\rm disk}}\overset{|x|\to\infty}{\to}(y_1\cdot y_2)(y_1\cdot v)^{L_1-1}(y_2\cdot v)^{L_2-1}\left(\frac{8\pi^2}{\lambda}\right)^{\frac{L_1+L_2}{2}}\frac{\sqrt{L_1L_2}}{f_{\pi}^2x^2}\period
\end{align}
We will later confirm this structure at one loop and conjecture that $f_{\pi}$ is not renormalized.

\subsection{Two-point functions at leading order}
The leading term in the expansion of $g_{L_1,L_2|\ell}$ can be computed by simple Wick contractions of operators.

	In what follows, we depict the color index corresponding to the unbroken $U(1)$ by a different color in the double line notation:
	\begin{center}
		\begin{tikzpicture}
			\draw[] (0,0+0.1) -- (1.5,+0.1) node[xshift=0.5cm, yshift=0.1cm] {\scalebox{0.8}{$a  \neq 1$}};
			\draw[ultramarineblue,thick] (0,-0.1) -- (1.5,-0.1) node[xshift=0.2cm, yshift=-0.07cm] {\scalebox{0.8}{$1$}};
			%-%
			\draw[ultramarineblue,thick] (0+5,+0.1) -- (1.5+5,+0.1) node[xshift=0.2cm, yshift=0.07cm] {\scalebox{0.8}{$1$}};
			\draw[ultramarineblue,thick] (0+5,-0.1) -- (1.5+5,-0.1) node[xshift=0.2cm, yshift=-0.07cm] {\scalebox{0.8}{$1$}};
		\end{tikzpicture}
	\end{center}
\paragraph{$\boldsymbol{\ell=1}$.} 
The leading order contribution for $\ell=1$ is given entirely by the exchange of the dilaton multiplet. For chiral primaries, it reads
\begin{align}
    g_{L_1,L_2|1}^{(0)}=\frac{1}{N}\left(\frac{8\pi^2}{\lambda}\right)^{\frac{L_1+L_2}{2}-1}\frac{\sqrt{L_1L_2}}{v^2 x^2}\period
\end{align}
Here are below the superscripts $(0)$ and $(1)$ mean the leading and the subleading corrections in $\lambda$.

	This tree-level dilaton contribution can be generalized to arbitrary single-trace operators. It takes a particularly simple form when written using a spin-chain representation:
	\begin{equation}\label{eq:treel1}
		\left\langle \mathcal{O}_1^{\dagger} (x)\mathcal{O}_2(0)\right\rangle_{v,{\rm disk}}^{(0)}\overset{\ell=1}{\supset}\,\,\,\frac{1}{N}\,\left(\frac{8\pi ^2v^2}{\lambda }\right)^{\frac{L_1+L_2}{2}-1}\,\frac{\left\langle \Psi _1\right|\Pi_1^{\dagger}\Pi_2\left|\Psi _2\right\rangle}{x^{2}\sqrt{L_1L_2\langle \Psi_1|\Psi_1\rangle\langle \Psi_2|\Psi_2\rangle}}\,,
	\end{equation}
	where $\Pi_j$ is a projector which projects each spin-chain state $|\Psi_j\rangle$ to a single site,
	\begin{equation}
		\Pi_j=\sum_{l=1}^{L_j}\frac{\langle v|}{v}\otimes\ldots \otimes\mathbbm{1}_l\otimes\ldots \otimes \frac{\langle v|}{v}\,,
	\end{equation}
	and the overlap is taken in the single-site Hilbert space.
	
	The result \eqref{eq:treel1} is consistent with the dilaton Ward identity \eqref{eq:dilatonward}. To see this, we just need to decompose $\mathbbm{1}_{l}$ into the direction parallel to $v$ and the directions orthogonal to $v$,
	\begin{align}\label{eq:onedecomp}
	    \mathbbm{1}_{l}=\frac{|v\rangle\langle v|}{v^2}+\sum_{I=1}^{5}|n_{I}\rangle\langle n_{I}|\comma
	\end{align}
	and use the relation
	\begin{align}\label{eq:actionbroken}
	    \hat{R}_{I}\left(\frac{|v\rangle}{v}\otimes\ldots\otimes \frac{|v\rangle}{v}\right)=\sum_{l=1}^{L}\frac{|v\rangle}{v}\otimes\ldots\otimes |n_I\rangle_{l}\otimes \ldots \otimes\frac{|v\rangle}{v}\period
	\end{align}
	The first term in \eqref{eq:onedecomp} gives $\Delta_1\Delta_2$ in \eqref{eq:dilatonward} while the second term gives $\sum_{I}\hat{R}_{I}^{(1)}\hat{R}_{I}^{(2)}$.
	Comparing with the general formula \eqref{eq:treel1}, we find that the dilaton decay constant at this order is given by
	\begin{align}\label{eq:dilatondecay}
	    f_{\pi}^2=\frac{8\pi^2}{g_{\rm YM}^2}\period
	\end{align}
	We will later see that this relation does not receive quantum corrections at one loop.
	\paragraph{$\boldsymbol{\ell>1}$.} For $\ell>1$, the leading contribution is given by a product of two massive propagators $D_v$ and $\ell-2$ massless propagators. Therefore, the result for chiral primaries reads
	\begin{align}
	    g_{L_1,L_2|\ell}^{(0)}=\frac{\sqrt{L_1L_2}}{N}\left(\frac{8\pi^2}{\lambda}\right)^{\frac{L_1+L_2}{2}-\ell}\left(\frac{1}{v^2 x^2}\right)^{\ell-2}\left(\frac{K_{1}(vx)}{v x}\right)^{2}\comma
	\end{align}
	where $K_1(vx)$ is the modified Bessel function. Its short distance expansion is given by
	\begin{align}
	    g_{L_1,L_2|\ell}^{(0)}=\frac{\sqrt{L_1L_2}}{N}\left(\frac{8\pi^2}{\lambda}\right)^{\frac{L_1+L_2}{2}-\ell}\left[\frac{1}{(vx)^{\ell}}+\frac{1}{(vx)^{\ell-2}}\left(\gamma-\frac{1}{2}+\ln\left(\frac{xv}{2}\right)\right)+o(1)\right]\period
	\end{align}
	\subsection{Two-point functions at subleading order for $\ell=1$}\label{subsec:genericD}
	We now discuss the subleading corrections. In this paper we focus on $\ell=1$ for simplicity. However, because of its important connection to the dilaton Ward identity, we perform the computation for arbitrary scalar operators that do not mix with fermions up to two loops. This assumption is valid for chiral primaries of arbitrary lengths as well as for the Konishi and the dimension three operators discussed in section \ref{subsec:example}.

	 For $\ell=1$, there are three diagrams:
	\begin{enumerate}
	\item The self-energy of the dilaton.
		\item The diagram with a cubic scalar vertex (see \eqref{eq:cubicvert}).
		\item The product of the tree-level dilaton exchange and the tadpole (see figure \ref{fig:tadpole}).
        \item The product of two massive propagators. We do not consider this contribution in this subsection, since it is exponentially suppressed at large distances, which is the case discussed here. We take it into account later on. 
	\end{enumerate}
	
	The self-energy correction is inserted into the dilaton propagator between two scalar fields. The rest of the operator is projected onto the VEV. This forces both color indices of the inserted propagator to be equal to $1$. As prescribed by the above topological considerations, the needed power of $N$ comes from setting the indices of fields in the loop to be $1a$, which makes these fields to be massive. We find that the result for the self-energy at positions $i,k$ is given by:
	\newcommand{\seik}{
		\begin{tikzpicture}[baseline={([yshift=-.5ex]current bounding box.center)}]
			\foreach \x in {-1,0}
			\draw[ultramarineblue] (\x+0.1,-1+0.1) --(\x+0.1,-1)   --(\x+0.9,-1) -- (\x+0.9,-1+0.1) ;
			\foreach \x in {-1,0}
			\draw[ultramarineblue] (\x+0.1,1-0.1) --(\x+0.1,1) --(\x+0.9,1) -- (\x+0.9,1-0.1) ;
			\draw[ultramarineblue] (-1-0.1,1-0.1) --(-1-0.1,1) --(-1.5,1) node [xshift=-0.35cm] {$\ldots$};
			\draw[ultramarineblue] (-1-0.1,-1+0.1) --(-1-0.1,-1) --(-1.5,-1) node [xshift=-0.35cm] {$\ldots$};
			\draw[ultramarineblue] (1+0.1,-1+0.1) --(1+0.1,-1) --(+1.5,-1) node [xshift=0.35cm] {$\ldots$};
			\draw[ultramarineblue] (1+0.1,1-0.1) --(1+0.1,1) --(+1.5,1) node [xshift=0.35cm] {$\ldots$};
			
			\node[] (i) at (0,1.25) {$i$};
			\node[] (i) at (0,-1.25) {$k$};

			\draw[ultramarineblue,rounded corners=2pt] (0-0.1,1-0.1) -- (0-0.1,1-0.6) arc (90:270:0.4) -- (0-0.1,-1+0.1);
			\draw[ultramarineblue,rounded corners=2pt] (0+0.1,1-0.1) -- (0+0.1,1-0.6) arc (90:-90:0.4) -- (0+0.1,-1+0.1);
			\draw[] (0,0) circle (0.28);
			\foreach \x in {-1,1}
			\filldraw[fill=blue] [bend right=45, looseness=1.25] (\x+0.1,-1+0.1) to (\x-0.1,-1+0.1);
			\foreach \x in {-1,1}
			\filldraw[fill=blue] [bend right=45, looseness=1.25] (\x-0.1,1-0.1) to (\x+0.1,1-0.1);
		\end{tikzpicture}
	}
	\begin{equation}\label{se}
		\seik= -\frac{\lambda^2}{N}\, S_{vv}(x)\, \delta^{ik}.
	\end{equation}
	Here the blue caps denote the vev and $S_{vv}(x)$ is the following Feynman integral:
	\begin{equation}
		S_{vv}(x)=\int\frac{d^dp}{(2\pi)^d} e^{ipx}S_{vv}(p),  \qquad S_{vv}(p)=\frac{1}{p^2}\int\frac{d^dq}{(2\pi)^d}\frac{1}{((p+q)^2+v^2)(q^2+v^2)}.
		\label{4.2}
	\end{equation}
	The subscript $vv$ indicates that both of the loop propagators are massive.
	Note, that the one loop correction to the $\phi^{i}_{11}$ fields propagator is SO(6) symmetric in the flavor indices. This is due to supersymmetric cancellation of unwanted contributions. The relevant loop integral can be calculated \cite{Boos:1990rg} and is given by:
	\begin{equation}\label{eq:Svv}
		S_{vv}(p)=   \frac{ v^{d-4} }{2^{d} \pi ^{\frac{d}{2}}p^2}\Gamma \left(2-\frac{d}{2}\right)  \, _2F_1\left(1,2-\dfrac{d}{2};\frac{3}{2};-\frac{p^2}{4 v^2}\right)\comma
	\end{equation}
	which, in the position space, reads
	\begin{equation}
\begin{aligned}\label{eq:besselsquared}
S_{vv}(x)&=- \dfrac{v^{d-4}}{4 x^{d-2} (2\pi)^d} \Gamma\left(\frac{2-d}{2}\right)\Gamma\left(\frac{d}{2}\right)+\dfrac{v^{d-4}}{2(d-3)x^{d-2} (2\pi)^d} \times  \\  &\left( (2-d+v^2x^2)\left(K_{\frac{d-2}{2}}(vx)\right)^2+(d-1)vx K_{\frac{d-2}{2}}(vx)K_{\frac{d}{2}}(vx) - (v^2x^2)\left(K_{\frac{d}{2}}(vx)\right)^2 \right).
\end{aligned}
\end{equation}  	
Expanding \eqref{eq:besselsquared} around $d=4-2\epsilon$, we obtain the following expressions for the divergent contribution,  the short distance expansion (UV) and the large distance expansion (IR):
	\begin{equation}\label{s2}
		\begin{split}
			\left.S_{vv}(x)\right|_{\mathrm{div}}&=\frac{1}{4\pi^2 x^2}\cdot \frac{1}{16 \pi ^2 \epsilon },%=\frac{1}{16 \pi ^2 \epsilon }D(x)
			\\
			\left.S_{vv}(x)\right|_{\rm UV}&=
			\frac{1}{32\pi^{4} x^2 }\left( \gamma +1+ \ln (\pi x^2) \right) + o(1),\\
			\left.S_{vv}(x)\right|_{\rm IR}&=\frac{1}{(2\pi)^3 x^2}\frac{ e^{-{2xv}}}{16xv}+o(1)\period
		\end{split}
	\end{equation}

	On the other hand the cubic vertex is inserted at three nodes and the rest of the operator is projected on to the VEV. The way we get the needed power of $N$ is by having massive fields in the loop. For a set of four adjacent vertices, the correction is given by
	\newcommand{\tripodd}{
		\begin{tikzpicture}[baseline={([yshift=-.5ex]current bounding box.center)}]
			\foreach \x in {-1,0,1}
			\draw[ultramarineblue] (\x+0.1,-1+0.1) --(\x+0.1,-1)   --(\x+0.9,-1) -- (\x+0.9,-1+0.1) ;
			\foreach \x in {-1,1}
			\draw[ultramarineblue] (\x+0.1,1-0.1) --(\x+0.1,1) --(\x+0.9,1) -- (\x+0.9,1-0.1) ;
			\draw[rounded corners=2pt,ultramarineblue] (1-0.1,-1+0.1) -- (0.5-0.1,0) -- (0-0.1,1-0.1) ;
			\draw[rounded corners=2pt,ultramarineblue] (1+0.1,-1+0.1) -- (0.5+0.1,0) -- (1+0.1,1-0.1) ;
			
			\draw[ultramarineblue] (2+0.1,1-0.1) --(2+0.1,1) --(+2.5,1) node [xshift=0.35cm] {$\ldots$};
			\draw[ultramarineblue] (2+0.1,-1+0.1) --(2+0.1,-1) --(+2.5,-1) node [xshift=0.35cm] {$\ldots$};

			\draw[ultramarineblue] (-1-0.1,1-0.1) --(-1-0.1,1) --(-1.5,1) node [xshift=-0.35cm] {$\ldots$};
			\draw[ultramarineblue] (-1-0.1,-1+0.1) --(-1-0.1,-1) --(-1.5,-1) node [xshift=-0.35cm] {$\ldots$};

			\node[] (k) at (0,1.25) {$k$};
			\node[] (k) at (1,1.25) {$l$};
			\node[] (k) at (,-1.25) {$j$};
			\node[] (k) at (0,-1.25) {$i$};

			\draw[rounded corners=2pt] (0+0.1,1-0.1) --(0+0.1,1) --(0+0.9,1) -- (0+0.9,1-0.1)-- (1-0.1,1-0.1)-- (0.5,0+0.18) --(0+0.1,1-0.1) ;
			\foreach \x in {-1,0,2}
			\filldraw[fill=blue] [bend right=45, looseness=1.25] (\x+0.1,-1+0.1) to (\x-0.1,-1+0.1);
			\foreach \x in {-1,2}
			\filldraw[fill=blue] [bend right=45, looseness=1.25] (\x-0.1,1-0.1) to (\x+0.1,1-0.1);
		\end{tikzpicture}
	}
	\begin{equation}\label{tripo}
		\tripodd =
		\frac{1}{4}\ \frac{\lambda^2}{N}\, S_{vv}(x)\, \Big[ v^iv^k\delta^{jl}+v^iv^l\delta^{jk}-2v^iv^j\delta^{kl}\Big],
	\end{equation}
	%Collecting contributions from all possible massive fields running in the loop we get:
	%The cubic vertex contribution is given by:
	\newcommand{\tripod}{
		\begin{tikzpicture}[baseline={([yshift=-.5ex]current bounding box.center)},scale=0.6]
			\node at (0,0.5) (A1) {$k$};
			\node at (1,0.5) (A2) {$l$};
			\node at (0,-2.5) (B1){$i$};
			\node at (1,-2.5) (B2){$j$};
			%\draw[very thick,double distance=2pt] (A1)--(B1);
			%\draw (0,0)--(1,-1.1)--(1,-2);
			%\draw (0.2,0)--(1,-0.9)--(1,0);
			%\draw (1.2,0)--(1.2,-2);
			\draw [double,double distance=3pt]
			plot  coordinates{(1,-2) (1,-1) (0,0) }
			plot [smooth] coordinates{(1,0) (1,-1)}
			plot [smooth] coordinates{(-1,0) (0,0) (1,0) (2,0)}
			plot [smooth] coordinates{(-1,-2) (0,-2) (1,-2) (2,-2)}
			plot [smooth] coordinates{(0,-2) (0,-1.5)};
		\end{tikzpicture}
	}
	
	%\begin{equation}\label{tri}
	\newcommand{\tri}{\begin{tikzpicture}[baseline={([yshift=-.5ex]current bounding box.center)},scale=0.25]
			\pgfdeclarelayer{nodelayer}
			\pgfdeclarelayer{edgelayer}
			\pgfsetlayers{nodelayer,edgelayer}
			\tikzstyle{none}=[]
			\begin{pgfonlayer}{nodelayer}
				\node [style=none] (19) at (-6, -16) {};
				\node [style=none] (22) at (0, -16) {};
				\node [style=none] (23) at (-6, -15) {};
				\node [style=none] (24) at (0, -15) {};
				\node [style=none] (25) at (0, -21.25) {};
				\node [style=none] (26) at (-1.5, -21.25) {};
				\node [style=none] (27) at (-5, -21.25) {};
				\node [style=none] (28) at (-6, -21.25) {};
				\node [style=none] (29) at (0, -22.25) {};
				\node [style=none] (30) at (-6, -22.25) {};
				\node [style=none] (32) at (-6, -20.5) {};
				\node [style=none] (33) at (-5, -20.5) {};
				\node [style=none] (34) at (-5, -16) {};
				\node [style=none] (35) at (-1, -16) {};
				\node [style=none] (36) at (-4.7, -15.825) {};
				\node [style=none] (37) at (-1.3, -15.85) {};
				\node [style=none] (38) at (1.5, -21.25) {};
				\node [style=none] (39) at (1.5, -22.25) {};
				\node [style=none] (40) at (-7, -21) {};
				\node [style=none] (41) at (-7.25, -22.25) {};
				\node [style=none] (42) at (-7, -16) {};
				\node [style=none] (43) at (-7, -15.25) {};
				\node [style=none] (44) at (1, -15) {};
				\node [style=none] (45) at (1.25, -16) {};
				\node [style=none] (46) at (-5.5, -19.5) {\(i\)};
				\node [style=none] (47) at (0, -19.5) {\(j\)};
				\node [style=none] (48) at (0, -17) {\(l\)};
				\node [style=none] (49) at (-5.5, -17) {\(k\)};
			\end{pgfonlayer}
			\begin{pgfonlayer}{edgelayer}
				\draw [ticklemepink,bend left=15] (23.center) to (24.center);
				\draw [ticklemepink,bend left=15] (29.center) to (30.center);
				\draw [ticklemepink,in=345, out=-75, looseness=0.50] (26.center) to (27.center);
				\draw [ticklemepink,bend left, looseness=1.25] (19.center) to (26.center);
				\draw [ticklemepink,in=165, out=-165, looseness=0.75] (22.center) to (25.center);
				\draw [ticklemepink,bend right=45, looseness=1.25] (32.center) to (33.center);
				\draw [ticklemepink,bend left=45] (32.center) to (33.center);
				\draw [ticklemepink,in=30, out=-90] (32.center) to (28.center);
				\draw [ticklemepink,in=150, out=-90, looseness=1.25] (33.center) to (27.center);
				\draw [ticklemepink,bend left] (34.center) to (35.center);
				\draw [ticklemepink,bend right=45, looseness=0.75] (36.center) to (37.center);
				\draw [ticklemepink,dotted] (43.center) to (23.center);
				\draw [ticklemepink,dotted, bend left, looseness=0.75] (42.center) to (19.center);
				\draw [ticklemepink,dotted, bend right=90, looseness=0.50] (40.center) to (28.center);
				\draw [ticklemepink,dotted, bend left=15] (41.center) to (30.center);
				\draw [ticklemepink,dotted, bend right=15] (24.center) to (44.center);
				\draw [ticklemepink,dotted, bend left=15] (22.center) to (45.center);
				\draw [ticklemepink,dotted, bend right=15, looseness=0.75] (25.center) to (38.center);
				\draw [ticklemepink,dotted, bend left=15] (29.center) to (39.center);
				\filldraw[fill=blue] [bend right=45, looseness=1.25] (32.center) to (33.center);
				\filldraw[fill=blue] [bend left=45] (32.center) to (33.center);
			\end{pgfonlayer}
	\end{tikzpicture}}
	\noindent Note that there is also a horizontally-flipped diagram in which the roles of the indices $k$ and $i$ are exchanged.
	
	%=
	%   \frac{1}{2}\ \frac{\lambda^2}{N}\, S(x)\, \Big[ v^iv^k\delta^{jl}+v^iv^l\delta^{jk}-2v^iv^j\delta^{kl}\Big],
	%\end{equation}
	%Here again, the $j$ scalar has color indices $11$, while the fields running in the loop are massive hence forming a hole. \\\\
	
	For the two-point functions of operators of lengths $L_1$ and $L_2$, there are $L_1L_2$ ways to draw the self-energy diagrams while there are $2L_1L_2$ ways to draw the cubic-vertex diagrams. Summing and rearranging their contributions, we find that the one-loop correction for the $\ell=1$ sector is given by a sum of the following insertions, where the index $j$ runs from $1$ to $L_1$ and the index $k$ runs from $1$ to $L_2$: 
	\begin{equation}\label{cbtwopoint}
		\left(V_{\mathrm{1-loop}}\right)^{j\, j+1}_{k\, k+1} =  -\frac{\lambda^2}{2N} \, S_{vv}(x)\, \left( v^j v^{j+1} \delta_{k\,k+1} + v_{k} v_{k+1} \delta^{j\,j+1} \right) \period
	\end{equation}
	This structure is reminiscent of the tadpole correction to the one-point function (see \eqref{fig:tadpole}) and we can rewrite this into the following spin-chain representation:
	\begin{align}
	\begin{aligned}
	    &(\text{self-energy})+(\text{cubic})=\\
	    &-\frac{\lambda^2v^{L_1+L_2}}{2N}\frac{S_{vv}(x)}{v^2}\Big(L_1\langle \Psi_1|B\rangle\langle B^{\prime}|\Psi_2\rangle+L_2\langle \Psi_1|B^{\prime}\rangle\langle B|\Psi_2\rangle\Big)\period
	    \end{aligned}
	\end{align}
	Here we are writing the result for the bare operators $\mathcal{O}_{1,2}^{b}$ and the definition of $|B^{\prime}\rangle$ can be found in  \eqref{eq:defBprime}. Using the relation $\langle B^{\prime}| \propto \langle B|\Gamma$ (see \eqref{eq:BprimeB}) with $\Gamma$ being the one-loop mixing matrix, we can simplify the expression to
	\begin{align}
	    \begin{aligned}
	    &(\text{self-energy})+(\text{cubic})=\\
	    &-\frac{8\pi^2\lambda v^{L_1+L_2}}{N}\frac{S_{vv}(x)}{v^2}\Big(\lambda\Delta_1^{(1)}L_2+\lambda\Delta_2^{(1)}L_1\Big)\langle \Psi_1|B\rangle \langle B|\Psi_2\rangle\period
	    \end{aligned}
	\end{align}
	
	Finally the contribution from the product of the tree-level dilaton exchange and the tadpole can be recast into the following expression:
\begin{comment}	
 \begin{align}
	\begin{aligned}
	&(\text{tadpole})\times (\text{dilaton})=\\
	&-\frac{3v^2\lambda^2 \left(\gamma+ \log (\pi x^2)\right)}{32\pi^4 x^2}
	   +\frac{\lambda}{8\pi^2Nv^2x^2}\left[\left((L_1-2)L_2+\sum_{I=1}^{5}\hat{R}_{I}^{(1)}\hat{R}_{I}^{(2)}\right)\langle \left(\mathcal{O}_1^{b}\right)^{\dagger}\rangle_v^{(1)}\langle \mathcal{O}_{2}^{b}\rangle_v^{(0)}\right.\\&\left.
	    +\left(L_1(L_2-2)+\sum_{I=1}^{5}\hat{R}_{I}^{(1)}\hat{R}_{I}^{(2)}\right)\langle \left(\mathcal{O}_1^{b}\right)^{\dagger}\rangle_v^{(0)}\langle \mathcal{O}_2^{b}\rangle_v^{(1)}\right]
		 \end{aligned}
	\end{align}
\end{comment}
\begin{equation}
    \begin{split}
        (\text{tadpole})\times (\text{dilaton})&= \frac{\lambda   \Gamma\left( \frac{d}{2} -1 \right)}{8\pi^{\frac{d}{2}}N v^2x^{2-d}}\left[\left((L_1-2)L_2+\sum_{I=1}^{5}\hat{R}_{I}^{(1)}\hat{R}_{I}^{(2)}\right)\langle \left(\mathcal{O}_1^{b}\right)^{\dagger}\rangle_v^{(1)}\langle \mathcal{O}_{2}^{b}\rangle_v^{(0)}\right.\\&\left.
	    +\left(L_1(L_2-2)+\sum_{I=1}^{5}\hat{R}_{I}^{(1)}\hat{R}_{I}^{(2)}\right)\langle \left(\mathcal{O}_1^{b}\right)^{\dagger}\rangle_v^{(0)}\langle \mathcal{O}_2^{b}\rangle_v^{(1)}\right],
    \end{split}
\end{equation}
where the prefactor is nothing but the dilaton propagator, where we keep it in dimensional regularization to further properly account for operator renormalization. 	To arrive at this expression, we used the decomposition of ${\bf 1}_{l}$  \eqref{eq:onedecomp}, and the action of the broken generator \eqref{eq:actionbroken}. The shift $L_j-2$ is because the tadpole diagram uses two fields and only the remaining $L_j-2$ fields can be connected to the dilaton propagator.
	
	Summing all the three contributions and taking into account the renormalization of operators given in \eqref{Orenormalized}, we find that the result up to one-loop for general operators is given by the following,
	\begin{align}\label{eq:1loop2ptel1}
	\begin{aligned}
	    \langle \mathcal{O}_{1}^{\dagger}(x)\mathcal{O}_{2}(0)\rangle_{v,{\rm red}}^{(0)+(1)}=&\frac{\left(\Delta_1\Delta_2+\sum_{I=1}^{5}\hat{R}_I^{(1)}\hat{R}_{I}^{(2)}\right)\langle \mathcal{O}_1^{\dagger}\rangle_v^{(0)+(1)}\langle \mathcal{O}_2\rangle_v^{(0)+(1)}}{f_{\pi}^2v^2|x|^2}\\&-\frac{\lambda^2}{N}S_{vv}^{\rm sub}(x)\left(\Delta_1^{(1)}L_2+\Delta_2^{(1)}L_1\right)\langle \mathcal{O}_1^{\dagger}\rangle_v^{(0)}\langle \mathcal{O}_2\rangle_v^{(0)}\comma
	    \end{aligned}
	\end{align}
	Here the first term corresponds to the exchange of the dilaton multiplet and $f_{\pi}$ is the dilaton decay constant \eqref{eq:dilatondecay}. On the other hand, the second term comes from the exchange of massive states and $S_{vv}^{\rm sub}$ is a subtracted and rescaled self-energy integral, which does not contain the divergent part or the leading IR part ($\propto 1/x^2$):
	\begin{align}\label{eq:IRfiniteSvv}
	\begin{aligned}
	S_{vv}^{\rm sub}(x)&=\lim_{\epsilon \rightarrow 0}  \left( \dfrac{2(2\pi)^{d-2} x^{d-4}}{v^{d-2}} S_{vv}(x) - \frac{1}{8 \pi ^2 v^2 x^2 \epsilon }  \right)
	&=\frac{v x K_1(v x){}^2-v x K_0(v x){}^2-K_1(v x) K_0(v x)}{4 \pi ^2 v x}.
	\end{aligned}
	\end{align}
	Its leading IR expansion is given by:
 \begin{equation}
    S_{vv}^{\rm sub}(x) \Big|_{\text{IR}} \sim  -\frac{e^{-2 v x}}{8 \pi  v^3 x^3}.
 \end{equation}
	There are two physical implications of the result \eqref{eq:1loop2ptel1}. First the result shows that the dilaton decay constant $f_{\pi}$ is not renormalized at one loop. Later in section \ref{sec:largecharge}, we provide an indirect argument based on the large-charge expansion that this renormalization might hold at all loops. It would be interesting to check it explicitly at higher loops, or to prove it directly from the superconformal Ward identities. Second, since the corrections are proportional to the anomalous dimensions, the $\ell=1$ part of the two-point function of CPOs does not receive one-loop corrections. This implies that the form factors of CPOs for a class of massive states on the Coulomb branch, which are responsible for the second term in \eqref{eq:1loop2ptel1}, must vanish. This is likely due to supersymmetry and it would be an interesting future problem to understand its detailed mechanism.

	\subsection{Testing the result against OPE}
	Having computed the renormalized one-loop correction to the two point functions on the Coulomb branch we can check that our calculations are consistent with the OPE expansion.

	From the conformal operator algebra we know the form of the short distance expansion
	\begin{equation}
		\O_1(x) \O_2(0) = \sum \dfrac{c_{\O_1 \O_2 \O_3} \O_3 (0)}{x^{\Delta_{\O_1}+\Delta_{\O_2}-\Delta_{\O_3}}}\period
	\end{equation}
 Since this is an operator equation, it should hold also in the Coulomb branch i.e.~
 \begin{equation}\label{twptexp}
		\left\langle \O_1(x) \O_2(x) \right\rangle_{v} =  \sum \dfrac{c_{\O_1 \O_2 \O_3} \left\langle \O_3 \right\rangle_v }{x^{\Delta_{\O_1}+\Delta_{\O_2}-\Delta_{\O_3}}}\period
	\end{equation}
	Note that, even if $\mathcal{O}_1$ and $\mathcal{O}_2$ are single-trace operators, the right hand side includes both single-trace and double-trace operators (and higher-trace operators at higher orders in $1/N$). However, all the double-trace operators have dimension larger than or equal to $ \Delta_{\mathcal{O}_1}+\Delta_{\mathcal{O}_2}$ at weak coupling. Thus by focusing on contributions with $\Delta_{\mathcal{O}_3}<\Delta_{\mathcal{O}_1}+\Delta_{\mathcal{O}_2}$, one can isolate the contributions from single-trace operators.
	The non-trivial check is to show that the one-loop corrections to the two-point function, OPE coefficients and the one-point functions are mutually consistent.
	Another intriguing question that can be studying using the expansion \eqref{twptexp} is the convergence and analyticity of the OPE series as well as the interplay between the short distance expansion and the asymptotic expansion of two point functions. We will discuss these points in section \ref{sec:OPE}. \\
	
	For simplicity we will deal with length two operators only, mainly the chiral primary operators (\ref{eq:cpo}):
	\begin{equation}
		\O^A (x) = \frac{\sqrt 2 (2\pi)^2    }{\lambda}C^{A}_{i_1 i_2} \operatorname{tr} \phi^{i_1} \phi^{i_2},
	\end{equation}
	where $C_{i_1 i_2}^{A}$ is symmetric, traceless and normalized $C_{i_1 i_2}^{A}C_{i_1 i_2}^{B}=\delta_{AB}$, and the Konishi operator $K$. Hence we have the following OPE's 
	\begin{equation}
		\begin{split}
			\O^A \times \O^B &\rightarrow \mathbbm{1} +  \sum P^{ABD}\O^D + K + \ldots\ ,
			\\
			K \times \O^A &\rightarrow   \O^A + \ldots\ ,
			\\
			K \times K &\rightarrow \mathbbm{1} +  K + \ldots\ .
		\end{split}
	\end{equation}
	We will use known results for one-loop OPE coefficients  \cite{Okuyama:2004bd,Georgiou:2012zj}:
	\begin{equation}\label{1ope}
		\begin{split}
			&c_{\O^A \O^B \mathbbm{1}} =\delta^{AB},  \qquad c_{KK \mathbbm{1}} =1,
			\qquad c_{K\O^A \mathbbm{1}} =0, \\
			&c_{\O^A \O^B K } = \delta^{AB}\frac{2}{\sqrt 3 N}\left(1-  \frac{3\lambda}{8 \pi^2}\right), \qquad c_{KKK}= \frac{2}{\sqrt 3 N}\left(1-\frac{9\lambda}{8\pi^2}\right).
		\end{split}
	\end{equation}
	Before we present the results let us mention the subtle issue of scheme dependence. Since the OPE coefficient is known by computing three or four point functions, all the quantities in \eqref{twptexp} if taken at hand are divergent at one-loop. After renormalization we are left with finite one-loop corrections, which might depend on the renormalisation scheme if taken carelessly. However, it's straightforward to show, that if the OPE coefficient is computed using the three-point function of renormalized operators, and so are the two-point function and the one-point function, all the scheme dependency cancels out. 
	\\\\
	Finally, we present the result.
	\begin{itemize}
		\item The simplest two-point function between two chiral operators, since they are not renormalized. As before we are interested in the connected part. As we have noticed the divergent loop-correction vanishes due to the traceless condition. Hence, only the finite contribution given by the square of the massive propagator survives, therefore giving:
		\begin{equation}
			\begin{split}
				\frac{\lambda^2}{32\pi^4}\langle \O^A(x) \O^B(y)\rangle_{\mathrm{conn}} &=\frac{\lambda}{N} \left( 2 v^{i}v^{j}\left( C^A_{ik}C^B_{kj} -\dfrac{1}{6}\delta_{ij}\delta^{AB} \right)\frac{1}{4\pi^2(x-y)^2}+\frac{v^2 \delta^{AB}}{3} \cdot \frac{1}{4\pi^2(x-y)^2} \right)\\
				&+\frac{\lambda^2}{N} \bigg(\delta^{AB}D^2_{v}(x-y)
				-\delta^{AB}D^2(x-y) \bigg)+\lambda^2 \frac{\delta^{AB}D^2(x-y)}{2} + o(\lambda^3).
			\end{split}
		\end{equation}
		In the lowest order have separated the contributions, coming from the Konishi operator and from the chiral primary, since one can decompose the emerging traceless tensor into the basis:
		\begin{equation}
			\left( C^A_{ik}C^B_{kj} -\dfrac{1}{6}\delta_{ij}\delta^{AB} \right) = \sum P^{ABD} C^D_{ij} , \qquad P^{ABD}=C^A_{ik}C^B_{kj} C^D_{ji}.
		\end{equation}
		Using short distance expansion of the massive propagator \eqref{eq:massiveshort}, we obtain:
		\begin{equation}\label{OO}
			\langle \O^A(x) \O^B(0)\rangle_{\mathrm{conn}}=\delta^{AB} \dfrac{\langle \mathbbm{1}  \rangle }{x^4}+ \frac{2^{3/2} }{N}\sum_{D}P^{ABD}   \dfrac{\langle\O^D\rangle}{x^2}  +\delta^{AB}\frac{2}{\sqrt 3 N}\left(1-\frac{\gamma_K}{2}\right)\frac{ \langle K\rangle}{x^{2-\gamma_K}} + \ldots \ ,
		\end{equation}
		which is in agreement with the one-loop OPE coefficients \eqref{1ope}. $\gamma_K=\frac{3\lambda}{4\pi^2}$ is one-loop anomalous dimension of Konishi.
		\item Next is the operator product of the Konishi with a chiral primary, which uses the same OPE coefficient. Now, however, the two point function is renormalized, while the emerging one point function is not. In fact the divergence comes from \eqref{s2} and is exactly canceled by considering the renormalized Konishi operator:
		\begin{equation}\label{KO}
			\left\langle K(x) \O^A(0) \right\rangle_{\mathrm{conn}} = \dfrac{2}{\sqrt{3}N}\left( 1-\dfrac{\gamma_K}{2} \right) \dfrac{\langle O^A \rangle}{x^{2+\gamma_K}}+\ldots\ .
		\end{equation}
		\item Finally for the product of two Konishi operators we also use the short distance expansion of the-loop correction \eqref{s2} to obtain:
		\begin{equation}\label{KK}
			\left\langle K(x)K(0) \right\rangle_{\mathrm{conn}} = \dfrac{ \left\langle \mathbbm{1} \right\rangle}{x^4}+ \dfrac{2}{\sqrt{3}N}\left( 1-\dfrac{3\gamma_K}{2} \right)\dfrac{ \left\langle K \right\rangle}{x^{2+\gamma_K}} + \ldots\ .
		\end{equation}
		The divergence from \eqref{s2} is again dealt with by considering renormalized operators. Notice that, the coefficient of the unity operator does not have the anomalous dimension. This happens, because the one loop corrections to this term appear at higher order in the two-point function. Mainly, the whole two-point function \eqref{KK} contains terms $O(1/\lambda)$ and $O(1)$, while the anomalous dimension in the unity operator piece would require calculation of order $O(\lambda)$. This is a general phenomenon: to restore a fixed loop order of a shorter operator would require computing more loops in the two-point function of a longer operator. This also work in reverse, vevs of longer operators in two-point functions of shorter operators appear with higher loop accuracy, that the initial two-point function.
	\end{itemize}

 %--%
	\section{Large charge and Coulomb branch}\label{sec:largecharge}
\subsection{Large charges in SCFT: review and general remarks}
\paragraph{Large charges in generic CFTs.} Conformal field theories in dimensions greater than $2$ are generally strongly coupled and do not admit perturbative descriptions. One way of making progress is to consider the limit of large quantum numbers, and use the inverse of the quantum numbers as expansion parameters. The idea itself is as old as the WKB expansion of quantum mechanics, but a systematic analysis in conformal field theories with $U(1)$ global symmetry was initiated only recently \cite{Hellerman:2015nra,Monin:2016jmo,Alvarez-Gaume:2016vff}. The simplest and the most important consequence of this analysis is a prediction for the lowest conformal dimension of operators with a given (large) charge $J$:
\begin{align}\label{eq:scalinggeneric}
    \Delta\overset{J\to\infty}{\sim} J^{\frac{d}{d-1}}\period
\end{align}
Here $d$ is the dimension of the spacetime.
To derive this scaling, one first puts the theory on a cylinder $R_t\times S^{d-1}$, where the radius of $S^{d-1}$ is set to $r$. On the cylinder, the operator with charge $J$ and dimension $\Delta$ is mapped to a state with charge $J$ and energy $E=\Delta/r$. One then takes the large volume ($r\to \infty$) limit, in which the cylinder decompactifies and becomes flat space. Generically one expects that the limit is described by a state in flat space with finite energy and charge densities ($\epsilon$ and $j$),
\begin{align}\label{eq:finiteje}
    \epsilon\sim\frac{E}{r^{d-1}}\sim \frac{\Delta}{r^{d}}\sim O(1)\comma\qquad \qquad j\sim \frac{J}{r^{d-1}}\sim O(1)\period
\end{align}
Comparing the two, one obtains the scaling \eqref{eq:scalinggeneric}. The $1/J$ corrections to the scaling \eqref{eq:scalinggeneric} can also be determined using the effective field theory (EFT) methods as shown in \cite{Hellerman:2015nra,Monin:2016jmo,Alvarez-Gaume:2016vff}. In three dimensions, this leads to a prediction of the universal $O(1)$ term,
\begin{align}
    \Delta = \# J^{3/2}+ \# J^{1/2}\underbrace{-0.0937256}+\cdots\period
\end{align}
This structure was verified both in the lattice simulation \cite{Banerjee:2019jpw} and the large $N$ analysis \cite{Alvarez-Gaume:2019biu,Giombi:2020enj}.

\paragraph{Large charges in SCFT.} Although the scaling \eqref{eq:scalinggeneric} applies to generic conformal field theories, it is violated in superconformal field theories by the BPS operators, whose dimensions are linear in $J$. There the argument above fails because of the presence of the vacuum manifold i.e.~zero-energy states in flat space in which scalars charged under the $R$-symmetry acquire nontrivial VEVs. Owing to the vacuum manifold, the large volume limit is given {\it not} by a state with finite energy density \eqref{eq:finiteje}, but instead by a state with zero energy and a finite charge density. 

In a series of papers \cite{Hellerman:2017sur, Hellerman:2018xpi}, Hellerman et al.~turned this expectation into a concrete computational framework: they considered the four-dimensional rank-1 $\mathcal{N}=2$ SCFTs (a canonical example being $\mathcal{N}=4$ SYM with $SU(2)$ gauge group), and analyzed the two-point function of chiral primaries with large charge using the Coulomb branch EFT, which describes the dynamics on the vacuum manifold. The results computed by the EFT are in perfect agreement with the exact results from supersymmetric localization \cite{Hellerman:2017veg,Bourget:2018obm}.

So far most of the analysis in the literature was for correlation functions that can be analyzed by supersymmetric localization. However the relationship between the large charge sector at the conformal point and the theory on the Coulomb branch is expected to hold more generally. Consider a correlation function of two chiral primaries with large charge and $n$ (non-BPS) light operators.  In the radial quantization, this corresponds to the expectation value of $n$ light operators on a large-charge state on $R_t\times S^3$. In general, the state created by the large charge operator produces a complicated profile of scalar fields. However, by taking an appropriate double-scaling limit in which one sends the charge to infinity while bringing the light operators close to each other, one can effectively ``freeze" the scalar VEV and relate it to the correlation functions in the Coulomb branch. Later in this section, we will propose a concrete formula which realizes this physical picture. 

The main goal of this section is to establish such a relation in the {\it planar limit} of $\mathcal{N}=4$ SYM. However, a generalization from the rank-1 SCFT to theories with $U(N)$ gauge group is rather nontrivial since some of the arguments in \cite{Hellerman:2017sur, Hellerman:2018xpi} use special properties of the rank-1 theories:
\begin{itemize}
\item The Coulomb branch moduli is one-dimensional per each scalar field and is parametrized at weak coupling by its eigenvalue:
\beq
\Phi \sim \pmatrix{cc}{a&0\\0&-a}\period
\eeq 
Relatedly, the only allowed Higgsing pattern is $SU(2)\to U(1)$.
\item For a given R-charge $J$, there is an unique operator in the chiral ring\footnote{This follows from the trace relation for the $SU(2)$ matrices
\beq
Z^2=\frac{1}{2}{\rm tr}\left(Z^2\right){\bf 1}\period
\eeq};
\beq
\left({\rm tr}\left(Z^{2}\right)\right)^{J/2}\period
\eeq
\end{itemize}
Thanks to these two properties, it was straightforward to write down a map between the large charge operators and the Coulomb branch moduli in these theories. By contrast, establishing a dictionary between the operator and the moduli in $U(N)$ theories is much more complicated because 
\begin{itemize}
\item The Coulomb branch moduli are multi-dimensional and there exist a variety of Higgsing patterns; $U(N)\to U(1)\times U(N-1)$, $U(N)\to U(2)\times U(N-2)$ etc. 
\item For a given $R$-charge $J$, there are multiple operators in the chiral ring;
\beq
{\rm tr}\left(Z^{J}\right)\comma\ldots\comma \left({\rm tr}\left(Z^2\right)\right)^{J/2}\period
\eeq
\end{itemize}
Obviously, the problem becomes worse at large $N$. Thus, in order to apply the idea of the large charge expansion to the setup discussed in this paper, the first question to be addressed is
\begin{enumerate}
 \item[]{\it Which operator corresponds to the Higgsing pattern $U(N)\to U(1)\times U(N-1)$}?
 \end{enumerate}
 Below, we first study the Gaussian matrix model to gain intuition into this question. We then discuss $\mathcal{N}=4$ SYM and propose a dictionary relating the one-point function on the Coulomb branch and the three-point functions involving large charge operators at the conformal point. Our strategy is to first compare them at tree level and establish a map between the charge and the Coulomb branch VEV. Since both the one-point function and the three-point function are tree-level exact for CPOs, the map is expected to be valid at finite coupling even though the computation is performed at  tree-level. We then conjecture that the dictionary is valid also for non-BPS observables and higher-point functions and provide an argument based on holography.
\subsection{Intuition from Gaussian matrix model}
Consider the Gaussian matrix model,
\beq
Z\equiv \int dM \, e^{-\frac{8\pi^2}{g_{\rm YM}^2}{\rm tr}\left(M^2\right)}\period
\eeq
Here $M$ is an $N\times N$ Hermitian matrix and we denote eigenvalues by $u_k$. In the 't Hooft limit $N\to \infty$ with $g_{\rm YM}^2N=\lambda$ fixed, the eigenvalues of $M$ forms a branch cut with the eigenvalue density
\beq
\rho(u)\equiv \frac{1}{N}\sum_{k}\delta (u-u_k)=\frac{\sqrt{4g^2-u^2}}{2\pi g^2}\qquad\qquad  g^2\equiv \frac{\lambda}{16\pi^2}\period
\eeq
The corresponding resolvent is given by
\beq
R(u)\equiv \frac{1}{N}\sum_{k}\frac{1}{u-u_k}=\frac{u^2-\sqrt{u^2-4g^2}}{2g^2}\period
\eeq
\paragraph{Single- and multi-trace operators with large charge.}
Let us now discuss how the eigenvalue distribution gets deformed by an insertion of some ``large-charge'' operator. The simplest possibility is to insert ${\rm tr}(M^{J})$, which leads to the following eigenvalue integral
\beq
\langle {\rm tr}\left(M^{J}\right)\rangle=\frac{1}{Z}\int \left(\prod_{k=1}^{N}du_k\right)\Delta (u) \left(\sum_k u_k^{J}\right) e^{-\frac{N}{2g^2}\sum_k u_k^2} \comma
\eeq
where $\Delta (u)$ is the Vandermonde determinant $\Delta(u)=\prod_{i<j}(u_i-u_j)^2$. To proceed, we use the fact that the integrand is symmetric under the perumtations of $u_k$'s and rewrite it as
\beq
\begin{aligned}
\langle {\rm tr}\left(M^{J}\right)\rangle=\frac{N}{Z}\int \left(\prod_{k=1}^{N}du_k\right)\Delta (u) \,u_1^{J}\, e^{-\frac{N}{2g^2}\sum_k u_k^2}\period
\end{aligned}
\eeq
In this second expression, the only eigenvalue which feels the effect of the insertion directly is $u_1$. The saddle-point equation for $u_1$ is given by
\beq
\begin{aligned}
\frac{u_1}{g^2}-\frac{2}{N}\sum_{k\neq 1}\frac{1}{u_1-u_k}=\frac{j}{u_1}\comma
\end{aligned}
\eeq
with
\beq
j\equiv \frac{J}{N}.
\eeq
The second term on the left hand side coincides with the definition of the resolvent, and using its large $N$ expression, we obtain
\beq\label{eq:largeNsaddleeq}
\sqrt{u_1^2-4g^{2}}=\frac{jg^2}{u_1}\comma
\eeq
In the large charge limit $j\to \infty$, the solution to \eqref{eq:largeNsaddleeq} reads
\beq\label{eq:matrixfinal}
u_1\sim g\sqrt{j}\period
\eeq
The result \eqref{eq:matrixfinal} shows that the insertion of the large charge operator ${\rm tr}(M^{J})$ shifts one of the eigenvalues ($u_1$) from the branch cut to $u_1\sim g\sqrt{j}$. This effectively breaks the gauge symmetry from $U(N)$ to $U(1)\times U(N-1)$. 

A similar analysis can be performed also for the multi-trace operators. For instance, a double-trace operator ${\rm tr}(M^{J_1}){\rm tr}(M^{J_2})$ can be expressed in terms of eigenvalues as
\beq
{\rm tr}(M^{J_1}){\rm tr}(M^{J_2})\sim \sum_{k=1}^{N}u_k^{J_1+J_2}+\sum_{k\neq l}u_k^{J_1}u_l^{J_2}\period
\eeq
This means that the insertion of this operator corresponds to a superposition of two different patterns of eigenvalue shifts: a shift of a single eigenvalue proportional to $\sqrt{J_1+J_2}$ and shifts of two eigenvalues proportional to $\sqrt{J_1}$ and $\sqrt{J_2}$ respectively. If we want to disentangle these two and realize the latter, we need to consider a linear combination of single-trace and double-trace operators ${\rm tr}(M^{J_1}){\rm tr}(M^{J_2})-{\rm tr}(M^{J_1+J_2})$. One can generalize this argument to find operators corresponding to more general Higgsing patterns.

\paragraph{Symmetric Schur polynomial.} In view of generalization to $\mathcal{N}=4$ SYM, it turns out to be  more convenient to use a different basis of operators called the Schur-polynomial basis  $\chi_{R}(M)$, rather than the single- and multi-trace basis. The Schur-polynomial operator is defined by a sum over permutations (see for instance \cite{Bissi:2011dc})
\beq
\chi_{R}(M)\equiv \frac{1}{J!}\sum_{\sigma\in S_J}\chi_{R} (\sigma) \sum_{i_1,\ldots,j_J}M_{i_1 \sigma_{i_1}}\cdots M_{i_J\sigma_{i_J}}\comma
\eeq
where $R$ is a Young tableau specifying the irreducible representation of $U(N)$ and $\chi_{R}(\sigma)$ is a character for the representation $R$ in $S_J$ for the element $\sigma$. 
In particular, the relevant Schur polynomials for us are the ones that correspond to totally symmetric representations, which we call symmetric Schur polynomials. The symmetric Schur polynomial can be obtained from the generating function
\beq\label{eq:symgenerating}
\frac{1}{\det (t{\bf 1}- M)}=\sum_{k=0}^{\infty}t^{-(N+k)}\chi_{{\rm sym}_k}(M)\period
\eeq
At finite $N$, the symmetric Schur polynomials are given by non-triviel linear combinations of single- and multi-trace operators but, at large $N$, it effectively plays the same role as the single-trace operator as we see below.

 Using the representation \eqref{eq:symgenerating}, we express the symmetric Schur-polynomial operator of charge $J$ as an integral
\beq\label{eq:schurdef}
\chi_{{\rm sym}_J}(M)=\oint \frac{dt \,t^{N+J-1}}{2\pi i }\frac{1}{\det (t{\bf 1}- M)}\period
\eeq
Inserting \eqref{eq:schurdef} into the Gaussian matrix model and taking the large $N$ limit, we obtain
\beq\label{eq:integralsymchi}
\left<\chi_{{\rm sym}_J}(M)\right>\simeq \oint \frac{dt\,t^{N+J-1}}{2\pi i }\exp \left[-\left<{\rm tr}\log (t{\bf 1}-M)\right>\right]\comma
\eeq
where the large $N$ expectation value of ${\rm tr}\log ({\bf 1}-t M)$ is given by
\beq
\left<{\rm tr}\log (t{\bf 1}- M)\right>=N\oint \frac{du}{2\pi i}\log (t- u)R(u)\comma
\eeq
with $R(u)$ being the large $N$ resolvent.
When the charge $J$ is of order $N$, the integral of $t$ in \eqref{eq:integralsymchi} can be approximated by the saddle point
\beq
\begin{aligned}
\left(1+j\right)\frac{1}{t}=\oint \frac{du}{2\pi i}\frac{1}{t-u} R(u)\period
\end{aligned}
\eeq
Evaluating the right hand side by deforming the contour, we get
\beq
(1+j)\frac{1}{t}=\frac{t-\sqrt{t^2-4g^2}}{2g^2}\quad \iff \quad j= \frac{t-\sqrt{t^2-4g^2}}{t+\sqrt{t^2-4g^2}}\period
\eeq
In the limit $j\gg 1$, we find that the saddle-point of the $t$-integral is given by\footnote{Precisely speaking, there is another saddle point $t\sim g/\sqrt{t}$, but its contribution is sub-dominant.}
\beq\label{eq:largechargT}
t\sim  g\sqrt{j}\period
\eeq
Let us now discuss the physical interpretation of this result. The generating function $1/\det (t{\bf 1} -M)$ is divergent when an eigenvalue of $M$ is at the position $t$. Thus, if we insert it in the matrix integral, it essentially pins one\footnote{The factor $1/\det (t{\bf 1}-M)$ is divergent also when several eigenvalues are at $t$, but such contributions vanish owing to the Vandermonde factor $\prod_{j<k}(u_j-u_k)$.} of the eigenvalues at $u=t\sim g\sqrt{j}$. As we saw in \eqref{eq:matrixfinal}, this is essentially the same as what we get from the insertion of the single-trace operator ${\rm tr}[M^{J}]$, which means that  the symmetric Schur polynomial realizes the symmetry breaking pattern $U(N)\to U(1)\times U(N-1)$. 
\paragraph{Summary.} In the Gaussian matrix model, both the single-trace operator ${\rm tr}(M^{J})$ and the symmetric Schur polynomial effectively shifts one of the eigenvalues to a ``Coulomb branch'' $u\sim g\sqrt{j}$.
\subsection{Comparison of tree-level correlation functions} We now compare the one-point function in the Coulomb branch and the three-point functions with two symmetric Schur polynomial operators at tree level, and establish a map between the Coulomb branch VEV and the charge of the Schur polynomial operators. The three-point function of our interest is 
\beq
\left<\chi_{{\rm sym}_J}((y_1\cdot\phi) (x_1))\,\,\chi_{{\rm sym}_J}((y_2\cdot\phi) (x_2))\,\,\mathcal{O}(x)\right>\period
\eeq
where $\mathcal{O}$ is a single-trace operator, not necessarily BPS. For now, we assume that it consists only of scalar fields. (A generalization to other operators will be discussed shortly.) To compute it, we use a semiclassical approach\footnote{See \cite{Budzik:2021fyh} for the application of the method to twisted holography.} developed in \cite{Jiang:2019xdz,Jiang:2019zig,Chen:2019gsb,Chen:2019kgc,Yang:2021hrl}, which we review below. 

Let us first consider the two-point functions of $\chi_{{\rm sym}_J}$'s. As in the Gaussian matrix model, the starting point is to  express the Schur polynomials as
\beq\label{eq:tintegralprojection}
\begin{aligned}
\chi_{{\rm sym}_J}((y_{1,2}\cdot \phi)(x_{1,2}))&=\oint \frac{dt_{1,2} \,t_{1,2}^{N+J-1}}{2\pi i }\frac{1}{\det (t_{1,2}{\bf 1}- (y_{1,2}\cdot \phi))}\comma
\end{aligned}
\eeq
To proceed, we write $1/\det$'s in terms of Gaussian integrals of auxiliary bosons $\varphi$ and $\bar{\varphi}$
\beq
\begin{aligned}
\frac{1}{\det (t_{1,2}{\bf 1}-(y_{1,2}\cdot \phi))}&=\int d\bar{\varphi}_{1,2} d\varphi_{1,2} \exp\left[-\bar{\varphi}_{1,2} (t_{1,2}{\bf 1}-(y_{1,2}\cdot \phi))\varphi_{1,2}\right]\period
\end{aligned}
\eeq 
Inserting this expression into the path integral of $\mathcal{N}=4$ SYM and integrating out $\mathcal{N}=4$ SYM fields at tree level, we get the quartic action of $\varphi$'s (see section 3 of \cite{Jiang:2019xdz} for details):
\beq
\int d\bar{\varphi}_{k}d\varphi_k \exp \left[\frac{g^2}{N}\sum_{i\neq j}\frac{y_i\cdot y_j}{x_{ij}^2}(\bar{\varphi}_i\varphi_j)(\bar{\varphi}_j\varphi_i)\right]\exp\left[-\sum_{k}t_{k}(\bar{\varphi}_k\varphi_k)\right]\period
\eeq
Here again $g=\sqrt{\lambda}/(4\pi)$.
We then perform the Hubbard-Stratonovich transformation by integrating-in auxiliary fields $\rho_{ij}$. As a result we get
\beq
\int d\rho d\bar{\varphi}_kd\varphi_k\exp \left[-\frac{N}{g^2}\sum_{i\neq j}\rho_{ij}\rho_{ji}-2\sum_{i\neq j}(d_{ij})^{\frac{1}{2}}\rho_{ij}(\bar{\varphi}_j\varphi_i)\right]\exp\left[-\sum_{k}t_{k}(\bar{\varphi}_k\varphi_k)\right]\comma
\eeq
with $d_{ij}\equiv \frac{(y_i\cdot y_j)}{x_{ij}^2}$. Finally, integrating out $\varphi$'s, we get the effective action for $\rho$:
\beq
\int d\rho \exp \left[-N\left(\frac{1}{g^2}{\rm Tr}\left[\rho^2\right]+2{\rm Tr}\log (T+2\hat{\rho})\right)\right]\comma
\eeq
with $\hat{\rho}_{ij}\equiv (d_{ij})^{\frac{1}{2}}\rho_{ij}$ and $T={\rm diag}(t_1,t_2)$.
Note that here we regarded $\rho$ as a $2\times 2$ matrix and the trace ${\rm Tr}$ is over these $2\times 2$ matrices. Since the action is proportional to $N$, we can evaluate the integral at the saddle point. The saddle point equation reads
\beq
1-\frac{g^2}{2}\frac{1}{\frac{t_1t_2}{4 d_{12}}-\rho \bar{\rho}}=0\comma
\eeq
where $\rho\equiv \rho_{12}$ and $\bar{\rho}\equiv\rho_{21}$. Solving this equation, we find that the saddle point is given by
\beq\label{eq:rhosaddle1}
\rho^{\ast}\bar{\rho}^{\ast}=\frac{t_1t_2}{4d_{12}}-\frac{g^2}{2}\period
\eeq
Here and below we put asterisks to the solutions to the saddle-point equations.
The saddle-point value of the action is
\beq\label{eq:saddlerhointe}
\begin{aligned}
&\int d\rho \exp \left[-N\left(\frac{1}{g^2}{\rm Tr}\left[\rho^2\right]+2{\rm Tr}\log (T+2\hat{\rho})\right)\right] \sim e^{-NS_{\rm eff}}\comma\\
&S_{\rm eff}\equiv \frac{ t_1t_2}{2g^2 d_{12}}-1+2\log \left(2g^2d_{12}\right)\period
\end{aligned}
\eeq
The first line of \eqref{eq:saddlerhointe} gives the correlation functions of two inverse determinant operators in the large $N$ limit. In order to extract the correlation functions of symmetric Schur polynomial operators, we then need to perform the integrals of $t_{1,2}$.
Plugging \eqref{eq:saddlerhointe} into the integrals of $t_{1,2}$ in \eqref{eq:tintegralprojection}, we find that the integrals can be approximated by the saddle point,
\beq\label{eq:saddlet1t2}
t_{1}^{\ast}t^{\ast}_{2}=2g^2d_{12}(1+j)\qquad \qquad j\equiv \frac{J}{N}\period
\eeq
Plugging this into \eqref{eq:rhosaddle1}, we obtain
\beq
\rho^{\ast}=\bar{\rho}^{\ast}=g\sqrt{\frac{j}{2}}\period
\eeq

As shown in \cite{Jiang:2019xdz}, by repeating these steps in the presence of the single-trace operator $\mathcal{O}={\rm tr}(\cdots \phi^{I_1}\phi^{I_2}\cdots)$, we obtain a matrix-product representation for the three-point function:
\beq
\mathcal{O}\mapsto {\rm Tr}\left[\cdots M^{I_1}M^{I_2}\cdots\right]\comma
\eeq
with
\beq
\begin{aligned}
M^{I}\equiv& \left.{\rm diag}\left(\frac{g^2 y_1^{I}}{|x-x_1|^2},\frac{g^2 y_2^{I}}{|x-x_2|^2}\right)\cdot \left(-\frac{2}{T+2\hat{\rho}}\right)\right|_{\text{at the saddle point}}\\
=&\frac{1}{d_{12}}{\rm diag}\left(\frac{ y_1^{I}}{|x-x_1|^2},\frac{ y_2^{I}}{|x-x_2|^2}\right)\cdot \pmatrix{cc}{-t_2^{\ast}&2\rho^{\ast}\sqrt{d_{12}}\\2\bar{\rho}^{\ast}\sqrt{d_{12}}&-t_1^{\ast}}\period 
\end{aligned}
\eeq
To proceed, we specialize the positions and R-symmetry polarizations of operators to be
\beq
x_{1}=-x_2\equiv x_0\comma\qquad y_1=\bar{y}_2\equiv y_0\comma\qquad x=0\period
\eeq
We then get
\beq
M^{I}=\frac{2\sqrt{2}g}{|x_0|}{\rm diag}\left(\frac{ y_0^{I}}{\sqrt{y_0\cdot \bar{y}_0}},\frac{ \bar{y}_0^{I}}{\sqrt{y_0\cdot \bar{y}_0}}\right)\cdot \pmatrix{cc}{-e^{i\theta}\sqrt{1+j}&\sqrt{j}\\\sqrt{j}&-e^{-i\theta}\sqrt{1+j}}\period
\eeq
Here we introduced $e^{i\theta}\equiv \sqrt{t_2/t_1}$. Note that the saddle point \eqref{eq:saddlet1t2} does not constrain the ratio $e^{i\theta}$. In other words, $\theta$ parametrizes the flat direction. Therefore, the correct resolved involves the integration over this direction,
\beq
\frac{\langle \chi_{{\rm sym}_{J}}\chi_{{\rm sym}_{J}}\mathcal{O} \rangle}{\langle \chi_{{\rm sym}_{J}}\chi_{{\rm sym}_{J}} \rangle}=\int_{0}^{2\pi}\frac{d\theta}{2\pi}{\rm Tr}\left[\cdots M^{I_1}M^{I_2}\cdots \right]\period
\eeq
This can be viewed as a weak-coupling analog of the orbit average, which was discussed in \cite{Yang:2021kot,Holguin:2022zii} in the computation of correlation functions of operators dual to D-branes at strong coupling.

In the large charge limit $j\gg 1$, the matrices $M^{I}$'s can be simultaneously block-diagonalized by a change of basis:
\beq\label{eq:thetaaverage}
M^{I}\sim \frac{2\sqrt{2j}g}{|x_0|}\pmatrix{cc}{0&e^{-i\theta}\frac{y_0^{I}-\bar{y}_0^{I}}{\sqrt{y_0\cdot \bar{y}_0}}\\0&-\frac{e^{i\theta} y_0^{I}+e^{-i\theta}\bar{y}_0^{I}}{\sqrt{y_0\cdot\bar{y}_0}}}\period
\eeq
Thus in the limit, we can evaluate \eqref{eq:thetaaverage} explicitly as\footnote{Here we removed the minus sign in front of $\frac{e^{i\theta}y_0^{I}+e^{-i\theta}\bar{y}_0^{I}}{\sqrt{2(y_0\cdot \bar{y}_0)}}$ since it can be absorbed into an integral over $\theta$.}
\beq\label{eq:beforecrucial}
\frac{\langle \chi_{{\rm sym}_{J}}\chi_{{\rm sym}_{J}}\mathcal{O} \rangle}{\langle \chi_{{\rm sym}_{J}}\chi_{{\rm sym}_{J}} \rangle}=\int_{0}^{2\pi} \frac{d\theta}{2\pi}\prod_{k=1}^{L}v^{I_{k}}(\theta)
\eeq
with
\beq
v^{I}(\theta)\equiv\frac{4g\sqrt{j}}{|x_0|} \frac{e^{i\theta}y_0^{I}+e^{-i\theta}\bar{y}_0^{I}}{\sqrt{2(y_0\cdot \bar{y}_0)}}\period
\eeq

Now comes the crucial observation: the integrand on right hand side of \eqref{eq:beforecrucial} can be interpreted as the tree-level one-point function on the Coulomb branch with the scalar VEV $v^{I}=v^{I}(\theta)$. The absolute value of the scalar VEV $v$ is related to the charge by
\beq\label{eq:identification}
|v|=\frac{4g\sqrt{j}}{|x_0|}=\frac{\sqrt{j\lambda}}{\pi |x_0|}\period
\eeq
In other words, the three-point function in \eqref{eq:beforecrucial} can be identified with an average (over $\theta$) of a one-point function on the Coulomb branch under the identification \eqref{eq:identification}. Since we performed the computation at tree level, one may worry that the map \eqref{eq:identification} gets modified once we include loop corrections. This however is not the case. Both the three-point function at the conformal point and the one-point function in the Coulomb branch are tree-level exact if $\mathcal{O}$ is half-BPS, and one can verify that the tree-level dictionary \eqref{eq:beforecrucial} and \eqref{eq:identification} correctly maps one to the other. Thus, assuming that there exists a universal relation like \eqref{eq:beforecrucial} between the two, we conclude that the map \eqref{eq:identification} should not be modified by the loop corrections.
\subsection{Formula relating large charge sector and Coulomb branch}
\paragraph{Conjecture.}Based on the tree-level computation we performed in the preceding subsection, we now propose a concrete formula relating the large charge sector and the Coulomb branch:
\begin{align}\label{eq:finalformula3pt}
    \lim_{\substack{r,j\to\infty\\ \frac{\sqrt{j\lambda }}{\pi r}:\text{ fixed}}}\int \frac{d^{3}\vec{n}}{2\pi^2} \,\frac{\langle \chi_{{\rm sym}_{J}}\left((y\cdot \phi)(r\vec{n})\right)\chi_{{\rm sym}_{J}}\left((\bar{y}\cdot \phi)(-r\vec{n})\right)\mathcal{O}(0) \rangle}{\langle \chi_{{\rm sym}_{J}}\left((y\cdot \phi)(r\vec{n})\right)\chi_{{\rm sym}_{J}}\left((\bar{y}\cdot \phi)(-r\vec{n})\right) \rangle}=\int\frac{d\theta}{2\pi}\langle\mathcal{O} \rangle_{v^{I}(\theta)}\period
\end{align}
Here $\mathcal{O}$ is an arbitrary non-BPS operator and $\langle \mathcal{O}\rangle_{v^{I}(\theta)}$ is a one-point function on the Coulomb branch corresponding to $U(N)\to U(N-1)\times U(1)$ with the scalar VEV
\begin{align}\label{eq:dictionaryvI}
    v^{I}(\theta)\equiv \frac{ \sqrt{j\lambda}}{\pi r}\frac{e^{i\theta}y^{I}+e^{-i\theta}\bar{y}^{I}}{\sqrt{2(y\cdot \bar{y})}}\comma \qquad\qquad  |v|=\frac{\sqrt{j\lambda}}{\pi r}=\frac{g_{\rm YM}\sqrt{J}}{\pi r}\period
\end{align}
As compared to the result in the previous subsection \eqref{eq:beforecrucial}, the formula \eqref{eq:finalformula3pt} contains an extra integral of $\vec{n}$ over a unit sphere. 
The integral sets the expectation value of $\mathcal{O}$ with spin to zero, which is required in order for the left hand side to match with the right hand side.
We also conjecture that a similar relation holds for higher-point functions:
\begin{align}\label{eq:finalformulahigherpt}
    \lim_{\substack{r,j\to\infty\\ \frac{\sqrt{j\lambda}}{\pi r}:\text{ fixed}}}\int \frac{d^{3}\vec{n}}{2\pi^2} \,\frac{\langle \chi_{{\rm sym}_{J}}\left((y\cdot \phi)(r\vec{n})\right)\chi_{{\rm sym}_{J}}\left((\bar{y}\cdot \phi)(-r\vec{n})\right)\prod_{j}\mathcal{O}_j(x_j) \rangle}{\langle \chi_{{\rm sym}_{J}}\left((y\cdot \phi)(r\vec{n})\right)\chi_{{\rm sym}_{J}}\left((\bar{y}\cdot \phi)(-r\vec{n})\right) \rangle}=\int\frac{d\theta}{2\pi}\left\langle\prod_{j}\mathcal{O}_j(x_j) \right\rangle_{v^{I}(\theta)}\period
\end{align}

Let us also discuss the range of applicability of these formulas. The conjecture for the three-point function \eqref{eq:finalformula3pt} was based on the computation in the planar limit. Thus, in the most conservative viewpoint, one expects the formulas to hold only in the planar limit. However, the holographic argument that we present below suggests that it holds also at order by order in the (perturbative) $1/N$ expansion. At the moment, we do not have other supporting arguments, and we leave it as an important open problem to check and verify the applicability of the formulas.

\paragraph{Comments on four-point functions.} For the correlation function of two light operators and two large-charge operators, we can restate the limit \eqref{eq:finalformulahigherpt} in terms of the conformal cross ratios. To simplify the analysis, we insert the two light operators at positions $\vec{x}/2$ and $-\vec{x}/2$. Using the standard definition of the conformal cross ratios
\begin{align}
    z\bar{z}=\frac{x_{12}^2x_{34}^2}{x_{13}^2x_{24}^2}\comma\qquad (1-z)(1-\bar{z})=\frac{x_{14}^2x_{23}^2}{x_{13}^2x_{24}^2}\comma
\end{align}
and setting 
\begin{align}
\begin{aligned}
    &\mathcal{O}_1 (x_1)=\chi_{{\rm sym}_J}((y\cdot \phi)(r\vec{n}))\comma\qquad \mathcal{O} _2(x_2)=\mathcal{O}_2(\vec{x}/2)\comma\\
    &\mathcal{O}_3(x_3)=\mathcal{O}_3(-\vec{x}/2)\comma\qquad \mathcal{O}_4(x_4)=\chi_{{\rm sym}_J}((\bar{y}\cdot \phi)(-r\vec{n}))\comma
    \end{aligned}
\end{align}
we find
\begin{align}\label{eq:crossratiolimit}
    z\simeq 1-\frac{2|x|e^{i\varphi}}{r}=1-\frac{2\pi v|x|e^{i\varphi}}{g_{\rm YM}\sqrt{J}}\comma\qquad \bar{z}\simeq 1-\frac{2|x|e^{-i\varphi}}{r}=1-\frac{2\pi v|x|e^{-i\varphi}}{g_{\rm YM}\sqrt{J}}\comma
\end{align}
with
\begin{align}
    \cos\varphi\equiv \vec{n}\cdot \frac{\vec{x}}{|x|}\period
\end{align}
In the second equalities in \eqref{eq:crossratiolimit}, we used the relation between $r$ and $\sqrt{J}$ given by \eqref{eq:dictionaryvI}. 
In the large charge limit $J\to\infty$, both $z$ and $\bar{z}$ approach $1$ at an appropriate rate dictated by \eqref{eq:crossratiolimit}. Such a limit was called the ``macroscopic limit"\footnote{In general, the definition of the macroscopic limit depends on theories. It is defined by \begin{align}
    z,\bar{z}\simeq 1-\frac{\bullet}{J^{\beta}} \qquad (J\to\infty)\comma
\end{align}
with some theory-dependent power $\beta$. For theories without the vacuum manifold, $\beta$ is expected to be given by $1/d$ (with $d$ being the dimension of the spacetime) while, for the free complex scalar, it is given by $1/(d-2)$. Our result shows that $\beta$ for $\mathcal{N}=4$ SYM is identical to that of the free complex scalar in $d=4$.
} 
in \cite{Jafferis:2017zna}. Our conjecture implies that the ratio of correlation functions (on the left hand side of \eqref{eq:finalformulahigherpt}) remains finite in the macroscopic limit. As emphasized also in \cite{Jafferis:2017zna}, this is far from obvious, and by assuming the existence of the macroscopic limit, one can derive constraints on the CFT data, some of which we discuss below.

The four-point function at the conformal point admits expansions into conformal blocks. The two channels that are of particular interest are the $s$-channel ($12\to 34$) and the $t$-channel ($23\to 14$). Let us first discuss the $t$-channel. The $t$-channel expansion gives a series in $(1-z)$ and $(1-\bar{z})$. At large $J$, contributions from descendants are suppressed as was shown in \cite{Jafferis:2017zna}. In addition, the integral over $\vec{n}$ projects out primaries with spin. As a result, we obtain a sum over scalar primaries
\begin{align}\label{eq:lightlightlargecharge}
\begin{aligned}
    \int \frac{d^3\vec{n}}{2\pi^2}\frac{\langle \mathcal{O}_1\mathcal{O}_2\mathcal{O}_3\mathcal{O}_4\rangle}{\langle \mathcal{O}_1\mathcal{O}_4\rangle}&=\frac{1}{|x|^{\Delta_{\mathcal{O}_2}+\Delta_{\mathcal{O}_3}}}\sum_{\mathcal{O}^{\prime}:\text{ scalar}}C_{14\mathcal{O}^{\prime}}C_{23\mathcal{O}^{\prime}}|1-z|^{\Delta_{\mathcal{O}^{\prime}}}\\
    &=\frac{1}{|x|^{\Delta_{\mathcal{O}_2}+\Delta_{\mathcal{O}_3}}}\sum_{\mathcal{O}^{\prime}:\text{ scalar}}C_{14\mathcal{O}^{\prime}}C_{23\mathcal{O}^{\prime}}\left(\frac{2\pi }{g_{\rm YM}\sqrt{J}}\right)^{\Delta_{\mathcal{O}^{\prime}}}\left(v|x|\right)^{\Delta_{\mathcal{O}^{\prime}}}
\end{aligned}
\end{align}
The series can be readily identified with the operator product expansion of the two-point function on the Coulomb branch \eqref{twptexp}. Comparing the two, one obtains the following expression for (an average of) the one-point function of operator $\mathcal{O}^{\prime}$ on the Coulomb branch:
\begin{align}
    \int \frac{d\theta}{2\pi} c_{\mathcal{O}^{\prime}}=\lim_{J\to\infty}\left[C_{14\mathcal{O}^{\prime}}\left(\frac{2\pi}{g_{\rm YM}\sqrt{J}}\right)^{\Delta_{\mathcal{O}^{\prime}}}\right]\period
\end{align}
For the limit on the right hand side to exist, the structure constant of two large-charge operators and $\mathcal{O}^{\prime}$ must scale like\footnote{The scaling in $J$ was already derived in \cite{Jafferis:2017zna}. Here we made it more quantitative by including the dependence on the coupling constant $g_{\rm YM}$.}
\begin{align}
    C_{14\mathcal{O}^{\prime}}\overset{J\to\infty}{\sim}c_{\mathcal{O}^{\prime}}\left(\frac{g_{\rm YM}\sqrt{J}}{2\pi}\right)^{\Delta_{\mathcal{O}^{\prime}}}\period
\end{align}
 This asymptotic behavior can be verified explicitly at tree level using the result in the previous subsection. At finite coupling, this predicts a nontrivial scaling of the OPE coefficient in the large charge limit. We leave it as an interesting future problem to verify this asymptotic behavior at finite coupling, and/or derive it from CFT axioms without assuming our conjectural formula \eqref{eq:finalformulahigherpt}.
 
 Let us next discuss the  $s$-channel expansion. The $s$-channel expansion is applicable when the two light operators are far apart. Thus, one would expect that its large charge limit gives the long-distance expansion of the two-point function on the Coulomb branch, namely the Kallen-Lehman spectral representation. However, it is not so obvious how this is realized in practice: the $s$-channel expansion is a series in $z,\bar{z} \sim 1-2\pi v |x|/(g_{\rm YM}\sqrt{J})$ while the integrand in the Kallen-Lehmann representation is of the form $e^{-m|x|}$. In principle, one can get a function of the form $e^{-\#|x|}$ from the $s$-channel expansion by using the relation
 \begin{align}
    \lim_{J\to\infty} \left(1-\frac{2\pi v|x|}{g_{\rm YM}\sqrt{J}}\right)^{\gamma_{\mathcal{O}^{\prime}}}=\exp\left[-\frac{2\pi v }{g_{\rm YM}}\left(\lim_{J\to\infty}\frac{\gamma_{\mathcal{O}^{\prime}}}{\sqrt{J}}\right)|x|\right]  \comma
 \end{align}
 with $\gamma_{\mathcal{O}^{\prime}}$ being an anomalous dimension of an operator exchanged in the $s$-channel. However, since the masses of particles in the Coulomb branch are $\sim v$, one needs to have an operator with
 \begin{align}\label{eq:explargeJ}
\gamma_{\mathcal{O}^{\prime}}\sim g_{\rm YM}\sqrt{J}\qquad (J\to\infty)\comma
 \end{align}
  in order for this to work. This seems in conflict with what we expect from perturbation theory,
  \begin{align} \label{eq:exppert}
\gamma_{\mathcal{O}^{\prime}}\sim g_{\rm YM}^2 \#+\cdots\period
\end{align}
For $\mathcal{N}=4$ SYM with SU(2) gauge group,
a resolution to this apparent paradox can be found in the analysis of \cite{Caetano:2023zwe}. The paper studied the so-called large charge 't Hooft limit, in which the charge $J$ is sent to infinity with $g_{\rm YM}^2 J$ fixed. There, it was found that there exist operators with dimensions
  \begin{align}
\gamma_{\mathcal{O}^{\prime}}\sim 2\sqrt{1 +\# g_{\rm YM}^2J}-2\period
  \end{align}
  This produces the expected perturbative expansion at finite $J$ \eqref{eq:exppert} while giving the correct asymptotics \eqref{eq:explargeJ} at large $J$. The discussion in this subsection suggests that similar operators should exist also in  theories with higher-rank gauge groups if we use the symmetric Schur-polynomial operators as large charge operators.

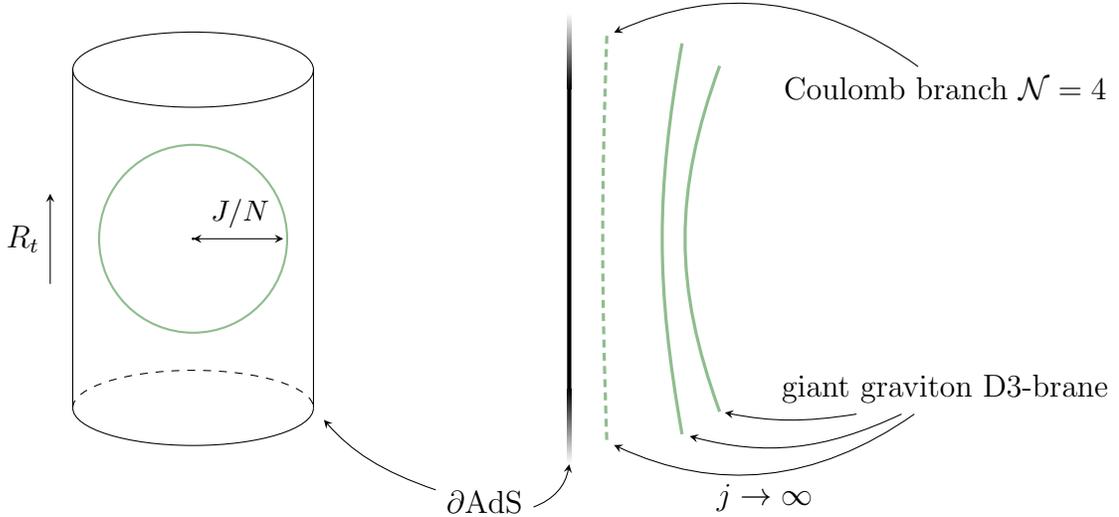
\begin{figure}[t]
    \centering
\definecolor{turquoisegreen}{rgb}{0.63, 0.84, 0.71}
\definecolor{grannysmithapple}{rgb}{0.66, 0.89, 0.63}
\definecolor{darkseagreen}{rgb}{0.56, 0.74, 0.56}
\tikzfading[name=fade out,
            inner color=transparent!0,
            outer color=transparent!100]
            \vspace{0.5cm}
    \begin{tikzpicture}[>=stealth]
    %right part of the picture
        \draw[line width=1.6pt] (0,2) -- (0,-2);
      %   \path[top color=white,bottom color=white,middle color=black]
    (-1,2.3) rectangle ++(-1pt,-4.6);%
     \fill [color=black,path fading=north] (-0.8pt,2)[yshift=-0.1pt] rectangle +(1.6pt,1);
      \fill [color=black,path fading=south] (-0.6pt,-2)[yshift=0.1pt] rectangle +(1.6pt,-1);
        \draw[very thick,color=darkseagreen] (2,2.3)  to[bend right =20] (2,-2.3);
      %  \draw[thick, dashed] (1.5,2)  to[bend right =15] (1.5,-2);
        \draw[very thick, color=darkseagreen] (1.5,2.6)  to[bend right =10] (1.5,-2.6);
        \draw[very thick, densely dashed,color=darkseagreen] (0.5,2.7)  to[bend right =2] (0.5,-2.7);
        \fill[pattern=north west lines, path fading = fade out] (0,3) rectangle ++(-0.2,-6); 
        \node (A) at (-1.125,-3.5) {$\partial$AdS};
        \node (B) at (5,-2) {giant graviton D3-brane};
        \node(C) at (5,2) {Coulomb branch $\mathcal{N}=4$};
        \draw[->] (A) to[bend right=30] (0,-3);
        \draw[->,shorten >=3pt] (B) to[bend left=30]  node[below] {$j \rightarrow \infty$} (0.5,-2.7);
        \draw[->,shorten >=3pt] (B) to[bend left=20] (1.5,-2.6);
        \draw[->,shorten >=2pt] (B) to[bend left=10] (2,-2.3);
        
        \draw[->,shorten >=3pt] (C) to[bend right=30] (0.5,2.7);
    %left part of the picture
    \draw (-5,2.25) ellipse (1.6 and 0.5);
\draw (-5-1.6,2.25) -- (-5-1.6,-2.25);
\draw[->] (-5-1.6-0.3,-0.6)--  node[left] {$R_t$} (-5-1.6-0.3,+0.6);
\draw (-5-1.6,-2.25) arc (180:360:1.6 and 0.5);
\draw [dashed] (-5-1.6,-2.25) arc (180:360:1.6 and -0.5);
\draw (1.6-5,-2.25) -- (1.6-5,2.25); 
\draw[thick, color=darkseagreen] (-5,0) circle (1.25);
\draw[fill=black](-5,0) circle (0.3 pt);
\draw[<->,shorten >=1.5pt] (-5,0) -- node[above] {\small{$J/N$}} (-5+1.25,0);
\draw[->,shorten >=6pt] (A) to[bend left=15]  (1.6-5,-2.25);
    \end{tikzpicture}
        \vspace{0.5cm}
    \caption{\label{fig:D3brane} Schematic of the dual giant graviton in the $j \rightarrow \infty$ limit}
\end{figure}

  Let us also make word of caution: the $s$-channel expansion is only marginally convergent in the large charge limit, in which both $z$ and $\bar{z}$ approach $1$. Thus, in order to make these discussions quantitative using the conformal bootstrap, more works are needed to justify the use of the $s$-channel expansion in the large charge limit.  This is an important open problem, which potentially sheds light on the structure of the OPE data in CFTs with moduli.
\paragraph{Holographic argument.} Holography provides another derivation of the relation between the large-charge limit and the Coulomb branch. The symmetric Schur operator in $\mathcal{N}=4$ SYM is dual to the so-called {\it dual giant graviton} \cite{Grisaru:2000zn}, which is a probe D3-brane solution with the worldvolume $R_t\times S^3$ embedded entirely in AdS$_5$. In the global coordinates
\begin{align}
    ds^2=-\cosh^2\rho dt^2+d\rho^2+\sinh^2\rho \,\,d\Omega_3^2\comma
\end{align}
the solution is static and localized at $\rho=\rho_{\ast}$ satisfying
\begin{align}\label{eq:relchargerho}
    \sinh^2\rho_{\ast}=j=\frac{J}{N}\comma
\end{align}
while it is extended in the three-sphere ($d\Omega_3^2$) directions. This is a spherical solution in the global AdS and has a nontrivial profile in the radial direction $z$ when viewed in the Poincare patch (see figure \ref{fig:D3brane}). However, in the limit $j\to\infty$, the D3-brane approaches the boundary of AdS and the spherical-brane solution can be well-approximated by a flat D3-brane solution localized at constant $z$ in the Poincare patch, which is precisely the solution describing the Coulomb branch used in  section \ref{sec:strong}. This provide  intuitive and geometric understanding of the relation between the Coulomb branch and the large charge sector.

To make the argument more quantitative, let us map the solution to the Poincare patch by the coordinate transformation \cite{Janik:2010gc} (see also \cite{Bissi:2011dc}),
\begin{align}\label{eq:poincaremap}
\begin{aligned}
    z&=\frac{r}{\cosh \rho \cosh t_{E}+n_1\sinh\rho }\comma\\
    x^1&=\frac{r\cosh\rho \sinh t_{E}}{\cosh \rho \cosh t_{E}+n_1\sinh\rho }\comma\quad  x^{2}=\frac{r\,n_2\sinh\rho }{\cosh \rho \cosh t_{E}+n_1\sinh\rho }\comma
    \end{aligned}
\end{align}
where $z$ is the radial coordinate, $t_{E}$ is the Wick-rotated global AdS time, and $n_{1,2,3}$ are the embedding coordinates of $S^3$. After the transformation, it describes the two-point function of dual giant gravitons inserted at $(\pm r,0,0,0)$. Now, to relate this to the solution describing the Coulomb branch, we need to measure the distance between the D3-brane and the AdS boundary near the origin on the boundary $x^{\mu}=0$. This can be done by setting $t_{E}=0$ and $n_1=1$ in \eqref{eq:poincaremap}, which gives $z_{\ast}=re^{-\rho_{\ast}}$. This can be rewritten in terms of the charge of the dual giant graviton using the relation \eqref{eq:relchargerho} as follows:
\begin{align}
z_{\ast} =re^{-\rho_{\ast}}\quad \overset{j\to\infty}{\sim} \quad \frac{r}{2\sqrt{j}}\period
\end{align}
Now, comparing this expression with the relation between the Coulomb branch VEV $v$ and the radial position \eqref{eq:coulombandz0}, we conclude that the dual giant graviton in the large charge limit can be approximated by a Coulomb branch D3-brane solution with
\begin{align}
    |v| = \frac{g_{\rm YM}\sqrt{J}}{\pi r}\period
\end{align}
This is in precise agreement with the gauge-theory analysis \eqref{eq:dictionaryvI}, providing justification of our conjectures \eqref{eq:finalformula3pt} and \eqref{eq:finalformulahigherpt} from holography.

\paragraph{$\mathcal{N}=4$ SYM with SU(2) gauge group.} The discussions so far were for the planar limit. We now propose that the formula similar to \eqref{eq:finalformula3pt} and \eqref{eq:finalformulahigherpt} holds also in the rank-1 theory, namely $\mathcal{N}=4$ SYM with the gauge group SU(2). As mentioned earlier, in the rank-1 theory, there is only a single operator in the chiral ring for a given charge. Thus we can use single-trace operators as the large-charge operators $\mathcal{C}_{J}(x,y)\sim {\rm tr}\left((y\cdot \phi)^{J}\right)(x)$, instead of the Schur-polynomial operators. Then the formula reads
\begin{align}\label{eq:su2formula0}
    \lim_{\substack{r,J\to\infty\\ \frac{g_{\rm YM}\sqrt{J}}{\pi r}:\text{ fixed}}}\int \frac{d^{3}\vec{n}}{2\pi^2} \,\frac{\langle \mathcal{C}_{J}(r\vec{n},y)\mathcal{C}_{J}(-r\vec{n},\bar{y})\prod_{j}\mathcal{O}_j(x_j) \rangle}{\langle \mathcal{C}_{J}(r\vec{n},y)\mathcal{C}_{J}(-r\vec{n},\bar{y}) \rangle}=\int\frac{d\theta}{2\pi}\left\langle\prod_{j}\mathcal{O}_j(x_j) \right\rangle_{v^{I}(\theta)}\comma
\end{align}
where the scalar VEV on the right hand side is given by
\begin{align}\label{eq:su2formula}
  \phi_{\rm cl}^{I}=\pmatrix{cc}{v^{I}(\theta)&0\\0&-v^{I}(\theta)} \comma\qquad \qquad  v^{I}(\theta)\equiv \frac{g_{\rm YM}\sqrt{J}}{\pi r}\frac{e^{i\theta}y^{I}+e^{-i\theta}\bar{y}^{I}}{\sqrt{2(y\cdot \bar{y})}}\comma \qquad  |v|=\frac{g_{\rm YM}\sqrt{J}}{\pi r}\period
\end{align}
Note that the relation between the charge and the scalar VEV is identical to that in the planar limit; the difference is that here we are not taking the 't Hooft limit. Instead we are sending both $J$ and $r$ to infinity with the ratio $\sqrt{J}/r$ fixed while the Yang-Mills coupling $g_{\rm YM}$ is held constant.

The formula \eqref{eq:su2formula0}  can be motivated from semiclassics. Solving the equation of motion of $\mathcal{N}=4$ SYM in the presence of operator insertions $\mathcal{C}_{J}(x_1,y)$ and $\mathcal{C}_{J}(x_2,\bar{y})$, we find a half-BPS solution (cf.~(2.28) of \cite{Hellerman:2017sur})
\begin{align}
    \phi^{I}(x) =\frac{g_{\rm YM}\sqrt{J}}{2\pi}\frac{|x_1-x_2|}{\sqrt{2(y\cdot \bar{y})}}\left(\frac{e^{i\theta}y^{I}}{(x-x_2)^2}+\frac{e^{-i\theta}\bar{y}^{I}}{(x-x_1)^2}\right)\pmatrix{cc}{1&0\\0&-1}\period
\end{align}
Here $\theta$ parametrizes the moduli of solutions and one needs to average over it in order to obtain a correct answer. (It is sometimes called the ``orbit average". See e.g.~\cite{Yang:2021kot,Bajnok:2014sza}.)
Setting $x_1=-x_2=r\vec{n}$ with $r\gg 1$, we find
\begin{align}
\phi^{I}(x)\sim \frac{g_{\rm YM}\sqrt{J}}{\pi r}\frac{e^{i\theta}y^{I}+e^{-i\theta}\bar{y}^{I}}{\sqrt{2(y\cdot \bar{y})}}\pmatrix{cc}{1&0\\0&-1}\period
\end{align}
This can be identified with the scalar VEV in the Coulomb branch given by \eqref{eq:su2formula}, which leads to our formula \eqref{eq:su2formula0}.

Let us end with a couple of comments. First, the argument above relies on semiclassics. For general $\mathcal{N}=2$ SCFTs, there is no reason to trust such an approximation since the coupling is $O(1)$ in the large charge limit and the theory remains strongly coupled. By contrast, in $\mathcal{N}=4$ SYM, one can justify it using the non-renormalization of the three-point functions of half-BPS operators. Namely, once we assume that there is a universal relation between the large charge sector and the Coulomb branch like \eqref{eq:su2formula}, the map between the scalar VEV and the large charge can be determined from semiclassics by analyzing the BPS three-point functions, which are tree-level exact. It is an important future problem to establish a similar formula for general $\mathcal{N}=2$ theories. For that, one may need some non-perturbative input such as  supersymmetric localization \cite{Gerchkovitz:2016gxx}. Second, it should be possible to generalize the  formula to theories with higher-rank gauge groups such as SU(3). However in those theories, one first needs to establish a map between operators and Higgsing patterns. We leave this as an important open problem. Third, our formula relates correlation functions at the conformal point and the Coulomb branch, and bears conceptual similarities to the relation between the superconformal indices and the BPS particle indices on the Coulomb branch, found in \cite{Cordova:2015nma}. To understand the similarities and the differences of the two formulae, it would be helpful to analyze the large charge sector of the superconformal indices and understand in detail its relation to the BPS particles on the Coulomb branch. See \cite{Caetano:2023zwe} for extended discussions on this point for $\mathcal{N}=4$ SYM with SU(2) gauge group.

\section{Convergence and asymptotic growth of OPE}\label{sec:OPE}
	In this section, we discuss the convergence of the operator product expansion. We first discuss lessons from the large charge expansion in the previous section, and discuss the one-loop results to argue that the radius of convergence of the OPE is infinite. We then derive the asymptotic growth of the CFT data.
	\subsection{Convergence of OPE}
	\paragraph{Lessons from large charge expansions.} As we saw in \eqref{eq:lightlightlargecharge}, the light-light OPE of the four-point function reproduces the OPE on the Coulomb branch term by term. In addition, it stays within the radius of convergence as we send positions of large-charge operators to infinity. These facts suggest that the radius of convergence of OPE on the Coulomb branch may be infinite. However to make this argument rigorous, one needs to prove that the limit of $J\to\infty$ commutes with the conformal block expansion. One possible way of achieving this is to derive a universal ($J$-independent) bound on the combination that appears in \eqref{eq:lightlightlargecharge}, 
	\begin{align}
	    C_{14\mathcal{O}^{\prime}}C_{23\mathcal{O}^{\prime}}\left(\frac{2\pi }{g_{\rm YM}\sqrt{J}}\right)^{\Delta_{\mathcal{O}^{\prime}}}\comma
	\end{align}
	by using a conformal block expansion in the crossed channel and the tauberian theorem (see \cite{Qiao:2017xif,Mukhametzhanov:2018zja} for derivations of similar bounds).  We will not pursue this question further and leave it for future work.
	
	\paragraph{Result at one loop.} Instead of presenting a general argument, we may look at the perturbative results to see if they are consistent with the infinite radius of convergence. A simple but instructive example is the product of elementary fields at tree level. In this case the OPE boils down to expanding the massive propagator (\ref{massiveGF}) in $x$, and since the Bessel function has an infinite radius of convergence, so does the OPE.
	
Correlators of generic operators are of course more complicated once the interactions are switched on. However, the one-loop calculation in 
sec.~\ref{subsec:genericD} again results in a  product of the Bessel functions, explicitly given by \eqref{eq:1loop2ptel1},  \eqref{eq:IRfiniteSvv}. Their radius of convergence is manifestly infinite. We expect this to be a generic feature of the Coulomb-branch correlation functions. Anomalous dimensions may generate logarithms in the loop integration, and those should be first absorbed into the $1/|x|^\Delta $ factors, after that we expect the radius of convergence to stay infinite to all loop orders.

	\paragraph{Position space vs.~momentum space.} Let us also make a remark on the convergence of the OPE in momentum space. Unlike the OPE series in position space, the radius of convergence of OPE in momentum space is at best finite, and is very likely to be zero. This simply follows from the fact that the massive propagator in the momentum space $1/(p^2+m^2)$ has a finite radius of convergence around $p\to\infty$. Since we expect to have states with arbitrarily heavy center of mass in the Coulomb branch, it is natural to expect that the radius of convergence is actually zero. 
	
	This demonstrates a clear advantage of using the OPE in position space rather than in momentum space. Historically, the OPE analyses in massive QFTs were performed often in momentum space because of the simplicity of computing Feynman diagrams. In view of our results, it might be interesting to revisit those results from the position space OPE.
	\subsection{Asymptotic growth of OPE data} If the radius of convergence is indeed infinite as conjectured above, we should be able to extract from OPE the long-distance behavior of two-point functions in a spontaneous broken phase. This in turn constrains the asymptotic behavior (or equivalently the ``tail") of the OPE data. Below we present simple sum rules which manifest such a relation. Our results only provide crude estimates for the asymptotics and can potentially be made more rigorous by the use of the Tauberian theorem \cite{Qiao:2017xif,Mukhametzhanov:2018zja}. We will leave it for future investigations.

  At long distance, the two-point function in a symmetry-broken vacuum can be expanded as
 \beq
 \langle \mathcal{O}_i (x)\mathcal{O}_j (0)\rangle_{\rm vac} =\underbrace{\langle \mathcal{O}_i\rangle_{\rm vac} \langle \mathcal{O}_j\rangle_{\rm vac}+\frac{\#}{x^2}}_{=:(\text{\tt power})}+\cdots
 \eeq
 The first term does not depend on the distance $x$ while the second term comes from the dilaton exchange. In general, there are also subleading power-law corrections from multi-dilaton exchanges. To make the discussion simple, below we consider the leading connected contribution in the large $N$ limit (the disk correlator discussed in section \ref{subsec:generalstructure}). For such correlators, multi-dilaton contributions are suppressed by $1/N$ and the next correction comes from a massive particle exchange $\sim e^{-m_0|x|}$ with $m_0$ being the mass of the lightest particle. We thus have
 \beq\label{eq:toprove}
 \langle \mathcal{O}_i (x)\mathcal{O}_j (0)\rangle_{\rm vac}-(\text{\tt power})\overset{|x|\to\infty}{\sim}e^{-m_0|x|}\period 
 \eeq
On the other hand, the two-point function can be expanded at short distance as
\beq\label{eq:OPEremind}
\langle \mathcal{O}_i (x)\mathcal{O}_j (0)\rangle_{\rm vac}=\sum_{k}C(\Delta_k)(vx)^{\Delta_k-\Delta_i-\Delta_j}\comma
\eeq
with 
\beq\label{eq:whatisC}
C(\Delta_k)=c_{ijk}\langle \mathcal{O}_k\rangle_{\rm vac}
\eeq

As is the case with four-point functions of single-trace operators\footnote{See e.g. discussions in \cite{Aharony:2016dwx, Fleury:2016ykk}.}, both single-trace and double-trace operators contribute to the leading connected correlator. We thus have
\beq
\begin{aligned}
&\left.\langle \mathcal{O}_i (x)\mathcal{O}_j (0)\rangle_{\rm vac}\right|_{1/N}=
&\sum_{k:\text{ single}}C(\Delta_k)(vx)^{\Delta_k-\Delta_i-\Delta_j}+\sum_{k:\text{ double}}\left.C(\Delta_k)\right|_{1/N}(vx)^{\Delta_k-\Delta_i-\Delta_j}\,.
\end{aligned}
\eeq
Here $\left.C(\Delta_k)\right|_{1/N}$ denotes the $1/N$ correction to the OPE data \eqref{eq:whatisC} for the double-trace operator\footnote{The $1/N$ correction comes from the correction to the one-point function $\langle \mathcal{O}_k\rangle$. The correction to the OPE coefficient is $O(1/N^2)$.}.
A crucial difference from four-point functions is that the leading connected correlator is $O(1/N)$ rather than $O(1/N^2)$. 
Because of this, there is no contribution from anomalous dimensions of double-trace operators, which produce a logarithmic term $\sim \log x$ and scale as $1/N^2$. Therefore, below we express the OPE at $1/N$ simply as \eqref{eq:OPEremind}
with the understanding that both single- and double-trace operators contribute and, for the double-trace operators, one needs to consider the $1/N$ corrections to $C(\Delta_k)$.

Substituting \eqref{eq:OPEremind} to \eqref{eq:toprove}, we obtain
\beq
\sum_{k}C(\Delta_k)(vx)^{\Delta_k-\Delta_i-\Delta_j}-(\text{\tt power})\overset{|x|\to\infty}{\sim}e^{-m_0|x|}\period
\eeq
Since each term on the left hand side grows like a power law at $|x|\to \infty$, we expect that the exponential suppression on the right hand side comes from a cumulative effect of operators with $\Delta\gg 1$. To capture this effect, we approximate the sum on the left hand side by an integral and assume that it can be approximated by a saddle point $\Delta_{\ast}$:
\beq
\frac{1}{(vx)^{\Delta_i+\Delta_j}}\int d\Delta \,e^{\Delta \log |vx| -f(\Delta)}-(\text{\tt power})\overset{|x|\to\infty}{\sim}\# e^{\Delta_{\ast} \log |vx| -f(\Delta_{\ast})} \,,
\eeq
where $e^{-f(\Delta)}=C(\Delta)$ and the saddle-point is determined by
\beq\label{eq:saddlepointdelta}
\log |vx|-\partial_{\Delta}f(\Delta_{\ast})=0\,.
\eeq
Equating this with the expected behavior at large distance $e^{-m_0|x|}$, we obtain the relation 
\beq\label{eq:legendre}
\Delta_{\ast}\log |vx| -f(\Delta_{\ast})=-\frac{m_0}{v} |vx|\,.
\eeq
A function $f(\Delta)$ satisfying \eqref{eq:saddlepointdelta} and \eqref{eq:legendre} turns out to be
\beq
f(\Delta)=-\Delta+\Delta \log \left(-\frac{v\Delta}{m_0}\right)\period
\eeq
We thus obtain the following estimate for the asymptotic growth of the OPE data:
\beq
C(\Delta)\,\,\overset{\Delta\to\infty}{\sim}\,\,\frac{e^{-\pi i \Delta}\left(m_0/v\right)^{\Delta}}{\Gamma(\Delta)}\,.
\eeq

It would be interesting to check this prediction by computing the OPE data using integrability. Another important direction is to compute the asymptotic growth of the spectral density from the OPE. We leave these for future studies.

	%---------------------%
	
	%---------------------%
	
	%---------------------%
	
	%---------------------%

	\section{Conclusions}\label{sec:conclusion}
    In this paper, we studied correlation functions on the Coulomb branch of planar $\mathcal{N}=4$ SYM, both at weak and strong couplings.

 At weak coupling, we showed that the one-point function can be expressed as overlaps on integrable spin chains between energy eigenstates and a boundary state. We found that the relevant boundary state at tree level is integrable; a hint that integrability of planar $\mathcal{N}=4$ SYM survives in the Coulomb branch. 
 %The spin-chain representation, which we established to the one-loop accuracy, is a natural starting point for the integrability analysis.
    Interestingly, the one-loop correction is structurally very simple, pointing towards underlying simplicity of the full non-perturbative answer. We also found that OPE is a highly efficient tool for computing the vacuum condensates, which effectively saves one order of perturbation theory --- the one-loop correction to the one-point function of Konishi can be extracted entirely from the tree-level diagrams.
	
	At strong coupling, we performed explicit supergravity calculation. The result reconfirms that the one-point functions of chiral primary operators are not renormalized \cite{Skenderis:2006uy};  another sign of underlying simplicity, which may facilitate non-perturbative integrability-based  methods.

    Our results shed light on the interplay between various aspects of planar $\mathcal{N}=4$ SYM. Below we list several future directions worth exploring:
\begin{enumerate}
    \item We proposed the concrete formula relating the correlation functions on the Coulomb branch and the correlation functions with large charge insertions at the conformal point. This strengthens the existing relation between the large charge limit and the physics on the Coulomb branch. It would be interesting to derive a similar formula for general $\mathcal{N}=2$ superconformal field theories.
    \item We provided a holographic argument relating the Coulomb branch and the large charge limit in planar $\mathcal{N}=4$ SYM using a large D3 brane in AdS. It would be interesting to make detailed comparison between the effective action on the large D3 brane in AdS and the large charge effective action in \cite{Hellerman:2017sur, Hellerman:2018xpi}.
    \item Our one-loop results suggest that the radius convergence of OPE on the Coulomb branch is infinite. It would be important to perform the analysis at higher orders and/or study other examples of spontaneous (or explicit) breaking of conformal symmetry to see if this is a universal phenomenon. One potentially interesting example is a fishnet CFT discussed in \cite{Loebbert:2020tje,Karananas:2019fox}.
    \item In this paper, we studied one-loop corrections to two-point functions only for the $\ell=1$ sector. It would be interesting to generalize the analysis to higher-$\ell$ sectors. (Technically, they are more difficult than the $\ell=1$ sector since one needs to compute loop diagrams with massive propagators but they are still tractable \cite{progress:2024}.)
    \item The results for the one-point functions suggest that the Coulomb branch preserves integrability of $\mathcal{N}=4$ SYM. More detailed analysis on the one-point functions from integrability will be presented in the upcoming work \cite{progress:2021}. Another interesting directon to explore is to compute the spectrum of massive states on the Coulomb branch and their scatttering amplitudes  using integrability.
    \item We provided crude estimates of the asymptotic growth of OPE data. It would be interesting to perform more rigorous analysis based on the Tauberian theorem \cite{Qiao:2017xif,Mukhametzhanov:2018zja}.
\end{enumerate}
	\subsection*{Acknowledgements}
	We are particularly grateful to Clay Cordova and Frank Coronado for the participation at the early stage of this project, numerous discussions and the collaboration on related projects.
	We would also like to thank Rob Klabbers, Istv\'an Sz\'ecs\'enyi, Edoardo Vescovi and Sasha Zhiboedov for interesting discussions. SK also thanks Vasco Goncalves and Zahra Zahraee for discussions. 

%\paragraph{Notations for chiral primaries}
%CPO:  $ 	\mathcal{C}_L(x,y),\mathcal{C}^A\right, \mathcal{O}_{L_1},\O^A $\\

\appendix

\bibliographystyle{nb}
\bibliography{CoulombRef}	
	
\end{document}